\documentclass[preprints,review,accept,moreauthors,pdftex]{Definitions/mdpi} 

\firstpage{1} 
\pubvolume{1}
\issuenum{1}
\articlenumber{1}
\pubyear{2022}
\copyrightyear{2022}
\datereceived{} 
\dateaccepted{} 
\datepublished{}
\hreflink{https://doi.org/10.48550/arXiv.2202.07104}
\pdfoutput=1

\usepackage{siunitx}
\usepackage{multicol}

\newcommand{\STAB}[1]{\begin{tabular}{@{}c@{}}#1\end{tabular}}

\usepackage{makecell}

\usetikzlibrary{calc}
\def\checkmark{\tikz\fill[scale=0.4](0,.35) -- (.25,0) -- (1,.7) -- (.25,.15) -- cycle;}

\usepackage[shortcuts]{extdash}
\usepackage[utf8]{inputenc}

\Title{The Hitchhiker's Guide to Fused Twins: A Review of Access to Digital Twins in situ in Smart Cities}

\TitleCitation{Access to Digital Twins in situ in Smart Cities: A Review}

\Author{Jascha Gr\"ubel $^{1,2,3,4,5\dagger}$*\orcidA{}, Tyler Thrash$^{6}$\orcidB{}, Leonel Aguilar $^{1,7}$\orcidC{}, Michal Gath-Morad\orcidD{} $^{8,9}$, Julia Chatain$^{2,10}$\orcidE{}, Robert W. Sumner$^{2}$\orcidF{}, Christoph H\"olscher$^{1}$\orcidG{} and Victor R. Schinazi$^{11}$\orcidH{}}

\AuthorNames{Jascha Gr\"ubel, Tyler Thrash, Leonel Aguilar, Michal Gath-Morad, Julia Chatain, Robert W. Sumner, Christoph H\"olscher and Victor R. Schinazi}

\AuthorCitation{Gr\"ubel, J.; Thrash, T.;Aguilar, L.;  Gath-Morad, M.; Chatain, J.; Sumner, R.W., H\"olscher, C.; Schinazi, V.R.}

\address{%
$^{1}$ \quad Chair of Cognitive Science, ETH Zurich\\
$^{2}$ \quad Game Technology Center, ETH Zurich\\
$^{3}$ \quad Visual Computing Group, Harvard University\\
$^{4}$ \quad Center for Sustainable Future Mobility, ETH Zurich\\
$^{5}$ \quad Chair of Geoinformation Engineering, ETH Zurich\\
$^{6}$ \quad Department of Biology, Saint Louis University\\
$^{7}$ \quad Data Science, Systems and Services Group, ETH Zurich\\
$^{8}$ \quad Department of Architecture, University of Cambridge\\
$^{9}$ \quad Bartlett School of Architecture, University College London\\
$^{10}$ \quad Professorship for Learning Sciences and Higher Education, ETH Zurich\\
$^{11}$ \quad Department of Psychology, Bond University}

\corres{Correspondence: jgruebel@ethz.ch}

\firstnote{Current address: Visual Computing Group, Harvard University} 

\abstract{\emph{Smart Cities} already surround us, and yet they are still incomprehensibly far from directly impacting everyday life.
While current Smart Cities are often inaccessible, the experience of everyday citizens may be enhanced with a combination of the emerging technologies Digital Twins (DTs) and Situated Analytics.
DTs represent their Physical Twin (PT) in the real world via models, simulations, (remotely) sensed data, context awareness, and interactions.
However, interaction requires appropriate interfaces to address the complexity of the city.
Ultimately, leveraging the potential of Smart Cities requires going beyond assembling the DT to be comprehensive and accessible.
Situated Analytics allows for the anchoring of city information in its spatial context.
We advance the concept of embedding the DT into the PT through Situated Analytics to form Fused Twins (FTs).
This fusion allows access to data in the location that it is generated in an embodied context that can make the data more understandable.
Prototypes of FTs are rapidly emerging from different domains, but Smart Cities represent the context with the most potential for FTs in the future.
This paper reviews DTs, Situated Analytics, and Smart Cities as the foundations of FTs.
Regarding DTs, we define five components (Physical, Data, Analytical, Virtual, and Connection environments) that we relate to several cognates (i.e., similar but different terms) from existing literature.
Regarding Situated Analytics, we review the effects of user embodiment on cognition and cognitive load.
Finally, we classify existing partial examples of FTs from the literature and address their construction from Augmented Reality, Geographic Information Systems, Building/City Information Models, and DTs and provide an overview of future directions.
}


\begin{document}


\section{Introduction}

\begin{figure*}[htbp!]
	\centering
	a) \includegraphics[width =0.69\textwidth]{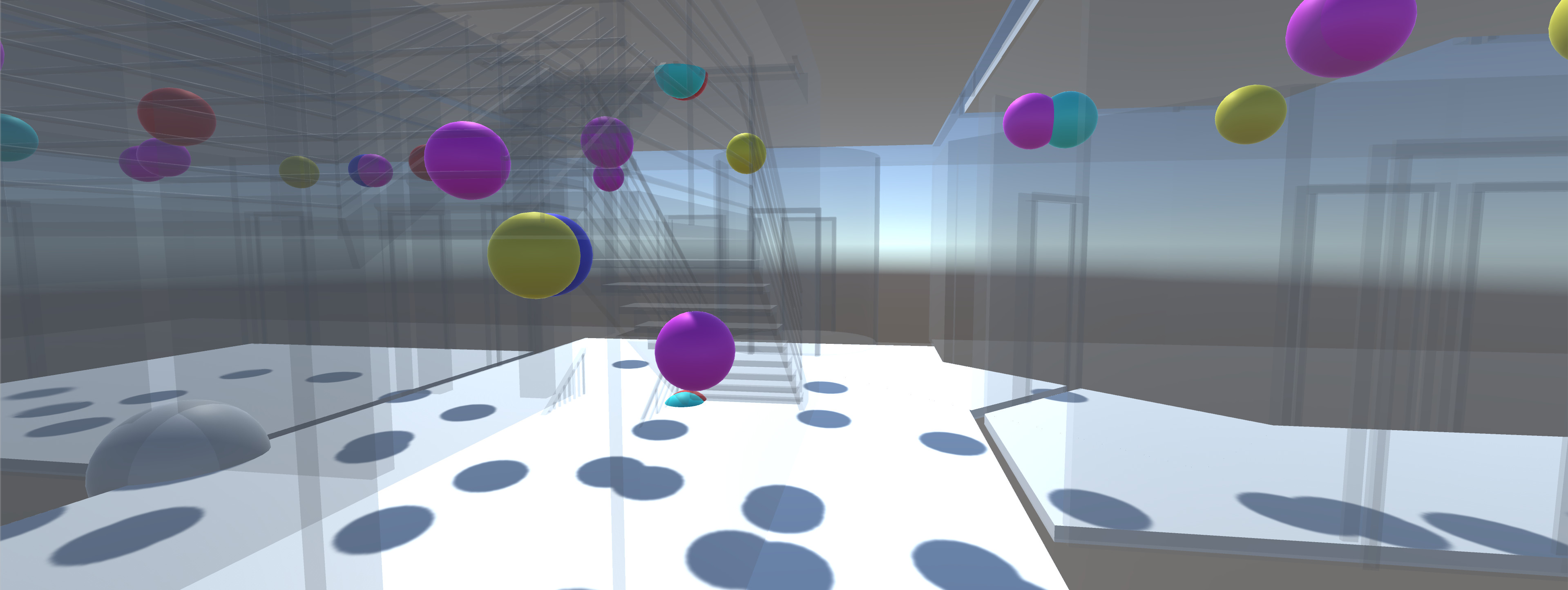}\\
	b) \includegraphics[width =0.69\textwidth]{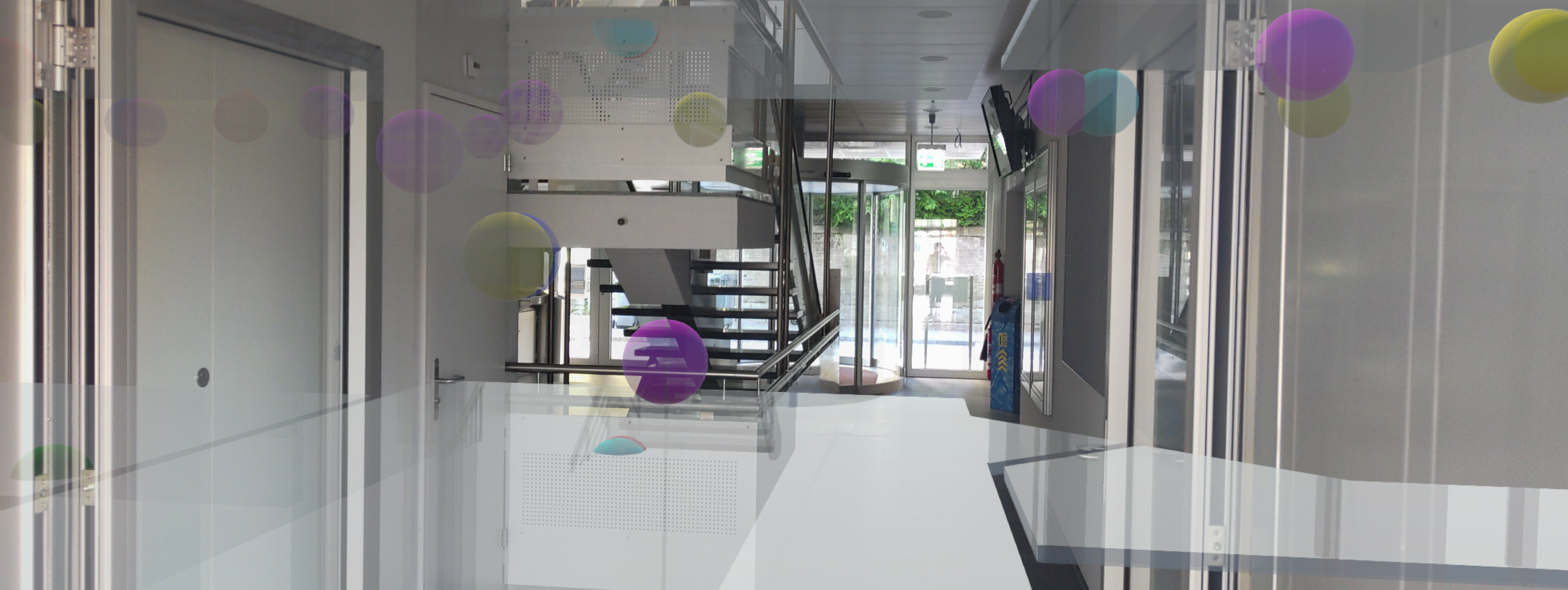}
	\caption{\textbf{Comparison of a Digital Twin  and Fused Twins.} In both rows, the same lobby of a building is shown. a)  the DT of a sensor system with 390 nodes is visualised in Virtual Reality (VR) ~\cite{gruebel2021feasibility, gruebel2016eve}. The white areas visualise the areas of observation and the coloured sphere encode the type of sensor that is located at the respective position in the building. b) the FTs \cite{gruebel2021fused} are visualised in Mixed Reality (MR)~\cite{speicher2019mixed} by embedding the DT \emph{in situ} in its PT thus fusing the two. The fused representation enables Situated Analytics~\cite{thomas2018situated} that facilitates the user understanding of analysis through spatial cues of the environment.}
	\label{fig:fused:twin:teaser}
\end{figure*}


Can Digital Twins (DTs)\cite{grieves2014digital,grieves2017digital} really make cities smart? 
The pressure to transform cities into \emph{Smart} Cities is mounting amongst urban policymakers worldwide~\cite{deakin2011intelligent}. 
Initially, the hope was that \emph{smartification} would be instilled in cities by simply creating citywide Internet of Things (IoT) installations and remote sensing applications~\cite{su2011smart,hung2017leading,deng2021systematic}. 
The idea behind this thinking is simple yet alluring: to have everything in the city attached to the Internet and make it proclaim its own state as if it were a smart object~\cite{sanchez2012adding} remotely observing everything~\cite{su2011smart,deng2021systematic}.
Once observed and connected, everything should become measurable.
In such a smart city, every minute detail, if measured, can be included in the analysis.
This would allow for the uncovering of new insights~\cite{tao2018digital} and give stakeholders a deeper understanding of processes that were previously thought to be fully understood.

However, the complexity of the collected data, the observed processes, and the city itself requires a unifying view. A DT~\cite{batty2018digital} provides a platform on which cities can become really smart by shifting attention from data collection to data use. Here, a digital representation of the city allows for modelling and predicting its ever-changing state via simulations~\cite{kaur2020convergence,melgar2014scalable,jacob2014agent,aguilar2017performance, aguilar2019mass}. Unfortunately, until now, the virtual and physical environments of cities remain insufficiently interconnected~\cite{solomon2021digital}. At best, DTs are remotely displayed in city administration offices to obtain insights~\cite{batty2018digital,lock2019holocity}. However, DTs have the potential beyond administration to inform decision-making on all levels from citizens to politicians and planners~\cite{gruebel2021fused}.
To realise such a wide-ranging implementation of a DT for a Smart City, its data must be accessible \emph{in situ} within its physical context.
Situated visualisations~\cite{white2009sitelens,thomas2018situated} enable sensing and actuating in the environment with which it is combined.  

We introduce the Fused Twins (FTs)\footnote{On a grammatical side note, FTs are always plural because they consist of both the Physical Twin and the Digital Twin and their interaction.} paradigm (see Figure~\ref{fig:fused:twin:teaser}) paradigm to enable the conceptualisation of an accessible Smart City.
We argue that there is a need to clarify the definitions of DTs, Situated Analytics, and Smart cities with an emphasis on the visualisation and development of a practical interface.
The FTs concept aims to embed the DT in its Physical Twin (PT), fusing them via Extended Reality (XR) applications~\cite{gruebel2021fused}.
Throughout this paper, we follow the working hypothesis that \emph{DTs combined with Situated Analytics enable lower cognitive load, enhanced cognition, and easier embodiment; and therefore, FTs can make Smart Cities more accessible to users ranging from managers and policy makers to citizens}.
Ultimately, only when data is accessible where it is most needed and understandable to those who consult it, can DTs fulfil the promise that IoT alone could not: to power Smart City developments by interlinking the physical and virtual environments.

\begin{figure}[htpb!]
	\centering
	\includegraphics[width =0.70\columnwidth]{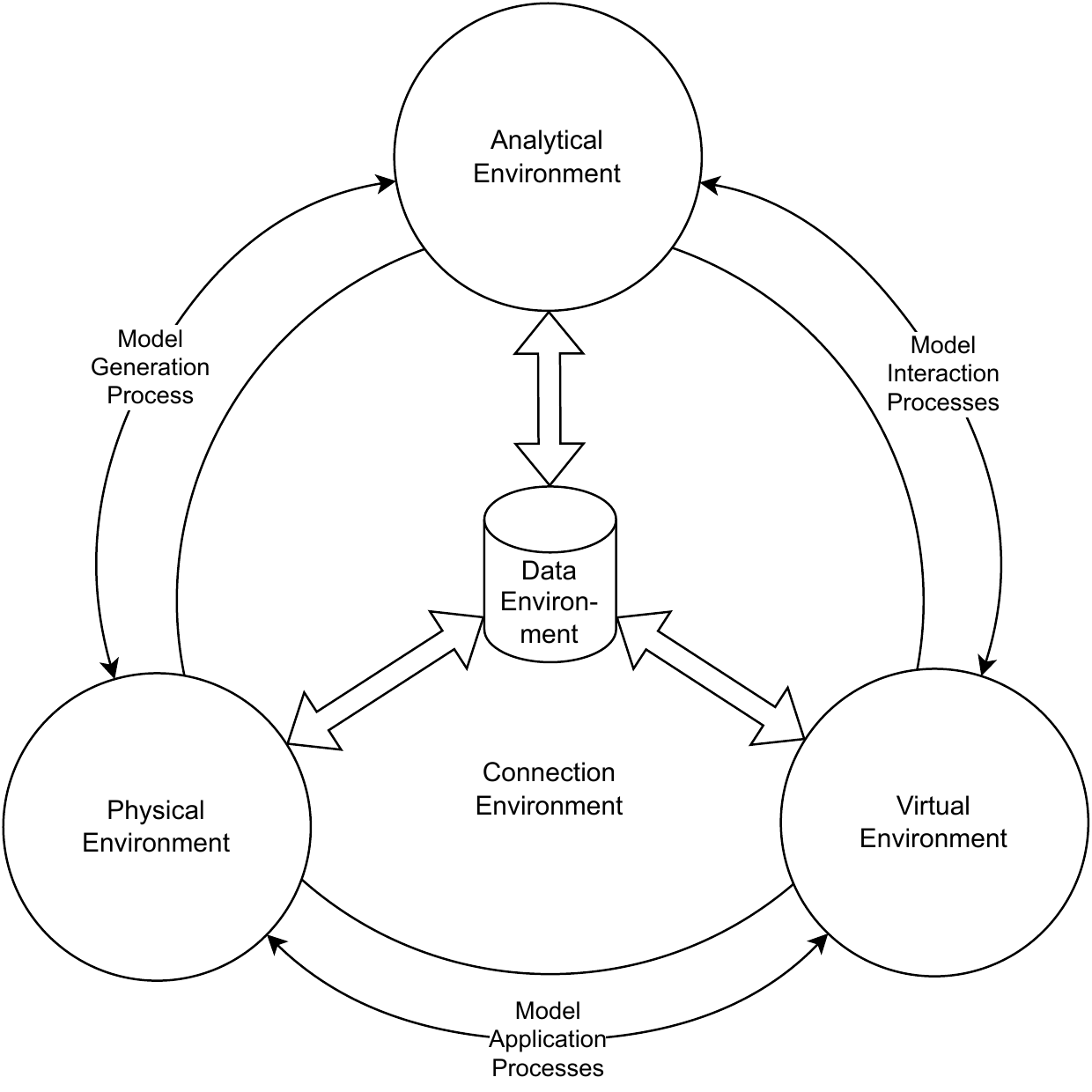}
	\caption{\textbf{Components of a Digital Twin.} A DT is often thought of as the interaction between the PT and the DT~\cite{grieves2017digital}, but any DT implements five core components, even if only partially. These components are the Physical Environment, the Data Environment, the Analytical Environment, the Virtual Environment, and the Connection Environment~\cite{tao2017digital}. The PT is measured through the Physical Environment of the DT, and the raw data is stored in the Data Environment. The Analytical Environment provides simulations, services and automatisation. Users often only perceive the DT represented in the Virtual Environment as a model of the PT without being aware of the involvement of all the other components. The Connection Environment is the invisible glue that holds the different components together and enables composition with other DTs if properly implemented. Generated with \url{http://draw.io}.}
	\label{fig:digital:twin:components}
\end{figure}

This review is structured in four main sections.
To define the term FTs, we need a thorough understanding of Smart Cities, Situated Analytics, and DTs.
The first three sections cover the necessary background to develop, understand, and apply the FTs paradigm.
We dedicate the Section~\ref{sec:smart:cities} to gain an understanding of how Smart Cities arose as a research field and why they converge towards DT research.
Section~\ref{sec:situated:analytics} covers Situated Analytics as a novel data exploration approach for Smart Cities.
The similarity in the implementation structure leads us to the following section on DTs.
The Section~\ref{sec:digital:twin} elaborates on DTs. We provide a definition following \citeauthor{tao2017digital}, where a DT is constructed using five components (see Figure~\ref{fig:digital:twin:components}).
DTs are a vague term that requires multiple clarifications in terms of composability (Subsection~\ref{sec:digital:twin:composition}), servicisation (Subsection~\ref{sec:digital:twin:servicisation}), and differences among cognate terms (i.e., functionally and historically related but slightly different terms \cite{carroll1992cognates}; Subsection~\ref{sec:digital:twin:cognates}). 
Finally, in Section~\ref{sec:fused:twins}, we cover the FTs concept in depth including the required steps to implement FTs and classify examples of \emph{de facto} FTs in the recent literature.
Section~\ref{sec:discussion} provides an in-depth discussion on Smart Cities (Subsection~\ref{sec:discussion:smart:city}), Situated Analytics  (Subsection~\ref{sec:discussion:situated:analyitcis}), DTs  (Subsection~\ref{sec:discussion:digital:twin}), and the newly introduced FTs paradigm (Subsection~\ref{sec:discussion:fused:twins}).

Beyond the definition for FTs, we also provide several new points on a wide range of topics. These topics have often been covered shallowly in shorter research papers which has left a great potential for confusion.
In this review, we try to discern some of these, especially in the context of DTs.



\section{Smart Cities}
\label{sec:smart:cities}

\begin{figure}[hbpt!]
	\centering
	\includegraphics[width =\columnwidth]{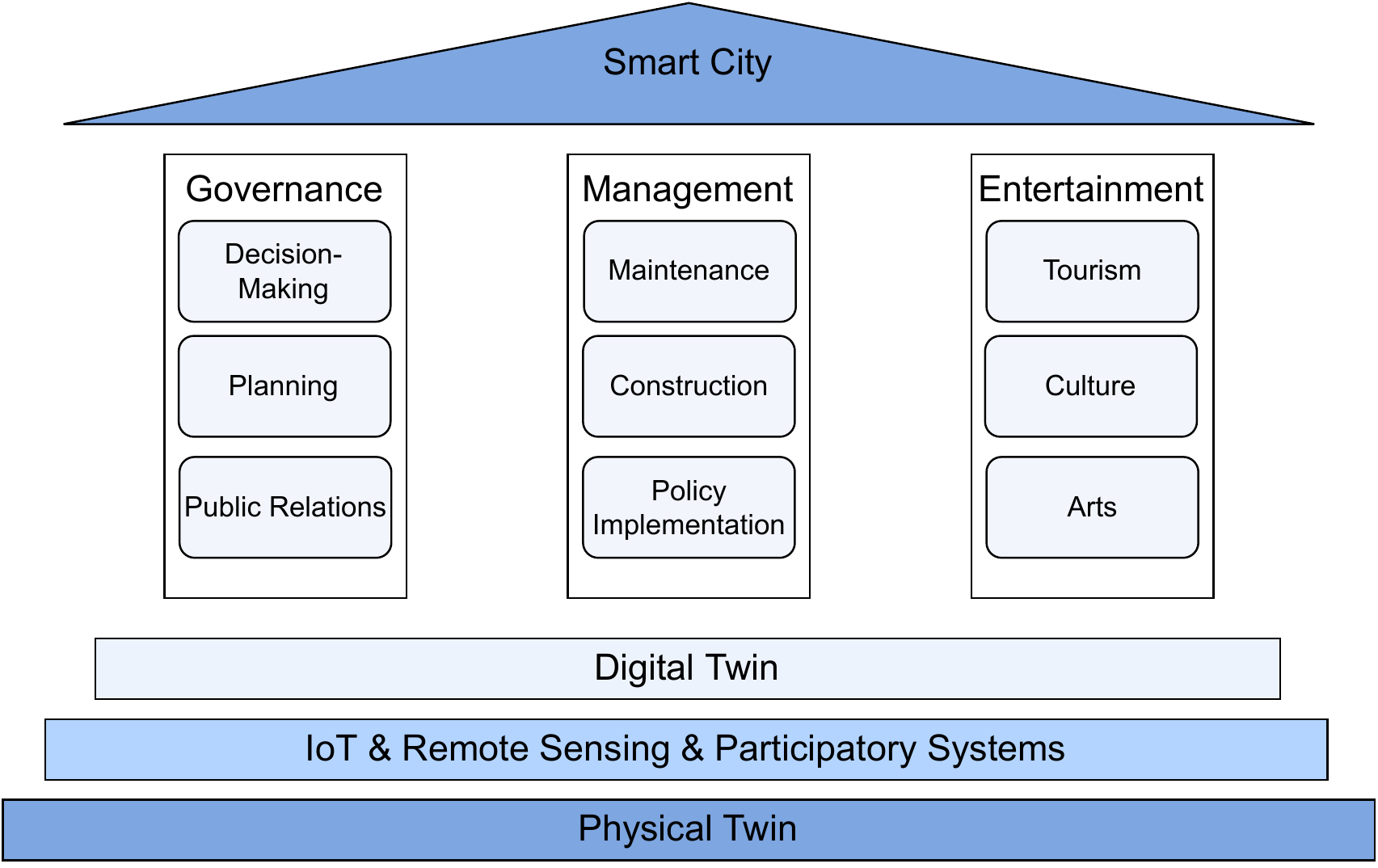}
	\caption{\textbf{The Smart City with a Digital Twin Foundation}. Smart Cities rely on effective data collection, analysis, and communication to work.
	A DT can provide the required processing of the data collection systems to formalise the PT.
	In a Smart City, a DT provides analysis for governance, management, and entertainment. While governance refers to directing the city's future, management refers to the present state of the city, and entertainment refers to the use of the city. Generated with \url{http://draw.io}.}
	\label{fig:twin:city:interplay}
\end{figure}

Smart Cities are an inherently technological approach to resolve urban issues that spawn economic, social, environmental, and governmental aspects~\cite{tompson2017understanding} (see Figure~\ref{fig:twin:city:interplay}).
Smart cities are constituted using bench-marking indicators that can be constructed from technological sources~\cite{vanolo2014smartmentality}.
This technologist approach only provides a narrow perspective~\cite{mattern2017city} that is often divided along disciplinary boundaries~\cite{luque2019developing}. 
The weakness in the Smart City concept is juxtaposed by its necessity because technological advances are one of the few options to manage the complexity of modern cities~\cite{tompson2017understanding}.
However, at the same time, it must be acknowledged that there are other perspectives or ``urban imaginaries'' \cite{huyssen2008other} beyond Smart Cities such as resilient cities~\cite{newman2009resilient}, intelligent cities ~\cite{vanolo2014smartmentality}, and sustainable cities~\cite{haughton2004sustainable}.
In the reminder of this section, we will expand on the evolution of ``smart cities'' from the nineteenth century all the way to its portrayal in contemporary urban discourse. 

Throughout their evolution, advances in ICT have repeatedly helped cities to cope with growing complexity~\cite{castells2011rise,townsend2013smart,gath2016smart}.
During the nineteenth century, industrial cities were faced with the management challenges of large industrial enterprises as well as new administration challenges due to unprecedented population growth.
The need to process information and communicate it quickly over great distances led to the invention of the telegraph, which enabled the industrial city to keep growing~\cite{castells2011rise,gottmann1977megalopolis}.

This transformation, from a society in which communication was impeded by distance to a society in which distance is addressed using ICTs, has forever changed humanity and cities ~\cite{gottmann1977megalopolis}.
On the one hand, critical obstacles to the growth of cities were removed with the integration of technologies that enabled businesses to flourish and municipalities to govern more effectively.
On the other hand, futurists such as  Toffler~\cite{toffler1981third}and Negroponte~\cite{negroponte1995being} predicted that these technologies will lead to the ``death of distance'' because it is possible that improved telecommunications technologies will make cities, and space itself, obsolete.
Their basic proposition was that electronics will eliminate the need for face-to-face interactions and that the cities at their core facilitate those interactions and thus would themselves become obsolete ~\cite{toffler1981third,negroponte1995being}.
Clearly, this is not yet the case.
Major cities have become points of intense investment in ICT because the production, exchange, and dissemination of information were and still are critical to their function.
Paraphrasing \citeauthor{abler1970makes}, the shift from goods handling to service and information industries has made cities into communication systems that are central to information flow in the global economy \cite{abler1970makes}.

Today, the integration of ICTs into Smart Cities has become a complex man-made creation that is unparalleled in its cascading effects on every aspect of the city and its citizens ~\cite{townsend2013smart,castells2011rise}.
Two innovations in the 20\textsuperscript{th} and 21\textsuperscript{st} centuries have enabled this evolution: the rise of the Internet infrastructure for networked computers and mobile personal computing.
In the words of  \citeauthor{townsend2013smart}, ``The democrati[s]ation of computing power that started with the PC in the 1970s and leap[t] onto the Internet in the 1990s is now spilling out into the streets~\cite{townsend2013smart}'' which can be interpreted as the rise of IoT as part of the Smart City landscape.

Contemporary ICTs have restored the importance of space to the previous discourse on urban telecommunication~\cite{picon2015smart}.
Everything and everyone in cities generate geo-tagable information which can reveal the position of a multitude of stationary or moving objects in real time~\cite{goodchild2007citizens,picon2015smart}. 
Geo-tags enrich space with contextualised electronic content and are a fundamental dimension of a sort of return to space, or \emph{spatial turn}, to use the expression coined by urban geographer Soja~\cite{soja1989postmodern}.
This spatial turn of digital technologies has been reinforced by the proliferation of electronic interfaces and wireless communications, which realise the notion of Ubiquitous Computing or Ambient Intelligence ~\cite{picon2015smart,townsend2013smart,flugel2009scientific,dohr2010internet,cubo2014cloud,darwish2018cyber} and ultimately underpin a DT.

Sometimes visible, but more often hidden, countless chips and sensors allow objects, people, and vehicles to be located; consumption levels and transactions to be recorded; and temperatures, pollution levels, population densities and flows to be measured~\cite{weiser1991computer, atzori2010internet,puiu2016citypulse}.
The possibility of transforming every Thing  in the built environment into something more ``smart'' appears more feasible than ever before.
Low cost, miniature sensors enable everything to sense and be sensed.
The ability to perform analytics over the data collected has become ubiquitous, enabling Things to ``think'', while wireless communication technology enables Things to ``talk'' to one another as part of the Internet of Things (IoT) ~\cite{goodchild2007citizens,negroponte1995being,picon2015smart,townsend2013smart}.
However, as note earlier, this dream of Smartification has not been realised due to the increasing complexity of cities.
It increasingly looks as if a DT is required to provide the necessary abstractions to enable the Smart City. 

Ambient Intelligence has taken on many forms that resemble the functionality of DTs, but they remain disconnected and often disjoint.
This leaves it up to the observer to connect the dots, forming the meta-analysis on the city level.
Examples of Ambient Intelligence range from trash cans and parking lots to streetlights.
Smart waste disposal containers in Barcelona are able to know when they have reached their maximum capacity and must be serviced~\cite{nikitin2016data}.
Smart parking uses sensors to know if a parking lot is free or not.
Citizens can then be directed to available parking in the London borough of Westminster~\cite{council2014westminster}.  
In the Future City Glasgow project~\cite{leleux2018delivering}, streetlights can be equipped to sense and analyse what is happening below them.
Sensors collect data on temperature, humidity, noise, and movement, and interpret the data into a functional ``information-scape''.
The streetlights can then respond in real time by adapting their light intensity or by communicating the interpreted data to the nearby police station, reporting any ongoing crime. 

There are also even more ambitions approaches towards constructing a DT.
The Spanish city of Santander with 180,000 inhabitants placed 20,000 sensors to monitor temperature, luminosity, ground humidity, and available parking slots~\cite{sanchez2013smartsantander}.
Paris’s 120,000 trees have each been equipped with a radio-frequency identification (RFID) chip~\cite{itrmanager2006ville}.
In a large number of towns, vehicular traffic is monitored in real time through technology ranging from induction loops, to wireless sensors, to video cameras.
In Singapore, this monitoring enabled the introduction of one of the first dynamic pricing systems, which consists of making drivers paying in proportion to the desirability and, more importantly, the congestion levels of thoroughfares~\cite{picon2015smart}.
This new capacity of cities to detect, measure, and record what is happening within them is truly unprecedented in their history~\cite{picon2015smart,townsend2013smart,goodchild2007citizens} and yet requires new infrastructure to manage the sheer complexity \cite{botin2022digital,bilal2020smart}.
In the words of \citeauthor{foth2008handbook}: ``If aerial photography revealed the skeletal structure of the city during the nineteenth century, the revolution brought about by contemporary ICTs is likely to reveal its circulatory and nervous systems – as it reveals the ``real-time city''. For the first time in history, cities can be understood as a whole, the way biologists see a cell – instantaneously and in excruciating detail, but also alive~\cite{foth2008handbook}''.

\begin{figure}[hbpt!]
	\centering
	\includegraphics[width =\columnwidth]{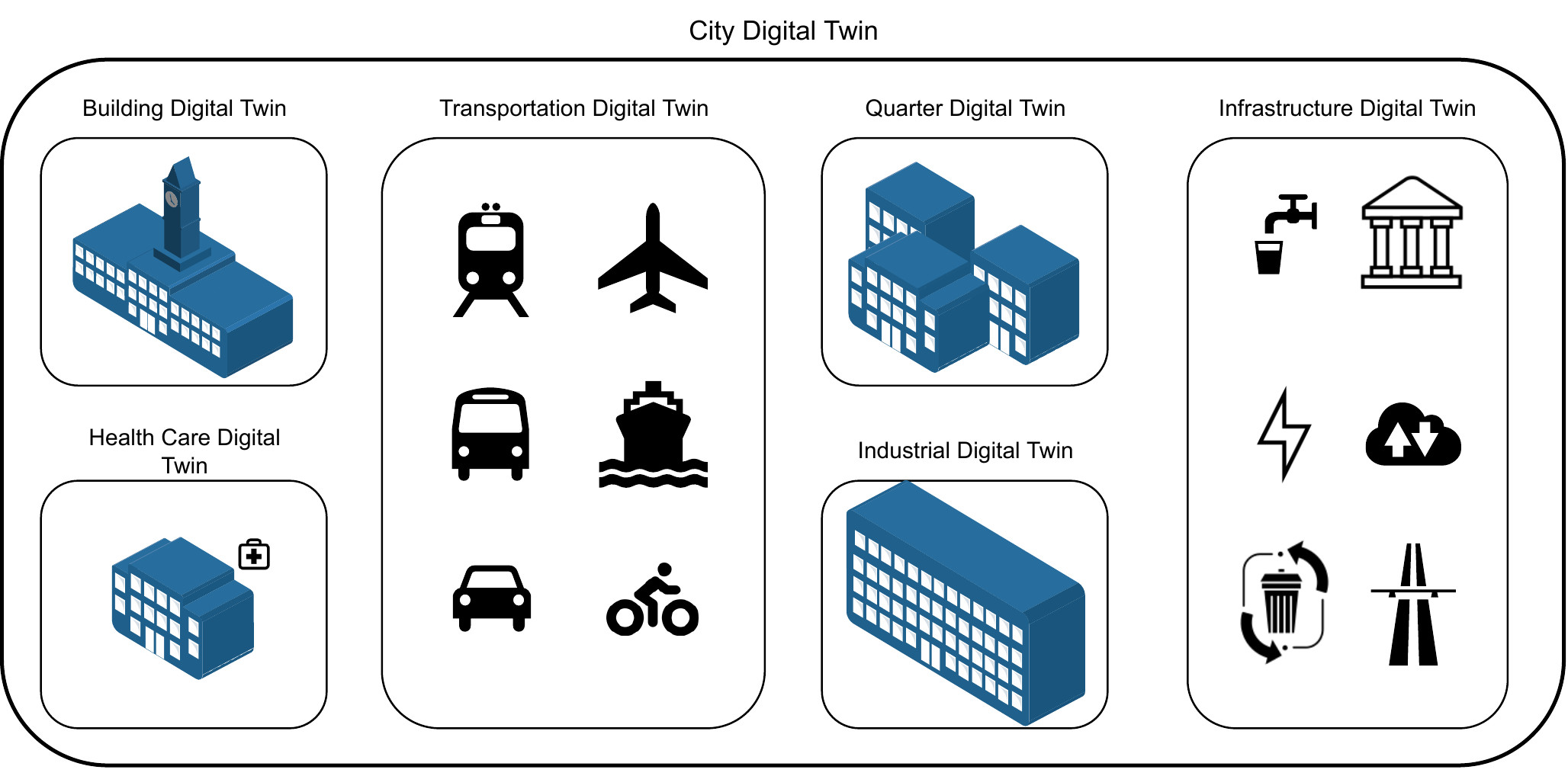}
	\caption{\textbf{Composition of Digital Twins to describe higher order complex systems.} Multiple DTs are composed into higher order systems such as cities. Here an open stationary city DT is composed of different open or closed and mobile or stationary DTs such as building DTs, transportation DTs and Infrastructure DTs.
	Each of the DTs contains complex nested DTs of different types such as for healthcare or industrial settings.
	The complexity level can be arbitrarily increased through nesting. Generated with \url{http://draw.io}.}
	\label{fig:digital:twin:composition}
\end{figure}

This new understanding of real-time urban dynamics is not (yet) centralised, nor is it ensured to be the property of city governments and planners.
If servicised properly and arranged as DT components, it may remain distributed in nature with a low barrier to entrance and can be accessed from nearly everywhere at nearly any time.
The ongoing rise in personal mobile computing, especially smartphones, is transforming how individuals perceive and interact with the urban space that surrounds them.
Situated Analytics powered by DT infrastructure would provide citizens, tourists, and other users unprecedented insights into the city to help them in their every day life and in planning how to use the city in the future.
Using DTs to power Smart Cities will make analytical tools available to everybody and will make them accessible in a spatial context that will democratise the understanding and use of the city. 
To realise this vision of a Smart City and deal with the complexity that is already plaguing current implementations of Smart Cities, DTs offer a bedrock.
The composability of separate DTs into a city-wide DT, underpinning the Smart City, is a promising approach (see Figure~\ref{fig:digital:twin:composition}).

\section{Situated Analytics}
\label{sec:situated:analytics}

\begin{figure*}[htbp!]
	\centering
    \includegraphics[width =\textwidth]{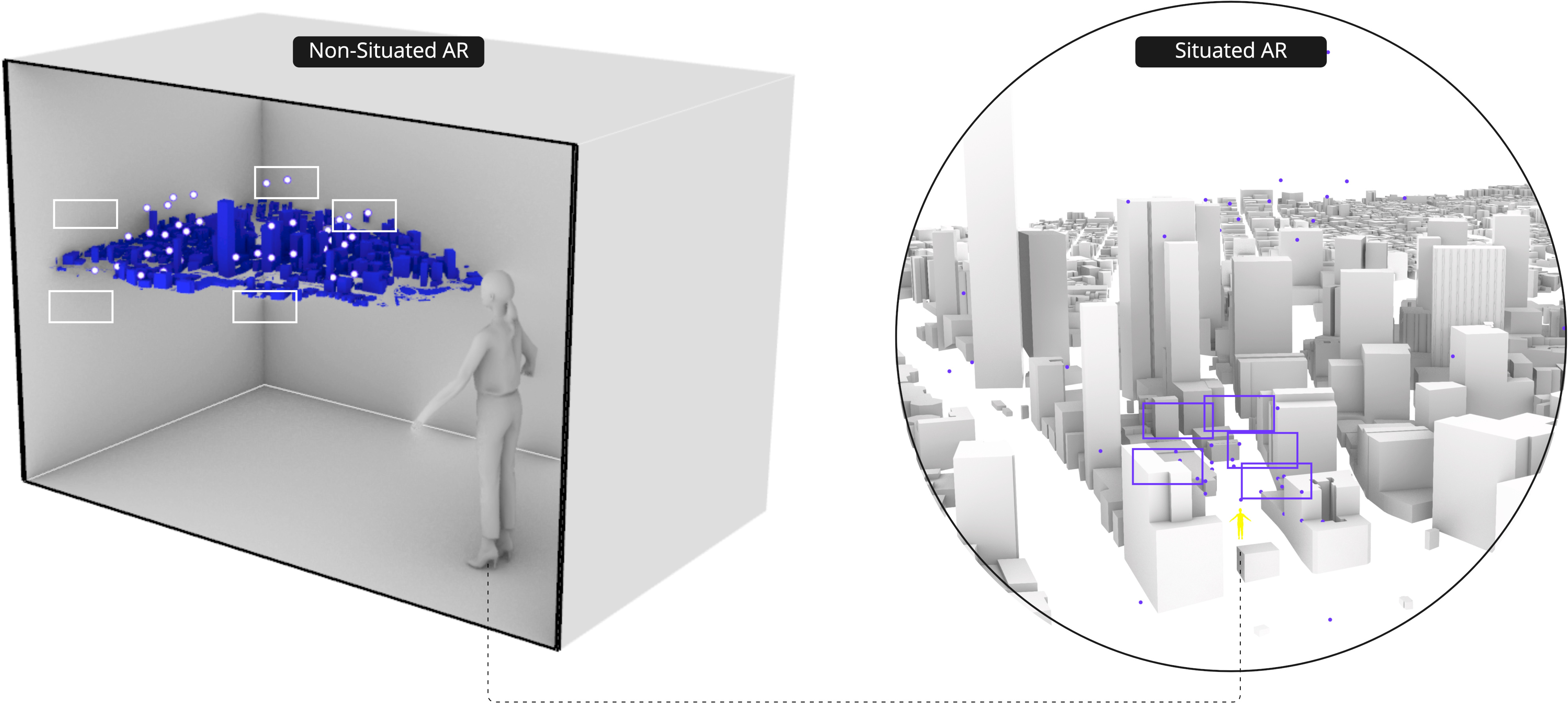}
	\caption{\textbf{Immersive Analytics compared to Situated Analytics}.
	    The real world is schematically represented in grey. AR content is shown in blue.
		On the left side, a user is displaying a smart city environment (blue model) in non-situated AR.
		Interactables are highlighted in white, and several pop-ups showing information (white squares) are opened.
		Immersive Analytics make use of 3D space to visualise content but can be independent of the spatial context surrounding them.
		On the right side, a user (in yellow) is walking through the physical environment of the smart city.
		In Situated Analytics, interactables are AR content that is displayed in situ of the physical environment that they describe.
		Similarly to a), pop-ups (blue squares) are showing content to the user in AR space.
		Generated with \url{https://miro.com} and \url{https://rhino3d.com} with royalty-free models from \url{https://cgtrader.com}.}
	\label{fig:comparison:analytics}
\end{figure*}

Augmented Reality (AR) combines computer-generated visual content in 3D with the real world by superimposing virtual content onto a view of the real world~\cite{azuma1997survey, milgram1995augmented}.
The principles of an AR system are three-fold, combining real and virtual elements, registering in three dimensions, and, importantly, providing interactions in real-time~\cite{azuma1997survey}.
There are at least two categories of approaches to produce AR.
First, screen-based approaches, or ``see-through AR'', can be divided into the optical combination of a screen with the user's perception of the real world or 2D displays that replay a live recording of the real world and superimposes virtual content~\cite{azuma1997survey, van2007augmented}.
Second, projector-based approaches, or ``spatial AR'', consist of projecting the virtual content directly onto the real-world, benefiting from projection mapping algorithms and hardware~\cite{van2007augmented, grundhofer2018recent,brock2018flymap}.
The last decades have refined see-through AR and, today, the main lines of AR systems are either based on the HoloLens for optical systems and ARKit or ARCore for display-based AR on smartphones and tablets.
With the systems maturing to the point where it is both easy to use and easy to generate content, AR applications are going mainstream~\cite{carmigniani2011augmented} and find applications across all industries~\cite{lavingia2020augmented}, commerce~\cite{scholz2016augmented},  entertainment~\cite{stapleton2002applying}, education~\cite{elmqaddem2019augmented}, and medicine~\cite{eckert2019augmented}.

As AR becomes more than a gadget, it opens the door for immersive analytics~\cite{marriott2018immersive} of data and information either in situ or remotely (see Figure~\ref{fig:comparison:analytics}).
Whereas the systems to display AR have been developed over the last decades, the visualisation of information in this new media is still in its infancy and yet rapidly evolving~\cite{dwyer2018immersive}.
For example, traditional visualisation on 2D screens and print is a well-established field with many well-known representations, toolboxes, and high literacy among producers and consumers.
However, most spatial 3D visualisation are still hand-coded and often imitate 2D visualisations.
3D visualisations also offer the opportunity for interaction and embedding in the augmented space~\cite{genay2021being,gervais2016tangible}, possibly easing cognitive load and enhancing cognition~\cite{willett2021superpowers} through externalisation~\cite{scaife1996external}.

The development of AR technologies also enables new types of interaction with digital content.
Specifically, embodied interaction is a concept that reconnects the fields of tangible and social computing and argues that interaction is fundamentally physical and social.
Thus, failing to consider these aspect for design interaction would be a fallacy~\cite{dourish2004action,kirsh2013embodied}.
Embodied interaction is defined as ``the creation, manipulation, and sharing of meaning through engaged interaction with artifacts'' and places the responsibility of meaning-making on users rather than designers~\cite{dourish2004action}.
The level of responsibility for meaning-making at the object's level is driven by different definitions of affordances across research fields that have yet to be reconciled~\cite{michaels2003affordances}.
There is an epistemological difference between affordances as properties of the object or interaction~\cite{gibson2014ecological} and affordances as properties of the users' mental representation~\cite{norman2013design}.
From both Normanian~\cite{norman2013design} and Gibsonian~\cite{gibson2014ecological} perspectives, AR can be viewed as deeply anchored in the physical world and therefore demonstrate strong affordances for embodied interaction~\cite{azuma1997survey}.
Moreover, AR supports a higher amount of embodiment compared to 2D displays by enabling high levels of sensorimotor engagement and  immersion~\cite{johnson2017embodied}.

For example, anchoring and embedding visualisations in the physical 3D world can offer tangible ways to explore one's physiological data~\cite{gervais2016introspectibles} and, to some extent, one's own ``personal'' DT.
For example, Tobe is an AR tangible humanoid DT to explore one's physiological data such as heart rate and breathing patterns and was used in a social context to let two users relax together via data observation~\cite{gervais2016tobe}.
Similarly, Teegi is an AR tangible humanoid DT used to explore one's brain activity via electroencephalogram visualisations~\cite{frey2014teegi}.
Non-humanoid representations have also been explored. Inner Garden is an augmented sandbox embodying one's physiological state, representing breathing patterns as waves and cardiac coherence as weather~\cite{roo2017inner}.
In the wider context of embodied interaction, users' bodies have also been explored as canvases for digital content~\cite{spiel2021bodies}, for example, by co-locating display and interaction on the users' hands~\cite{chatain2020digiglo} or utilising embodied metaphors, such as mimicking scissors with one's hand to cut through digital content~\cite{pei2022hand}.
Tests of embodied AR approaches have also been initiated for data exploration in the context of geographical data for visually-impaired users~\cite{thevin2018augmented, brule2018exploratory, albouys2018towards}.

\begin{figure*}[htbp!]
	\centering
	\includegraphics[width =
	\textwidth]{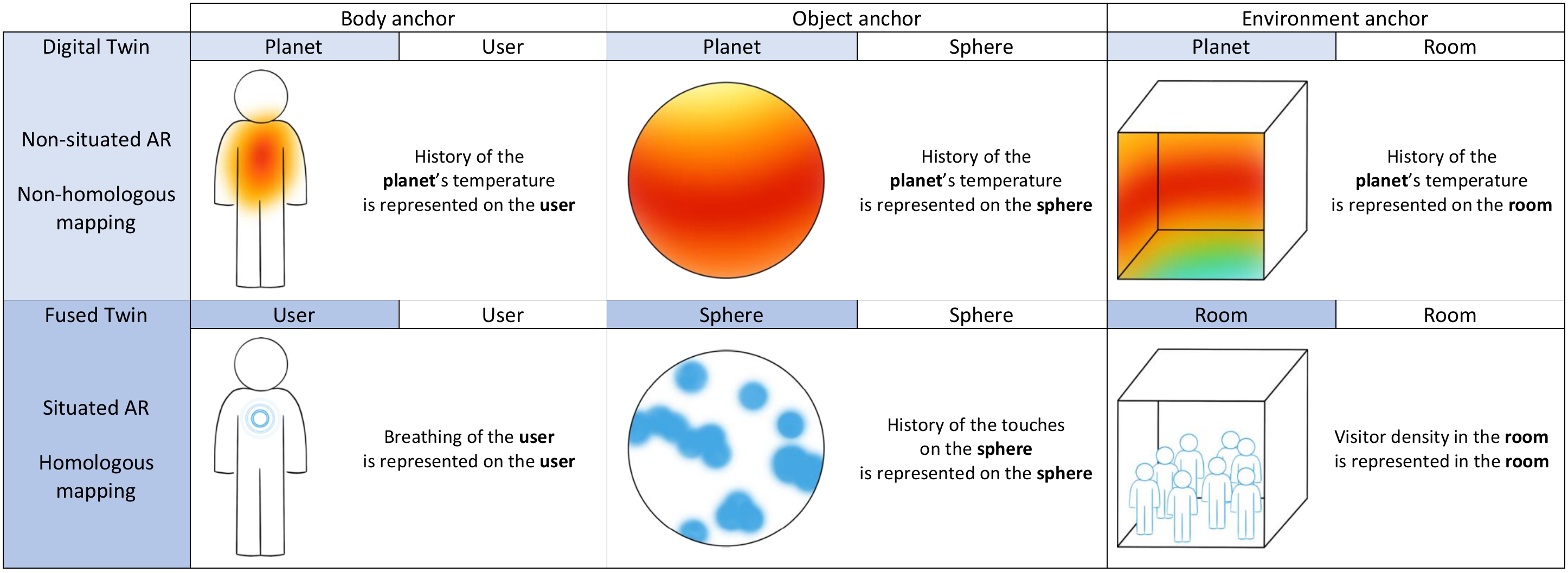}
	\caption{\textbf{Examples of Embodied Digital Twins and Fused Twins}.
	We observe the difference between non-situated AR (top row) and situated AR (bottom row) for anchoring content at the body of the user, an object, or the immediate environment.
	The distinction between body-, object-, and environment-centred AR is inspired by the classification of different physicalities of embodiment offered by~\cite{melcer2016,ottmar2019embodied}.
	With a body anchor, the DT is anchored on the users' bodies, similar to the directly-embodied approach.
	With an object anchor, the DT is anchored on an object with which the user can interact, similar to the manipulated approach.
	With an environment anchor, the DT is anchored in a room, similar to the augmented approach.
	In the situated environment (implementing the FTs paradigm), the relationship between the DT and the representation is homologous such that the body, object, or environment is used to anchor its own DT in a one-to-one correspondence.
	However, in the non-situated environment, this relationship is non-homologous because the mapping from the DT to the representation is transformation instead of a one-to-one correspondence.
	}
	\label{fig:embodied:twins}
\end{figure*}

The integration of users' bodies and their physical context has been conceptualised during interaction with digital content.
Towards that end, different physicalities of embodiment have been described: body-centred, object-centred, and environment-centred embodiment~\cite{melcer2016,ottmar2019embodied}.
We offer examples of embodied DTs and FTs following this categorisation in Figure~\ref{fig:embodied:twins}.
These categories themselves include direct-embodied, enacted, manipulated, surrogate, and augmented approaches~\cite{melcer2016,ottmar2019embodied}.
The direct-embodied approach considers the users' bodies as primary constituents of cognition and therefore the body state is used to convey information.
In the enacted approach, bodily actions are considered a form of enacting knowledge through movement.
The manipulated approach uses the body to directly manipulate a representative object.
The surrogate approach uses a physical avatar that performs an interaction with the object.
Finally, in the augmented approach, the body is used to manipulate an avatar to interact with the larger environment.
The impact of the physicality of embodiment on interaction and cognition is not yet clear.
However, these considerations are of importance because a mismatch between activity design and type of embodiment can result in decreased movements, persistence, and learning outcomes~\cite{chatain2022grasping}.

When discussing the spatial context of data visualisation and data embodiment, we also ought to consider the role of the users' bodies in the interaction.
Often, the users' bodies are solely considered as physical entities (i.e., \emph{Körper}) utilised to push buttons and perform actions, rather than feeling entities (i.e., \emph{Leib})~\cite{mueller2018experiencing}.
Novel approaches such as somaesthetic appreciation design~\cite{hook2016somaesthetic, hook2018designing} describe the manner in which users' bodies may be integrated in the design process as well as in the experience with digital content and data.
Empirical evidence suggests that bodies' representations impact the sense of embodiment~\cite{kilteni2012sense, nimcharoen2018me}, as well as the types of gestures performed to interact with a system and the time spent observing and interacting with data~\cite{trajkova2020move}.
In this regard, the challenge of FTs extends beyond meaningfully anchoring virtual data onto a physical context to include the design of the interaction with the augmentation from an embodied, somaesthetic perspective~\cite{dourish2004action,hook2016somaesthetic, hook2018designing}.

Two further problems arise when visualising embodied information in augmented space.
First, most traditional authoring tools are focused on 2D visualisations, and new toolkits such as DXR~\cite{sicat2018dxr} and RagRug~\cite{fleck2022ragrug} have only provided the basis to display 2D visualisations of data in 3D.
Second, the benefit of the third dimension is difficult to grasp and even more difficult to implement.
Most modern immersive analytics actually rely on embedding 2D displays into the real world~\cite{prouzeau2020corsican,fleck2022ragrug,gruebel2021fused,sicat2018dxr}.
Visualisations in 3D are typically derived from the previous 2D media, but there is a trend towards exploring new possibilities.
For example, multiple 2D visualisations can be arranged to use the third dimension to filter information and visualise this filtration~\cite{hubenschmid2021stream}.

\begin{figure*}[htbp!]
	\centering
	\includegraphics[width =0.9\textwidth]{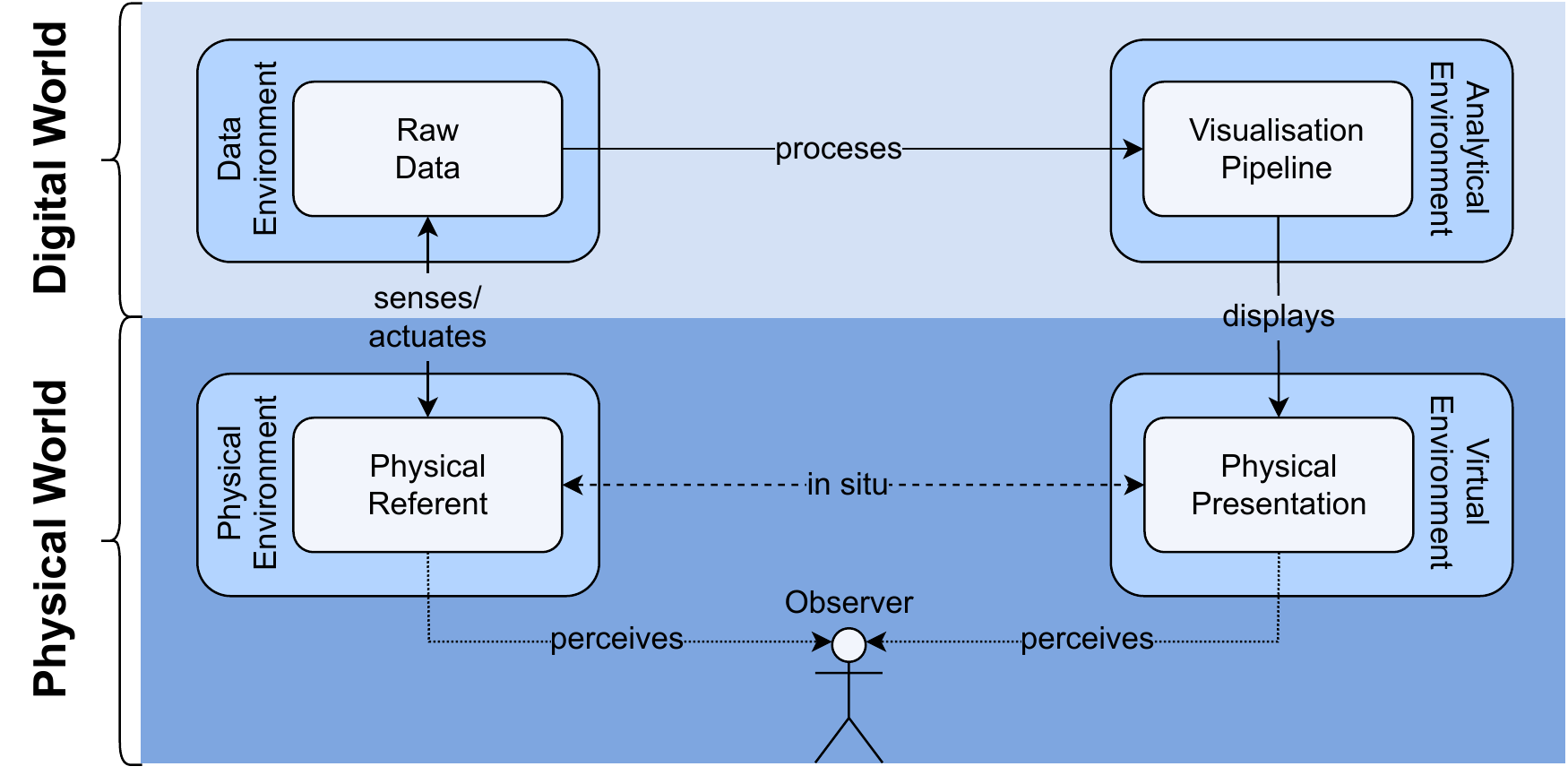}
	\caption{\textbf{Situated Analytics~\cite{thomas2018situated,willett2016embedded} within a DT context}.
		Situated Analytics use the Digital/Physical Divide to define the generative steps in situ immersive analytics in the physical world (bottom row) or digital world (top row).
		There is a remarkable similarity between the components of DTs (blue boxes) and the generation of situated analytics (white boxes). The mapping is visually nested to be highlight.
		A observer in the physical world sees the physical referent or PT (in the DT context).
		The observer also sees the physical presentation of information (Virtual Environment of a DT).
		In the digital world, the raw data is received from the referent, which covers the processes of gathering data by the Physical Environment and storing data by the Data Environment.
		The Visualisation Pipeline produces a visual output (Analytical Environment of a DT).
		Lastly, the data is placed into the physical presentation (Virtual Environment).
		Generated with \url{http://draw.io}.}
	\label{fig:situated:analytics}
\end{figure*}

Situated visualisations is less represented in the literature.
Recently, there has been some defining work that characterised the types of enhancements that immersive analytics may provide through enhanced vision and visual synesthesia~\cite{willett2021superpowers}.
Addressing the question of how situated visualisations might improve cognition is an ongoing research area.
The supposition has been put forth that cognition may be improved by enhanced attention, enhanced numeracy, enhanced recall, enhanced comparison, and enhanced prediction~\cite{willett2021superpowers}.
Moreover, theoretical frameworks such as embodied, situated, and socially situated cognition support the idea that cognition is deeply rooted in its physical and social context~\cite{wilson2013embodied,brown1989situated,smith2004socially}.
However, there are no empirical evaluations yet because, so far, researchers have not focused on the embodiment of the user in the analytics or understanding how the spatial context can reduce cognitive load or enhance cognition.

Some research is starting to address the impact of the third immersive dimension on visualisation possibilities~\cite{merino2020toward}.
For example, one small study demonstrated that a toolkit to create new immersive visualisations helped to keep users engaged for longer and produce more elaborate immersed visualisations. 
We have a clear understanding of how 2D visualisation improve cognitive load through externalisation~\cite{scaife1996external}, but for AR, the evidence is less clear~\cite{turan2018impact,ibili2019effect,thees2020effects,buchner2021systematic,buchner2022impact}.
Previous work has proposed the following mechanisms for how 3D visualisations can improve cognition by reducing task switching with visual cues~\cite{neumann1998cognitive,bonanni2005attention} and context-sensitive displays~\cite{kim2009simulated}.
Moreover, recent work also explores the potential of AR to reduce negative split-attention effects in multimedia activities~\cite{chatain2020digiglo,altmeyer2020use,altmeyer2021effect}.
Split attention effects~\cite{mayer1998split} occur when two related sources of information are spatially or temporally distant.
However, split-attention is not always negative.
For example, indexing (i.e., connecting multiple data representations through pointing gestures) is investigated as a mechanism for sense-making~\cite{hornecker2016and}.
The distance between information sources can also be reduced using AR, in particular the FTs paradigm, because it co-locates physical and virtual entities in real-time~\cite{sweller2011split,kapp2020effects,altmeyer2020use,altmeyer2021effect,azuma1997survey}.
In addition, overwhelming displays of information have been shown to decrease performance~\cite{tang2003comparative}.
There is no definite answer as to how immersive analytics ultimately improves users' performance, but theoretic arguments keep this research area active.

Furthermore, the spatial context in the augmented space has often been used as an empty canvas~\cite{sicat2018dxr} instead of a source of information~\cite{lobo2020opportunities}.
Most commonly, this spatial context has been used for geographic representations to show the built environment either in the past through remote sensing or as planned in the future~\cite{lobo2020opportunities}.
The subfield of Situated Analytics~\cite{thomas2018situated} combines data with its spatial context to achieve a form of virtual data physicalisation~\cite{dragicevic2020data} and embodied interaction~\cite{dourish2004action,kirsh2013embodied,chatain2020digiglo,genay2021being,pei2022hand} that enables humans to maximise the use of all senses to perceive the data (see Figure~\ref{fig:situated:analytics}).
Indeed, information on cities can be embedded in the situated context with meaningful 3D representations that users can explore.
The real world can become the canvas to help users make sense of the information \emph{in situ}~\cite{bressa2021s}.

\section{The Digital Twin}
\label{sec:digital:twin}

Before we explore how FTs can make DTs more accessible, we will review DTs, their historical background, and their relations to other adjacent technologies and the wide range of definitions that accompany them.
Historically, the conceptualisation of DTs originated from the industrial branch of IoT developments in 2003~\cite{grieves2014digital}.
The concept was loosely defined at the time but already included the most important components: a physical object,  a virtual object, and their connections~\cite{grieves2017digital}.
Throughout this section, we will expand this understanding to reach our definition of a DT based on composability and the five components the Physical Environment, the Data Environment, the Analytical Environment, the Virtual Environment, and the Connection Environment (see Figure~\ref{fig:digital:twin:components}).

Historically, there was a lag in the adoption of DTs because the technologies necessary to effectively implement a DT were not yet developed~\cite{tao2018digital}.
In the 2010s, the DT concept outgrew its industrial roots by capturing cyber-physical integration as well as representing the physical world virtually~\cite{lee2008cyber, rajkumar2010cyber}.
As DTs grow in size from factories and cities~\cite{lee2022geospatial} to the globe~\cite{nativi2021digital}, remote sensing with drones and satellites plays an increasingly important role to keep DT representations up-to-date but also offer an opportunity for interactions with the city through interactive mobile displays technology~\cite{brock2018flymap}.
For a full review on current DT applications, we refer to~\cite{botin2022digital}.
Here, we will elaborate the definitions of DTs to form a common basis from where we can develop the concept of FTs.

DTs have been ``hyped'' according to Gartner as one of the top-ten most promising technological trends of the previous decade~\cite{pettey2017prepare, cearley2017top, cearley2018top, panetta2018gartner, tao2018digital}.
Beyond the hype, across many different definitions, DTs fuzzily describe any software aiming to closely represent something in reality~\cite{boschert2016digital}.
These qualities are in line with common and older definitions such as Cyber-Physical Systems (CPS)~\cite{wiener1961cybernetics}, Ubiquitous Computing (UC)~\cite{weiser1991computer}, and Mirror Worlds~\cite{gelernter1993mirror}, which begs the question of whether we can (and should) draw a line between DTs and these concepts.
Here, we argue that DTs should be approached as an analogy to a mental model that helps to structure information and actions and to attach them to the real world through appropriate mechanisms such as the FTs concept.

In this review, we next situate DTs in the context of its predecessors.
DTs are a specialised form of Cyber-Physical Systems (CPS)~\cite{wiener1961cybernetics} because they focus on the accurate representation and manipulation of processes in the real world with a model.
DTs are similar to Ubiquitous Computing~\cite{weiser1991computer} because they attach computations to real world objects, but without the FTs paradigm, DTs are ultimately not Ubiquitous Computing because the computations are detached from the embodiment of the represented object.
Fusing the PT and DT~\cite{gruebel2021fused, prouzeau2020corsican} joins CPS properties with the PT to produce truly Ubiquitous Computing as envisioned by Weiser~\cite{weiser1991computer}.

Whereas Ubiquitous Computing has no clear requirements for the kind of computations to be performed, there is a common theme across different definitions of DTs with respect to the need for simulations to predict a system's response to critical events.
The aim to predict often goes beyond simulating known events towards understanding deviations between the actual observations and the predictions to indicate unknown issues~\cite{glaessgen2012digital,aguilar2016automatic}.
Furthermore, DTs are often expected, not only to report on the state of the system, but to also actively manipulate the system through actuators~\cite{grieves2017digital,botin2022digital,aguilar2016automatic}.
The tension between the accuracy of DT predictions and computational capacity is captured by the idea of Mirror Worlds~\cite{gelernter1993mirror}, which perfectly imitate reality similar to a mirror.
Early ideations of the DT therefore required the underlying model to support similitude~\cite{glaessgen2012digital}, while similitude requires the implementation of a multi-physics,  multi-scale, probabilistic, ultra-fidelity simulation that mirrors the PT based on historical data, real-time sensor data, and a physical model.
The ``ideal DT'' requires similitude to become as close as physically possible to mirroring the real world.
Indeed, DTs need to model the mutual dependency of hardware and software to tightly couple the PT and the DT, producing high-quality simulations.
Until recently, most simulations were feasible only for certain sub-components of complex systems that relied on scientific knowledge and real observations to produce a more accurate offline simulation.
Therefore, the requirements of similitude have only been satisfied in specialised cases such as with the components of flying vehicles~\cite{glaessgen2012digital}.
New computational methods increase simulation performance such that events in the DT can be simulated fast enough for predictive purposes in the real world~\cite{gabor2016simulation}.
However, even with respect to aircraft, there are only plans to produce a DT of a whole vehicle~\cite{chowdhury2019methodology,economist2022digital} which are expected to be completed in the coming decade.
Regarding the sheer scale of cities, the mirror worlds that would underlie a smart city seems to be a distant dream. 
Due to the simulation complexity of large-scale systems such as cities, researchers have since called for a less rigorous model definition to make DTs relevant to real world applications~\cite{batty2018digital}.

In other words, stepping away from ideal DTs to ``practical DTs'' can allow for more practical applications.
Easing any of the core aspects of similitude allows for the implementation of many useful concepts such as analytical assessment, predictive diagnosis, and performance optimisation~\cite{tao2018digital}, even in large-scale systems. Limiting the functionality of DTs to a synchronous representation of reality can also allow for the monitoring of complex production processes, adjust these processes, and possibly optimise them~\cite{weyer2016future}.
Ideally, the model underlying a DT is complex enough to enable a system to continuously react to dynamic changes in physical space~\cite{rosen2015importance}, including  unforeseen/unexpected changes~\cite{vachalek2017digital}. DTs also open the door to continuous design improvement through data feedback loops~\cite{canedo2016industrial} enabling data assimilation into a model~\cite{wikle2007bayesian,reichle2008data}. 
These incomplete but practical DTs can become especially useful if they are composable to cover more features of the PT. Next, we will introduce specialised, practical DTs which answer questions on particular topics and can be composed to resemble ideal DTs.

In our definition, we side-step the discussions surrounding ideal DTs by focusing on structural features of practical DTs.
We refine the definition of a DT based on earlier work~\cite{tao2018digital} as a five component system: the Physical Environment, the Data Environment, the Analytical Environment, the Virtual Environment, and the Connection Environment (see Figure~\ref{fig:digital:twin:components}).
In the Physical Environment, IoT systems~\cite{atzori2010internet} and remote sensing units~\cite{xiao2018towards} try to capture the PT and provide interactions with the PT through actuators.
The Data Environment contains the DT's memory and is usually located somewhere in the cloud but may also be stored locally or at the edge~\cite{shi2016edge,hossain2018edge}.
For the Analytical Environment, the processes and models of the DT are computed~\cite{schluse2018experimentable}, often in the cloud, and insights and predictions are communicated back to the Data Environment.
The Virtual Environment makes the DT accessible to human users and possibly other machines~\cite{gruebel2021fused,prouzeau2020corsican,swetina2014toward}.
Finally, the Connection Environment provides an Application Programming Interface (API) to transfer data as required between the components.
We consider two ways of designing the connection environment, 1) from a service perspective, where the Connection Environment is not an independent component but refers to the APIs exposed by the different components, and 2) from a compatibility/hypervisor layer perspective where the connection environment is an independent component that mediates between the other different components in the system.
Each component could be implemented in many different ways, and making a sensible choice may be overwhelming.
This review gives an overview to guide newcomers through the most important components and decisions that are required to implement the FTs concept with a DT.

To understand DTs, it is necessary to comprehend the function and goal of the core components in relation to the literature.
Specifically, we connect our definition of the Physical Environment to research on IoT and remote sensing, as well as our definitions of Analytical and Virtual Environments to research on simulation and interaction with DTs.
Our definitions of the Data and Connection Environments are not broadly covered in previous research because previous implementations have used case specific data solutions and had were not focused on interoperability.

The Physical Environment component captures the PT and associated interactions via IoT and remote sensing infrastructure.
The composition of IoT stack is regularly repeated but rarely analytically decomposed with respect to its trade offs.
For most applications, the stack is designed with the purpose of underlining where and why a specific new technology is needed~\cite{xu2014internet,lee2015cyber,knappmeyer2013survey,guinard2011internet}.
Technically, this is a useful approach to help experts situate a technology in the literature.
However, these stacks typically make many unspecified assumptions that we disentangle here.
We start with the notion that the DT perspective can be mapped to aspects of IoT and remote sensing.
Following the semantic decomposition of the term IoT~\cite{huang2010descriptive}, Things themselves are not actually \emph{in} the Internet, but information about them is perceived either with attached sensing units or with remote sensing and then shared via the Internet. 
The DT can be used as the necessary representation of the perceived Thing in the Internet.

Previous resource-oriented and service-oriented architectures have represented IoT hardware as practical DTs without using this particular nomenclature~\cite{spiess2009soa,guinard2010architecting, guinard2010interacting, guinard2010resource, guinard2011internet, kindberg2002people,blackstock2012iot}.
In contrast to IoT definitions, the DT perspective does not necessitate the digital information to be shared globally as may be implied by the term Internet.
The potential local scope of DTs is reflected in their origins in cyber-physical systems~\cite{wiener1961cybernetics,grieves2017digital} where the application scope is usually limited to an industrial production site.
Nonetheless, the DT perspective provides us with a useful definition of the properties that a Thing should have when represented digitally. 
A broad range of example IoT applications have been addressed in previous publications~\cite{atzori2010internet,sanchez2012adding,borgia2014internet,islam2015internet,wang2016towards,darwish2018cyber} and can be generalised to DT solutions.

The Analytical and Virtual Environments in DTs are difficult to precisely define because their development is multi-disciplinary and reaches beyond the simulation perspective.
These paradigms have been advanced by independent research in the areas of information science, production engineering, data science, and computer science~\cite{tao2018digital} as well as electrical engineering~\cite{akyildiz2002wireless}.
A DT is a highly complex system that combines infrastructure, simulation, data processing, human-computer interaction, and services ~\cite{tao2018digital}.
These approaches give rise to another understanding of DTs as an advanced digitalisation of a system that extends representation of the PT with operations.
This produces a rich digital model that has many desired properties such as scalability, interoperability, extensibility, and high fidelity~\cite{schleich2017shaping}.
To arrive at a digital model, digitalisation in a space can also be understood as a virtualisation process~\cite{moreno2017virtualisation}.
Here, the system is represented in 3D space via modelling and remote sensing. First, behaviours are extracted and knowledge representations are created.
Second, the interactions between its constituent elements are modelled and integrated into the knowledge representations.
Third, operations and their effects are modelled.
Lastly, everything is assembled in a simulation to make the system match the PT as close as possible.
Because digitalisation/virtualisation only creates the Analytical Environment (see Figure~\ref{fig:digital:twin:components}), most research has clearly focused on the analysis and not interaction with the DT.
Most research on DTs also follows the simple physical-digital split into PT and DT components~\cite{boschert2016digital, grieves2017digital}, but this split often fails to capture the difference between the Analytical Environment for computation and the Virtual Environment for interaction (but see~\cite{tao2017digital}).
The FTs paradigm offers a new perspective that highlights the unique contribution of the Virtual Environment, resulting in a more intuitive use of the DT in nearly every application domain.

\begin{figure}[hbpt!]
	\centering
	\includegraphics[width =\columnwidth]{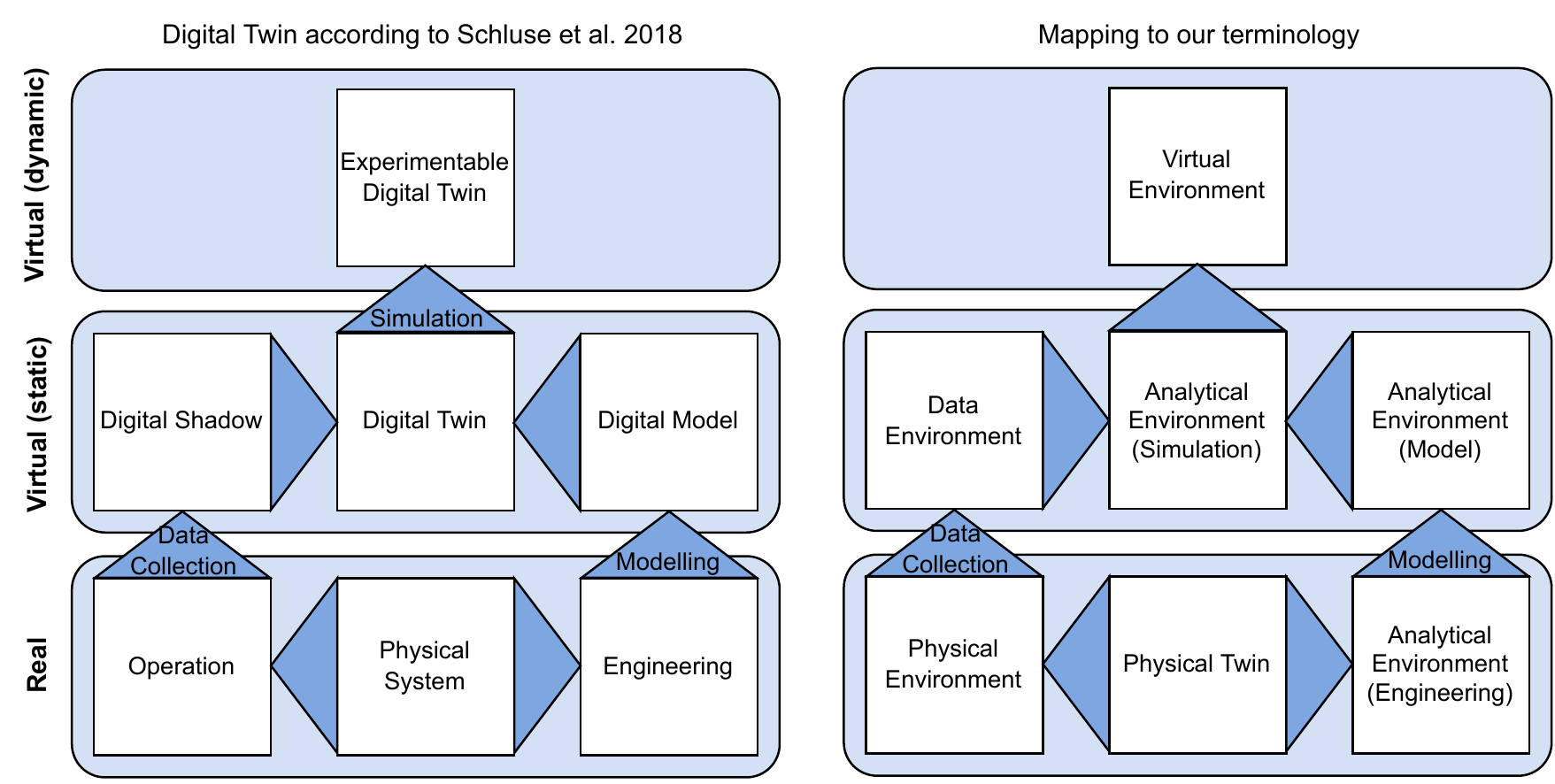}
	\caption{\textbf{Digital Shadows and Digital Models in the context of Digital Twins}.
	The left side demonstrates the relations between a digital shadow, a digital model and a DT~\cite{schluse2018experimentable}.
	The right side demonstrates the mapping to our definition. The Digital Shadow corresponds to the Data Environment and relies on the Physical Environment to collect data.
	The Engineering, Digital Model and their DT simulation corresponds to the Analytical Environment.
	The experimentable DT is akin to the Virtual Environment.
	In our definition, all blocks are components of the DT whereas in the other definition, the DT is a piece in the diagramm.
	The reduction of the DT to a simulations removes the HCI from the system and requires terms such as digital shadow and digital models to differentiate DTs.
	Under our definition, theses terms reflect different aspects of a DT and map to different components. 
	Generated with \url{http://draw.io}.}
	\label{fig:shadow:model}
\end{figure}

There may also be issues with many definitions of DTs because of the unrecognised difference between an universal ideal DT and a partial practical DT.
Ultimately, universal ideal DTs are unachievable because of the need to integrate the complex requirements of similitude and to comply with the many definitions across different literatures.
Therefore, any practical DT is only partially implemented compared to ideal DTs.
Confusion regarding this distinction has led to a heated debate and has resulted in the differentiation of DTs according to their degree of completeness from digital models and digital shadows to (ideal) DTs~\cite{schluse2018experimentable,sepasgozar2021differentiating}, see Figure~\ref{fig:shadow:model}.

A digital model is an independent representation of a real-world system that may be used in simulations but that is generally not fed real-time data.
According to our interpretation, a digital model reflects the implementation of the Analytical Environment.
A Digital Shadow is the assembly of data from the real-world system for descriptive purposes.
By our interpretation, a digital shadow reflects the Physical Environment of the DT.
Thus, our definitions can shift the discussion on digital shadows and digital models from whether something is a DT to what component of a DT is implemented.
While a completely implemented, ideal DT at the urban scale is probably out of scope for now, it is feasible to implement and compose partial practical DTs that are capable of performing important tasks for Smart Cities. 
To elucidate IoT, DTs, and Smart Cities, the present paper adopts the FTs perspective because it unites three views on DTs that are often confounded: Why to generate the DT (i.e., business case for the Analytical Environment), how to generate the DT (i.e., technological implementation of the Physical Environment), and how to use the DT (i.e., end user interaction of the Virtual Environment).

\subsection{Composition of Digital Twins}
\label{sec:digital:twin:composition}

\begin{figure}[hbpt!]
	\centering
	\includegraphics[width =\columnwidth]{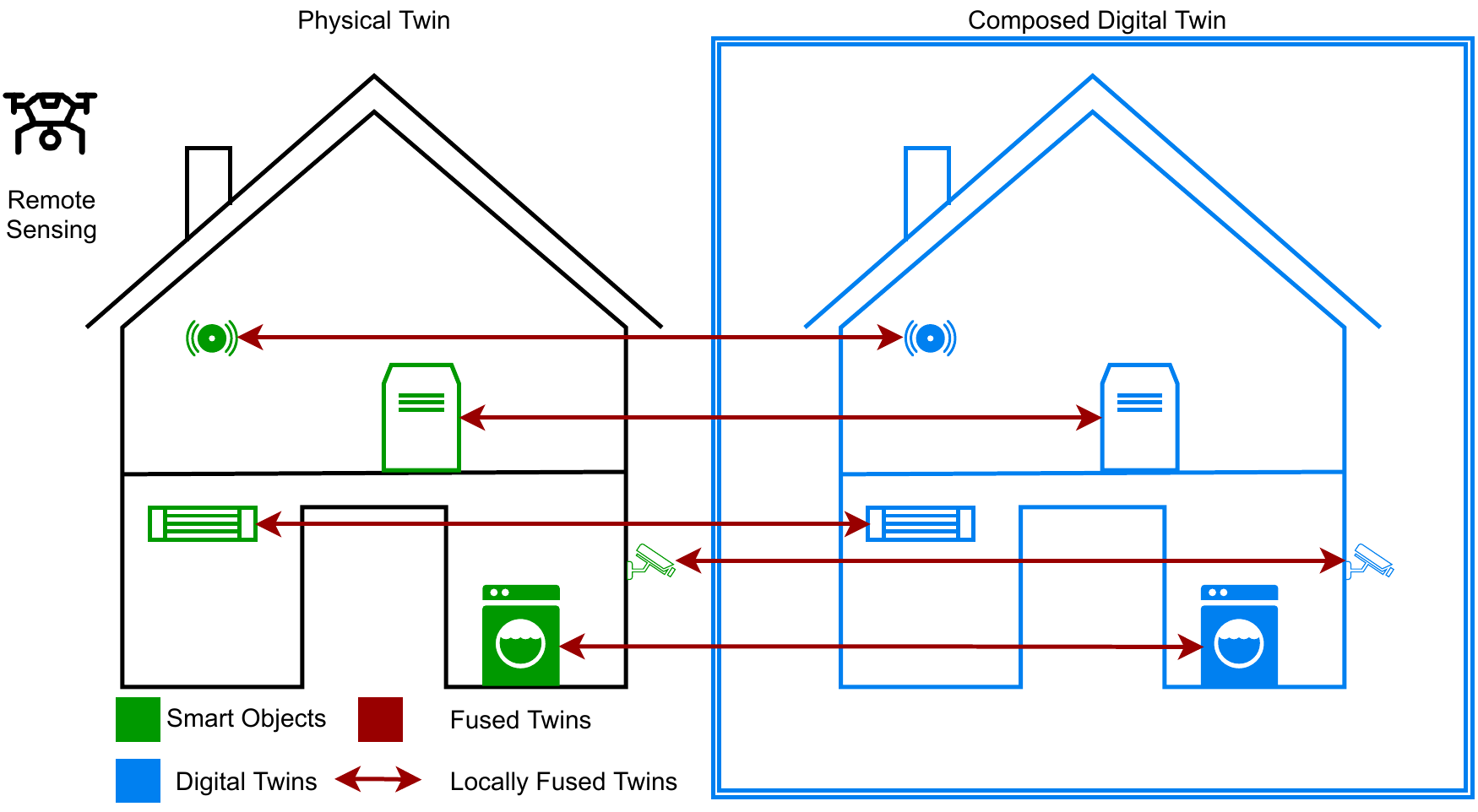}
	\caption{\textbf{Example Composition of DTs.} Multiple Smart Objects (SOs) and their DTs are composed into a DT of a Household. Remote sensing provides the external context to the building. Locally FTs (SOs) provide physical user interfaces enabling end users to directly interact with the DT in a natural way. Generated with \url{http://draw.io}.}
	\label{fig:digital:twin:smart:objects}
\end{figure}

Composed DTs offers an integrative perspective to understand and address the complexity of the PT by using multiple abstractions and performing different tasks on each abstraction (see Figure \ref{fig:digital:twin:smart:objects}). 
Towards this end, DTs are composable through their Connection Environment allowing both for the nesting and recombination of components.
This approach allows us to re-conceptualise any sensors in the Physical Environment as virtually belonging to all associated composites of the DTs.
Composites of the DTs can be produced to offer previously unavailable information via new virtual sensors~\cite{madria2014sensor} or virtual sensor networks~\cite{ocean2006snbench}~\cite{khan2016wireless}.

\begin{figure*}[hbpt!]
	\centering
	\includegraphics[width =0.8\textwidth]{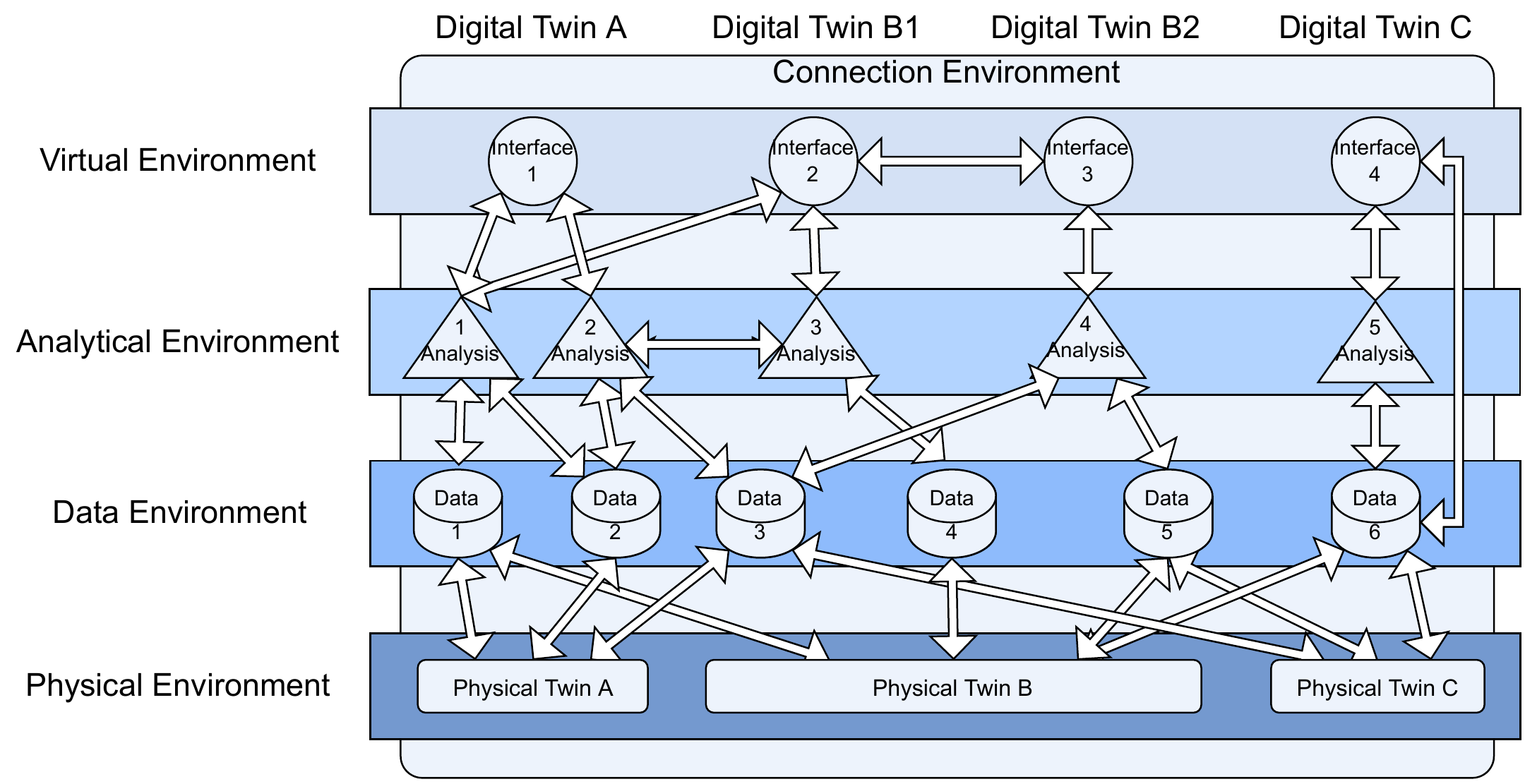}
	\caption{\textbf{Composition of Components to form Digital Twins.} Each DT is represented as a column. The Connection Environment (represented by arrows) connects instances in all environments across DTs. The other environments are aligned in rows to show the implementation across DTs. Instances are shown in each environment as rectangles, cylinders, triangles or circles. Data from the Data Environment can processed in the Analytical Environment or displayed directly in the Virtual Environment. DTs can also be siloed (e.g., DT C) but more often will use the Connection Environment with complex dependencies across all environments. For example, DT A uses Data 3 and Analysis 3 from PT B. PT B has two DTs where DT B2 is composed into DT B1 in the Virtual Environment by combining interface 2 and 3. DT B1 \& B2 are using data that is collected from all PTs. Despite DT C being siloed, PT C is evaluated in DT A through Data 3 and Analysis 2. Note that this abstraction makes no claims about the mode of data collection in the Physical Environment that may be with either local sensors or remote sensing. Generated with \url{http://draw.io}.}
	\label{fig:digital:twin:components:interaction}
\end{figure*}

For DTs to be composable, interoperability between components is the most important feature ~\cite{broring2017enabling, desai2015semantic,al2015toward,schleich2017shaping,raza2017critical}.
An example composition of the different components is shown across associated DTs in Figure~\ref{fig:digital:twin:components:interaction}.
The Physical Environment contains sensory equipment within arbitrary boundaries to capture properties of the PT, which is immutable and equally anchors all overlapping DTs.
The Data Environment is a heterogeneous composition of storage architectures that best befits the data requirements.  
The Analytical Environment is usually defined per task per DT.
In addition, it may also have feedback loops to other DTs via the Data Environment.
Outputs of the Analytical Environment are fed back into the Data Environment as virtual sensors to provide the information for display and interaction at a later time.
The Virtual Environment forms a local Mirror World~\cite{gelernter1993mirror} that allows for the visual inspection of information in the DTs.
The scope of the Virtual Environment is limited to the current composition of DTs and allows for access to and interaction with both the data from the Physical Environment and the data from the Analytical Environment.
The Connection Environment manages access rights between components and ensures that users (both humans and other DTs) have the appropriate access rights to use the respective environments, addressing concerns regarding trust, security, and privacy~\cite{das2018taxonomy,keoh2014securing,granjal2015security, kouicem2018internet,zhou2008securing,chen2009sensor,zhang2014iot, singh2016twenty,heer2011security,yaqoob2017rise,zhao2013survey,sarma2009identities,roman2011securing,sadeghi2015security,raza2017critical}.

In practice, DTs are applicable to all fields in which IoT has been applied~\cite{akyildiz2002survey,atzori2010internet,borgia2014internet}, including industry~\cite{spiess2009soa,bi2014internet}, health care~\cite{dohr2010internet, doukas2012bringing, amendola2014rfid, islam2015internet, catarinucci2015iot, kumari2018fog, cubo2014cloud}, urban management~\cite{vlacheas2013enabling,krylovskiy2015designing, puiu2016citypulse}, vehicular application~\cite{al2020intelligence}, and mobile end-user applications~\cite{kamilaris2016mobile}.
However, the composability of DTs suggests that, by extension, these applications should be composable, too. 
For example, DTs in industry, health care, and transportation could exchange vital information to improve their effectiveness, but, to the best of our knowledge, this has not been accomplished in practice.
To understand how DTs of these varied systems could be composed in the context of truly smart cities, we categorise each application by its system type and location type to form a taxonomy of DT contexts (see Table~\ref{tab:iot:application:taxonomy}).

\begin{specialtable}[!htbp]
	\caption{Taxonomy of Digital Twin Context Types} 
	\centering
	\setlength\extrarowheight{2pt}
	\resizebox{0.5\textwidth}{!}{
		\begin{tabular}{*{2}{r}*{2}{c}}
			\toprule
			&&\multicolumn{2}{c}{System Type}\\
			\cmidrule(lr){3-4}
			&&Open&Closed\\
			\midrule
			\multirow{2}{*}{\STAB{\rotatebox[origin=c]{90}{\makecell{Location\\Type}}}}& Stationary & Cities, Economies \& Countries & Industrial processes\\[5pt]
			&Mobile &End users, goods \& vehicles & Body (health) \& machines (internal)\\[5pt]
			\bottomrule
			\label{tab:iot:application:taxonomy}
		\end{tabular}
	}
\end{specialtable}

The system type of a DT is either open or closed~\cite{bertalanffy1969general}. Open systems are characterised by an exchange with the wider environment in which they are embedded, whereas closed system do not directly exchange with the environment and are usually observed in isolation.
In addition, closed systems usually supervise well-defined processes, whereas open systems try to characterise ill-defined processes.
Obviously, no system is entirely isolated, but industrial processes or health applications can be considered relatively closed systems with usually well-defined interactions with the environment.
In contrast, cities and traffic~\cite{raza2017critical} have no clear limit to what is entering from the surrounding environment, and congestion may be a side effect of this ill-defined boundary.

The location type of a DT describes the relation between the unit of observation and the sensors as active or passive sensing~\cite{moussaid2018virtual}. In stationary systems, the sensors are fixed in the environment and observe throughput. In mobile systems, the sensors are moving around either with the observed unit or independent of it. They may also capture environmental conditions around the observed unit such as in human navigation~\cite{kiefer2014investigating} and health monitoring~\cite{petajajarvi2017evaluation}. 

When composing different types of DTs, it is important to understand their characteristics.
Mobile DTs may only temporarily compose with Stationary DTs, while Closed DTs often form units within Open DTs.
Open Stationary DTs often form the overarching type of DT that contains other types of DTs, whereas Mobile Closed DTs usually describe atomic units in composed DTs.
Typically, Mobile Closed DTs can be part of Mobile Open DTs when a person has a Body Area Network (BAN)~\cite{singh2021wireless} for health care applications and a Personal Area Network (PAN)~\cite{kushalnagar2007ipv6} to interact with the environment.
In this example, the PAN will provide a wide range of interactions that can be composed with the Open Stationary DT of a Smart City, whereas the BAN will only share limited information.
Similarly, a Closed Stationary DT for a Smart Factory within Open Stationary DT of a Smart City will only share limited information. 

\subsection{Servicisation of Digital Twins}
\label{sec:digital:twin:servicisation}

While the word DT suggests some form of unitary entity, it would be more accurate to understand DTs as context-aware representations of Things in the cloud composed of services that act upon the abstraction.
Services can range from interacting with the environment to representing the sensed and form the bedrock of ubiquitous computing~\cite{weiser1991computer}.
Basic services again can be composed into higher-order services to address more complex issues.

With digital representation by the DT, it is possible to turn everything into a service---or \emph{servicise}~\cite{white1999servicizing,toffel2008contracting,plepys2015european,agrawal2016potential}.
In other words, the customer buys the temporary use of a service associated with a product instead of the product itself.
The cost of developing and maintaining a DT's infrastructure justifies this approach because it is more efficient to reuse resources rather than to duplicate functionalities~\cite{agrawal2016potential}.
Therefore, DTs can be freely composed of services that conform to the requirements of the Connection Environment .

\begin{figure*}[thbp!]
	\centering
	\includegraphics[width =\textwidth]{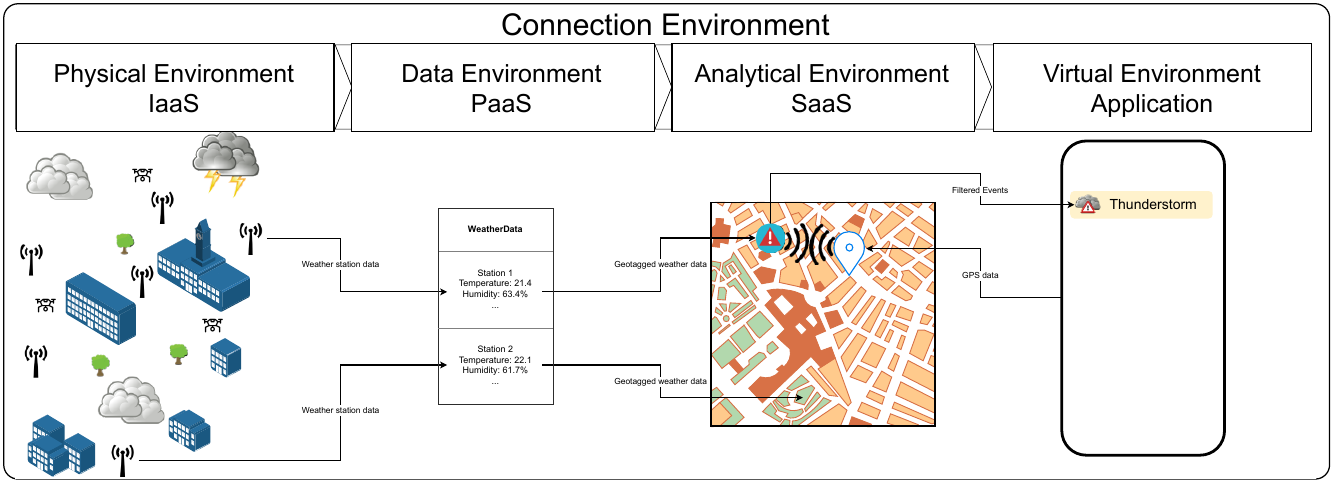}
	\caption{\textbf{Mapping of the ``as a Service'' paradigm to the components of a Digital Twin and the weather services example.}
		This example application is an alert for a nearby weather event.
		Presented from left to right, the PT forms the Physical Environment of the DT and is captured by local sensors and networks and remote sensing as IaaS. The Data Environment of the DT serves as a PaaS. The Analytical Environment of the DT models the weather and calculates the proximity to the predicted event as a SaaS. The Virtual Environment is implemented as a smartphone application that provides an alert on an abstract dashboard (see Figure \ref{fig:digital:twin:continuum} for more details). The Connection Environment spans across all services and environments providing an API that allows for one environment to be replaced by another. This makes the DT both modular and composable. Generated with \url{http://draw.io}.}
	\label{fig:digital:twin:services}
\end{figure*}

The three most common themes in ``as a Service'' (*aaS) are Infrastructure as a Service (IaaS), Platform as a Service (PaaS) and Software as a Service (SaaS).
Newer types of services can be subsumed in these categories, including Sensing as a Services (SenaaS)~\cite{alam2010senaas, distefano2015utility}, Semantics as a Service (SemaaS) ~\cite{desai2015semantic}, and Robots as a Service (RaaS)~\cite{chen2013internet}.
Notably, Infrastructure, Platform, and Software ``as a Service'' form an interesting interaction with the DT components because they can provide abstractions for particular components (see Figure~\ref{fig:digital:twin:components}).
From an abstraction perspective, each replaces the lower layers of the stack with an interface that hides implementation complexity.
In this section, we will use the example of weather services as part of a smart city to explain how IaaS, PaaS, and SaaS can map to the components of a DT (see Figure~\ref{fig:digital:twin:services}).

\subsubsection{Infrastructure as a Service (IaaS)} 

A DT requires the creation of a Physical Environment as the representation of the PT.
The Physical Environment provides access to and measurement of the PT using infrastructure to collect raw data (see Figure~\ref{fig:digital:twin:components}).
Creating and maintaining the DT infrastructure is the most physically laborious part of developing a DT.
This severely limits the number of entities that can afford to create a DT on a city scale.
However, provided as a service, it may be possible to finance the required infrastructure in aggregate and reuse it across multiple DTs.
When creating a DT, the IaaS approach allows for the reuse of a service provider's communication network, sensor network, and remote sensing units rather than having to build one's own system from scratch.
Therefore, IaaS enables to have alternative competing and composable DTs based on the same or different data source(s).
There are also more specialised intrastructure services that are available such as Sensing as a Services (SenaaS)~\cite{alam2010senaas, distefano2015utility}, Remote Sensing as a Service (RSaaS)~\cite{toulios2015remote}, and Robots as a Service (RaaS)~\cite{chen2013internet}.

In our weather service example (see Figure~\ref{fig:digital:twin:services}), the sensors, network, and remote sensing units in a smart city could be servicised.
The city government or a private provider could implement a policy to maintain such an IaaS to make digital services in a city more accessible, enabling the creation of different city DTs.
This would allow new entrants to easily develop their own DTs that provide large-scale digital services without having to build physical hardware with redundant functionality~\cite{toffel2008contracting}.

At the same time, IaaS may raise some concerns regarding security.
The IoT used for IaaS consists of billions of locatable, addressable, and readable devices~\cite{roman2011securing}.
There are three avenues by which security in IoT must be enforced: information, physical infrastructure, and management practices~\cite{zhao2013survey}.
Security breaches from unexpected behaviour may arise from data collection and management via privacy protection problems, heterogeneous network authentication, and access control problems~\cite{sarma2009identities,roman2011securing, zhao2013survey,ouaddah2017access}. 
The fundamental challenges for all of these topics are questions of trust: how to give whom access for what purpose and for what duration.
Over the last decade, the IEEE and IETF have laid the foundations to build a security infrastructure along the IEEE 802.15.4 standard~\cite{ieee2011ieee802.15.4, ieee2012ieee802.15.4e} and the IETF LLN stack~\cite{kushalnagar2007ipv6}.
These efforts include sensor network security~\cite{chen2009sensor},
wireless sensor network security~\cite{zhou2008securing}, and dedicated secure protocols such as CoAP/Lithe~\cite{bormann2012coap, raza2013lithe}, 6LowPSec~\cite{glissa20196lowpsec} and RPL~\cite{mayzaud2016taxonomy}.
Ultimately, such as with the Internet, security will be an ongoing process that must be attended at all times.

\subsubsection{Platform as a Service (PaaS)} 

PaaS extends access to hardware by simplifying interfaces. 
PaaS usually packages digital representations into a Data Environment and thus reduces the complexity of accessing the data.
Platforms vary largely and are usually tied to one of the five IoT application areas industry~\cite{spiess2009soa,bi2014internet}, health care~\cite{dohr2010internet, doukas2012bringing, amendola2014rfid, islam2015internet, catarinucci2015iot, kumari2018fog, cubo2014cloud}, urban management~\cite{vlacheas2013enabling,krylovskiy2015designing, puiu2016citypulse}, vehicular applications~\cite{al2020intelligence}, and end-user applications~\cite{kamilaris2016mobile}.
The concept of PaaS can be vague, and its use may sometimes be difficult to differentiate from IaaS or SaaS.
In the context of a DT, a PaaS ought to provide only \emph{access to data} as a Data Environment in contrast to an IaaS that collects \emph{raw data} from the physical world as a Physical Environment and a SaaS that \emph{derives higher-order data} from previously collected data as a Analytical Environment.

A DT probably cannot rely on a single PaaS to source all of the required data. More often, a DT will mix and match PaaS sources \cite{villegas2012cloud} to fulfil its task. The Data Environment is more complex than simply making data available and needs to be curated, indexed, and integrated into a knowledge graph based on context-awareness~\cite{knappmeyer2013survey}, ontologies~\cite{hachem2011ontologies,wang2012comprehensive,song2010semantic}, and middleware~\cite{ngu2017iot, razzaque2016middleware, mineraud2016gap}. In reality, IaaS and PaaS are often mixed and sometimes offer a subset of services that should be associated with SaaS \cite{linthicum2017paas}.

In our weather service example, the PaaS stores the information of the weather-related sensor data from sensing units throughout the city, including their spatiotemporal context.
Specifically, the PaaS does not include any analytical components but offers a raw data source upon which higher-order data services may be built.
A clear distinction is necessary between services to disambiguate their different responsibilities in the DT and ensure composability.

\subsubsection{Software as a Service (SaaS)} 

According to the classical SaaS literature, SaaS is often exposed directly to an end user~\cite{kamilaris2016mobile} and ought to follow a singular use case.
However, in the context of the DT, SaaS could be mapped to both the Analytical Environment and the Virtual Environment. 
Following the ``as a service'' paradigm, we suggest limiting the use of SaaS to the Analytical Environment in DTs to reduce redundant functionality and to enable more servicisation in end-user applications (see Figure~\ref{fig:digital:twin:components}).
Conventionally, one may consider SaaS as built upon PaaS, but the relationship between the Data and Analytical Environments may actually be bidirectional.
The results of the analyses may be written back to the Data Environment before being forwarded to the Virtual Environment for user interaction.
Thus, there may be an ambiguity that makes it difficult to separate SaaS from PaaS.
The difference lies in whether one considers the results of an analysis to be newly computed as a SaaS or retrieved from memory as a PaaS.

In our weather service example, the classification of the predictive weather data to an end user could be qualified as SaaS  if it is being calculated directly.
However, if the predictive weather data is provided as a component of a smart city environmental observation system, it would be qualified as a PaaS.
In general, a SaaS conducts some analytical task and represents the Analytical Environment of DTs because it generates the models to mirror reality in the DT.

\begin{figure*}[htbp!]
	\centering
	\includegraphics[width =\textwidth]{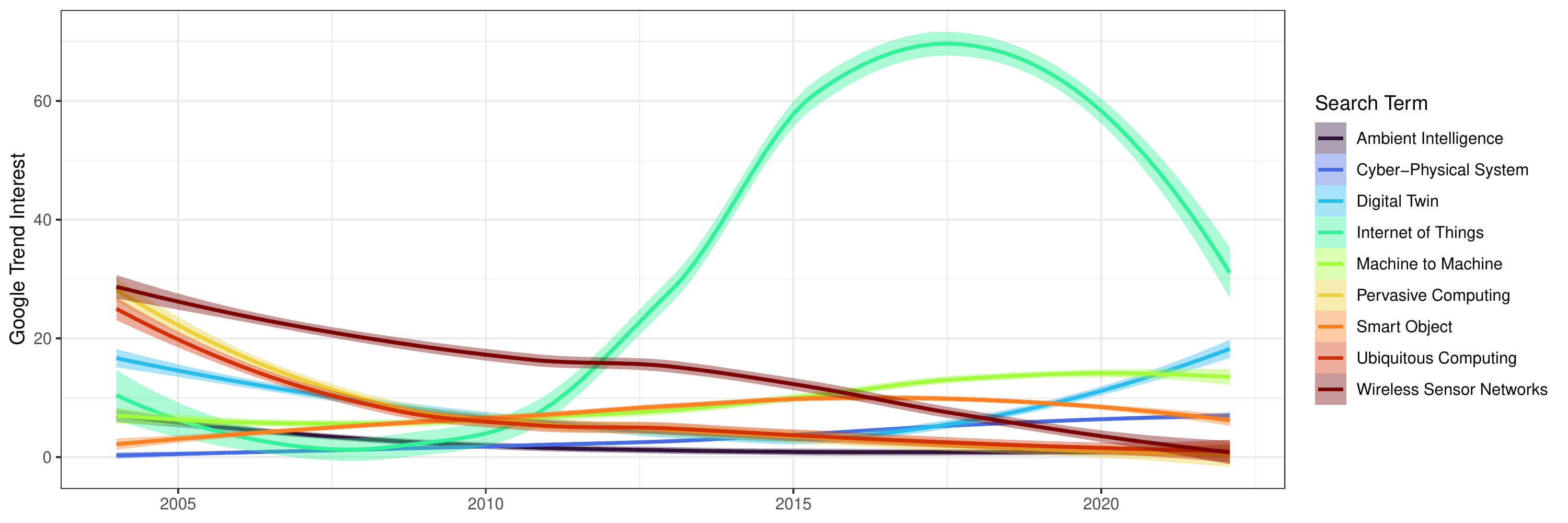}
	\caption{\textbf{Google search trends by search term in web searches}.
		Interests are averaged with LOESS across months from 2003 to early 2022. In general, Internet of Things has been falling since 2017, and Digital Twin has been steadily rising. This year, Digital Twin has become the most searched term among the cognates other than Internet of Things. Terms for underlying technologies and previous system names are stagnating or declining in use. Data collected from \url{https://trends.google.com} on the 8\textsuperscript{th} of February 2022.}
	\label{fig:iot:trend}
\end{figure*}

\subsection{The Cognates of Digital Twins}
\label{sec:digital:twin:cognates}

DTs are a relatively new perspective that shares a common but amorphous root with a variety of other perspectives . 
While this root may be difficult to trace precisely and may differ from the intent of the author, it may have originated with various thinkers such as Wiener~\cite{wiener1961cybernetics}, Weiser~\cite{weiser1991computer}, or even Tesla \cite[p. 122]{atzori2017understanding}.
Given this complexity, we borrowed the term ``cognate'' from linguistics~\cite{carroll1992cognates} that refers to words with a formal resemblance or a common historical root.
The confusing overlap among DT cognates is driven by developments across a diversity of research backgrounds such as information science, production engineering, data science, and computer science~\cite{tao2018digital}. 
The different components of the DT are prioritised differently in each field, which results in similar yet slightly different meanings.
For instance, the cognate term IoT is dominating the public sphere according to search word trends (see Figure~\ref{fig:iot:trend}).
However, as hype gives way to pragmatism, simply having the physical infrastructure necessary for the creation of DTs is insufficient by itself.
The two cognates IoT and DTs are often confused, but IoT may form the basis of DTs in a similar manner that the Internet forms the basis of the Web.
Although it is too early to determine, search word trends may eventually place DTs or another cognate as the ultimate concept to describe these systems.
The following sections clarify the nuances associated with DTs and their cognates across different disciplines.

\subsubsection{Cyber-Physical Systems}
\begin{quote}
A Cyber-Physical System (CPS) represents physical objects as virtual objects. Users can with either physical or virtual objects to change the state of both. A DT is a special instance of a CPS that includes a model of the PT for predictive operations.
\end{quote}
The terms DT is relatively new because the original idea of digital representation was first described in detail by Wiener~\cite{wiener1961cybernetics} in the field of control theory.
Later, the term Cyber-Physical Systems (CPSs) Theory~\cite{baheti2011cyber,lee2015past} was coined to describe a \emph{physical} object, homologous to a ``Thing''~\cite{huang2010descriptive} in the IoT context and the PT in the DT context.
This physical object has a \emph{cyber} correspondence, which is homologous to the ``Digital Twin''~\cite{grieves2017digital}.
\emph{A priori}, CPSs have no predetermined (physical) scope and are an idealised form to describe possible interactions with entities ~\cite{rho2016cyber,darwish2018cyber}.
Therefore, the CPS literature is rarely linked directly to DTs, remote sensing, or the IoT (but see~\cite{alam2017c2ps,um2019drones}).
CPSs are increasingly used to capture social data as well~\cite{sheth2016internet}.
Their high level of abstraction makes CPS a suitable paradigm to theorise DTs, and findings in the CPS literature are very relevant to DTs.

\subsubsection{(Wireless) Sensor (and Actuator) Networks}
\begin{quote}
	(Wireless) Sensor (and Actuator) Networks (WSN/SN/SAN/WSAN) provide sensing and actuating infrastructure that processes the signals to and from the physical world to enable a CPS or DT.
\end{quote}
A DT cannot be conceptualised without having first digitised the PT.
While the DT is usually presented as agnostic with respect to its underlying hardware, the DT cannot be constructed without it. 
Creating this representation requires sensing to comprehend the physical world and actuating to change the physical world~\cite{akyildiz2002survey, akyildiz2002wireless}. 
Sensor networks ground any Cyber-Physical System in the physical world by provided readings thereof.
For each device, questions of how to implement ``autonomicity, scalability, energy efficiency, heterogeneity in terms of user equipment capabilities, complexity and environments''~\cite{rawat2016cognitive} need to be answered.
Sensor networks have been researched long before the origination of the terms IoT or DT.
However, contemporary research focuses more on how sensor networks are seen as a component of an IoT system~\cite{mainetti2011evolution}.

In a sensor network context, there are active and passive devices that can be placed in the environment.
Passive devices called Radio Frequency Identification (RFID) harvest energy from the transmission signal via induction to return a signal~\cite{welbourne2009building}.
The signal can be used as an identification of an object in order to trigger a software-based service or to determine its location depending on the RFID reader~\cite{li2019review}.
Typically, RFID cannot return environmental information (but see~\cite{shao2018next}).
Active devices are usually battery powered and have actual computational infrastructure, sensory equipment, and possibly actuators to interact with the environment.
Whether active or passive devices are required depends on the task.
A DT will often require a combination of both active and passive devices to efficiently represent the PT and the processes therein.
A comparison of tasks and the technology that best captures them can be found in Table~\ref{tab:iot:devices:tasks}.

\begin{specialtable}[!htbp]
	\caption{Task Classification for Active and Passive Devices} 
	\centering
	\resizebox{\columnwidth}{!}{
		\begin{tabular}{rl*{3}{c}}
			\toprule
			Task & Description & Active & Passive\\
			\midrule
			Accessing &  \makecell{Accessing information on an object linked to the device.} & \checkmark & \checkmark\\[5pt]
			Actuating & Triggering an action of the object linked to the device. & \checkmark & \\[5pt]
			Locating &  \makecell{Identifying the location of an object linked to the device.} & \checkmark & \checkmark\\[5pt]
			Sensing & Measure characteristic of the object linked to the device. & \checkmark & \\[5pt]
			\bottomrule
			\label{tab:iot:devices:tasks}
		\end{tabular}
	}
\end{specialtable}

Without using the term DT, RFID implement a similar abstraction because a ``physical object is accompanied by a rich, globally accessible virtual object that contains both current and historical information on that object’s physical properties, origin, ownership, and sensory context~\cite{welbourne2009building}''. 
RFID helps to embed access to knowledge in the environment and can make access to the DT context-driven~\cite{welbourne2009building}.
For example, if users and environment are both tagged, the appropriate DT services can be called to facilitate the users' task.
At the same time, the system can track such interactions to help model activity for the DT.
However, RFID tags could reveal sensitive information, and it is important to manage access control~\cite{rastogi2008access}.

In contrast to passive devices, active devices usually have local computational power but may be constrained, which should be classified according to usability.
Resource constraints on active devices can be grouped into six categories: battery, communication range, bandwidth, memory storage, processing capability, half duplex communication~\cite{raza2017critical}.
The IETF has classified devices based on hardware capabilities~\cite{bormann2014terminology}, and each class requires different operating systems (OS) to function optimally~\cite{hahm2016operating, javed2018internet}.
They range from Class 0 devices to Class 2 devices.
Class 0 devices have a tailored OS that provides no functionality beyond the required features.
In contrast, Class 1 devices and above may already run a generalised if bare OS fulfilling the criteria of a small memory footprint, support for heterogeneous hardware, network connectivity, energy efficiency, real-time capabilities, security, and opportune scheduling policies~\cite{hahm2016operating, javed2018internet}.
Active devices can be fully integrated into DTs and can be managed from within the DT, whereas passive devices currently only provide anchor points for DTs.

\subsubsection{Smart Objects}
\begin{quote}
	Smart Objects are physical objects that have embedded computational power to provide interactions for its virtual object to make it more interesting and may represent the fusion of the PT and DT.
\end{quote}
A DT can be understood as a highly accurate model of a physical system~\cite{boschert2016digital,grieves2017digital, batty2018digital}, whereas a Smart Object (SO) offers interesting interactions with a particular object through embedded or remote computing power~\cite{sanchez2012adding}. Both DTs and SOs represent a CPS~\cite{lee2015past}.
It could be argued that the concept of an SO either incorporate or encapsulate the concept of a DT. 
However, a DT is usually required to mirror the role of an actual object in a larger context, whereas SOs only provide end-users with interactions with the particular object~\cite{sanchez2012adding}.
Nonetheless, many SOs will represent the underlying physical object sufficiently close to use them interchangeably for a DT.
Assuming that an SO qualifies as a DT and that the SO can offer physical access to its digital representation, it can be considered a dual \emph{embodiment} for both the object (i.e., PT) and the DT, providing a physical instance of the FTs paradigm (see Figure~\ref{fig:digital:twin:smart:objects}).

However, a DT may also describe an entity that is composed of multiple SOs, other Things, processes, and high-order entities.
These entities may entail buildings, cities, or regions~\cite{li2011smart, su2011smart, sanchez2014smartsantander,perera2014survey} modelled through remote sensing~\cite{lee2022geospatial,botin2022digital,kothari2022multi}.
Moreover, a DT may also be formed of Things that are immaterial or at least incomprehensible in the physical world such as phantom traffic jams~\cite{gazis1992moving}.
These can be considered purely virtual objects in contrast to virtualised objects~\cite{nitti2016virtual}, which would have a mapping back to a physical object.
Composition allows for the formation of even higher order virtual objects that form the basis of a DT in a complex system such as the processes or aggregates representing a city~\cite{rathore2016urban,kaur2020convergence}, traffic flow~\cite{anda2021synthesising}, or production line~\cite{tao2017digital}. 
Furthermore, a DT may also be produced exclusively by remote sensing without having local sensors attached to the PT~\cite{botin2022digital}.
A DT is thus a more encompassing concept than a SO.

\subsubsection{Ubiquitous and Pervasive Computing / Ambient Intelligence}
\begin{quote}
	Ubiquitous Computing is a paradigm that expects computation to be available for all objects to form a CPS, whereas Pervasive Computing specifies the computation to occur at the object or close to it. SO and Ambient Intelligence are instances of Ubiquitous Computing in CPSs in which the sensing and actuating capacity is directly embedded into the object that is described by the corresponding DT.
\end{quote}
Ubiquitous and pervasive computing are usually considered interchangeable.
The former is an older term to link computations to any object in the physical world~\cite{weiser1991computer}.
The latter is a younger term and associated with placing computing power within (or close to) any object in the physical world~\cite{satyanarayanan2001pervasive}.
Without loss of generality, we will only refer to Ubiquitous Computing in the remainder of the present paper.
For a DT, any data collected in the physical world must be represented in ``cyberspace'' and thus requires computation.
Many ubiquitous computing applications pre-date the IoT paradigm but have been subsumed by it (see~\cite{ma2005towards}).

The closeness between the Thing and computations performed on sensor data from the Thing has often resulted in them being regarded as equivalent to each other~\cite{atzori2010internet}.
Similarly, the concept of a DT is often used as a synonym for both the PT (homologous to the Thing) and the Physical Environment with WSAN (homologous to computations on sensor data from the Thing) from which the DT is created.
Ubiquitous computing offers us a lens to distinguish between PT and DT by considering the difference between computations on the state of an object and the actual object.
SOs~\cite{kortuem2010smart,fortino2012agent,sanchez2012adding,atzori2014smart} are a typical example where the distinctions between PT, DT, sensors, actuators, and computations are blurred.
SOs are expected to have awareness of their environment and have some model of the ongoing activities, policies, and processes around them.
At the same time, SOs should represent functions, rules, and flows that allow for interactivity~\cite{kortuem2010smart,sanchez2012adding}.
This formulation clearly indicates that the SOs have a Physical Environment in which they are embedded, an Analytical Environment to model their behaviour, a Data Environment to maintain their state, and a Virtual Environment in which their interactivity can be triggered.
When the computations on the object or Thing become unrecognisable to humans because they are ``woven into the background'', as predicted by Weiser~\cite{weiser1991computer}, the term Ambient Intelligence becomes more relevant ~\cite{flugel2009scientific,dohr2010internet,cubo2014cloud,darwish2018cyber}.
Making a clear distinction between a Thing, a SO, and Ambient Intelligence is probably neither possible nor desirable.

\subsubsection{Internet of Things}
\begin{quote}
    The IoT is an amalgam of competing technologies with the goal of digitally representing the physical world. It interconnects WSANs to be combined into local CPSs that in turn compose larger DTs.
\end{quote}
The idea behind IoT was first hyped~\cite{tan2010future,khan2012future,stankovic2014research}, then overestimated, and then slowly became practical~\cite{hurlburt2012internet}.
The term IoT was probably coined towards the end of the last millennia~\cite{ashton2009internet}, but its roots are older, albeit unnamed.
The functional similarity of CPSs~\cite{wiener1961cybernetics} and ubiquitous computing~\cite{weiser1991computer}, as well as stipulations from early thinkers such as Nikolai Tesla \cite[p. 122]{atzori2017understanding}, make its historical provenance a question of choosing a remotely related cognate.
Following Atzori and colleagues~\cite{atzori2017understanding}, we take the probable origin of the term ``Internet of Things'' by Ashton in 1999~\cite{ashton2009internet} as a starting point to limit ourselves to cases where the general idea of IoT was already the goal of research.

Nearly anything remotely related to Information and Communication Technologies (ICT)~\cite{hurlburt2012internet} is included in modern IoT research~\cite{atzori2010internet,gluhak2011survey,chen2012challenges,gubbi2013internet,borgia2014internet,singh2014survey,want2015enabling,al2015internet,whitmore2015internet,madakam2015internet,palattella2016internet,lin2017survey,qiu2018can,javed2018internet,shafique2020internet} because IoT promises to be the invisible glue to digitally represent the world and offers new means of analysing and understanding~\cite{weiser1991computer}.
In part, this may be due to the well-known term \emph{Internet} that immediately allows people to associate it with global information infrastructure~\cite{huang2010descriptive}.
On the other side, a \emph{Thing} is vague enough to label any entity that can be distinguished by some measure~\cite{huang2010descriptive}.
The Intranet of Things in a WSAN has been transformed through ICT~\cite{zorzi2010today, wu2017design,shafique2020internet} and cloud-centric processing~\cite{gubbi2013internet,alamri2013survey, misra2017theoretical} to the IoT.
This loose description is what allowed the IoT to soak up many kinds of ICT developments for the last decade (see Figure~\ref{fig:iot:trend}).

While the term IoT has recently become the most common term for the linkage between the virtual and the real, it was not always the case.
Previously, a set of competing ideas (e.g., CPS, WSAN, and Ubiquitous Computing) co-created the ecosystem that today is generously labelled IoT and which laid the hardware foundation for DTs.
These ideas were closely related but ultimately based on different assumptions because they originated from different fields.
According to public perception, the terminology has settled on IoT even as the underlying technology is still quickly diversifying and expanding.
With current trends, DTs appear to be a concept that is set to overtake the IoT in terms of public perception (see Figure~\ref{fig:iot:trend}).

Another issue is that IoT is a buzzword that is difficult to grasp because it is overloaded with meaning.
Simply tapping into the literature is overwhelming and, at the same time, highly specific.
The many technologies peddled under the umbrella-term IoT would be more distinctive if labelled specifically according to their function.
While the technological umbrella-term  is now well-established, if ill-defined, there is still a lingering sense that different fields of application are uniquely distinct and require special attention.
Research and development is thus heavily directed towards five distinct real-world applications that promise the biggest gains: industrial application, urban management, health care, vehicular applications, and end-user applications~\cite{atzori2010internet,borgia2014internet,kamilaris2016mobile}.
This ingrained belief regarding the distinction leads to difficulties in synthesising common knowledge across applications.
Overcoming this barrier could help the interdisciplinary fields of IoT and DTs in many ways.
Not only would it enable learning from each other to clarify their theoretical backgrounds, but it would also make the use of IoT as a basis for DTs more accessible.
Indeed, a clearly communicated and unified approach to IoT would lower the barriers of entry tremendously.

Weiser~\cite{weiser1991computer} postulated the goal of Ubiquitous Computing to make the computations disappear in the mundane of the environment.
To obtain services that hide in plain sight, three categories of design considerations have been brought forward that need to be considered for IoT to work: fundamental properties of a Thing, network communication properties, and real-time interaction properties.
The fundamental properties of a Thing include existence, sense of self, connectivity, interactivity, dynamicity and sometimes environmental awareness~\cite{roman2011securing,rawat2016cognitive, raza2017critical}.
Much research has also been conducted on properties to make network communication more efficient, including energy, latency, throughput, modular design for improved scalability, topology, security of information, safety~\cite{flugel2009scientific, rawat2016cognitive, raza2017critical}, deployment, mobility, cost, efficient resource usage, heterogeneity, communication modality, infrastructure, coverage, network size, and lifetime~\cite{rawat2016cognitive, raza2017critical}.
Real-time requirements in industrial and healthcare settings have brought forth another set of properties that need to be considered to enable real-time interaction with IoT networks: predictable delays and latencies, congestion free communication, accurate time synchronisation, real-time assurance, interoperability with existing infrastructure, and prioritised communication~\cite{raza2017critical}.

The historic development IoT and the evolution of definitions has been addressed in detail by \citeauthor{atzori2017understanding} ~\cite{atzori2017understanding}. The work identified three generations of IoT development with different foci that have resulted in the current diffusive definitions (see Table~\ref{tab:iot:generations}).
An additional new generation is identified in this review paper that exemplifies the most current trends of DT research in IoT that was not reported in~\cite{atzori2017understanding}.
Notably, while these four generations focus on distinctive features, they are effectively contemporary because these generations are still actively researched.
To make sense of these concurrent generations, it is best to understand each generation as a necessary condition to start work on the following generations.

\begin{specialtable}[!t]
	\caption{Waves of IoT generations} 
	\centering
	\resizebox{\columnwidth}{!}{
		\begin{tabular}{r*{4}{c}}
			\toprule
			&Generation I&Generation II&Generation III& Generation IV\\
			\midrule
			\makecell[r]{Earliest\\Mention}& 1999 & 2010 & 2011 & 2012\\[12pt]
			\multirow{3}{*}{\makecell{\\[-1pt]Technologies}} & RFID & Inter-networking &  Cloud computing & Fog Computing\\[4pt]
			& M2M & Web of Things & ICN & DTs\\[4pt]
			& Object integrations & Constrained devices&Semantic IoT & Opportunistic IoT \\[12pt]
			Focus & Digitalisation & Networking & Centralisation & Decentralisation \\[12pt]
			Source &\cite{ashton2009internet,atzori2017understanding} &\cite{pfisterer2011spitfire,atzori2017understanding} &\cite{atzori2017understanding,gubbi2013internet} & own definition, \cite{bonomi2012fog,guo2013opportunistic,grieves2017digital}\\
			\bottomrule
			\label{tab:iot:generations}
		\end{tabular}
		
	}
\end{specialtable}

The first generation is driven by the rise of RFID tags~\cite{ashton2009internet} to digitally distinguish Things.
At this time, IoT was often taken to be synonymous with RFID.
This generation is tightly coupled to Ubiquitous Computing~\cite{weiser1991computer}, Machine-to-Machine (M2M), and CPSs as the RFID technologies ultimately derive from these research fields.
However, by the late 2000s, most of these technologies had been tagged IoT.
The second generation is driven by interconnecting Things instead of just centralising the information flow.
It focuses on the network component of IoT devices and realistically brings the \emph{Internet} into the IoT~\cite{pfisterer2011spitfire}.
Control of information becomes more central as well, reinforcing the connections to its roots in CPSs~\cite{wiener1961cybernetics}.
The third generation is driven by a re-centralisation around cloud technologies that matured meanwhile~\cite{gubbi2013internet,alamri2013survey, misra2017theoretical}.
This generation focuses on processing data and generating knowledge.
The fourth generation is driven by a second wave of decentralisation due to time-sensitivity of applications, fog computing~\cite{bonomi2012fog}, and the rise of DTs~\cite{grieves2017digital} and servicisation \cite{agrawal2016potential}.
As questions of processing have been addressed, the newest generation brings more efficiency to IoT.
IoT is often mentioned as a multiplier that makes other technologies more attractive~\cite{atzori2017understanding}, contributing to the diffusion of the term.

\subsubsection{M2M}
\begin{quote}
	Machine-to-Machine (M2M) communication allows Things and network components to interact automatically with each other without human actions, thus enabling automated formation of CPSs or DTs on different scales.
\end{quote}
The complexity of an IoT system underpinning a DT requires much communication between each involved Thing and the network infrastructure to support a control system ~\cite{wiener1961cybernetics,paraskevakos1974apparatus,lee2015past}.
These communication protocols have been aggregated and named M2M communication.
M2M is an ongoing research field and is usually required for all other IoT-related and DT-related activities~\cite{datta2014iot,andreev2015understanding,gazis2017survey,patil2021systematic}.
Essentially, M2M describes how IoT components interact with each other and can be understood as part of the implementation of the Connection Environment of a DT.
As such, M2M, IoT, and DTs cannot be meaningfully separated, although they do not constitute the same concept. 
Before IoT became the defining word for ICT, M2M was a contender for being the umbrella-term for these technologies.

The number of devices connected to IoT today surpasses the number of devices manually connected to the Internet.
Thus, it is impossible to manually manage the communication between devices by connecting them or maintaining them~\cite{baronti2007wireless, yick2008wireless, khorov2015survey, raza2017critical}.
This aspect triggered the development of the M2M concept in the early 2000s~\cite{wang2017survey, gazis2017survey} followed by standardisations such as oneM2M~\cite{swetina2014toward} in the 2010s.
At the centre of M2M activities are the protocols that are used to exchange information between devices.
Protocols describe predefined procedures of how to initiate, maintain, and reestablish communication~\cite{swetina2014toward}.
While some protocols are high-level, they are often bound by the physical limitations of the communication capabilities of the constrained devices.
The complexity of the classical standard Internet Protocol (IP) was quickly disqualified as a platform~\cite{ishaq2013ietf} because of the assumptions that the computing infrastructure was powerful, wired, and connected and that communication was a negligible overhead. 
However, constrained devices have been highly optimised in terms of energy consumption, size, and costs such that communication has become the major bottleneck~\cite{sheng2013survey}.

The physical infrastructure to connect constrained devices to the network is usually referred to as physical / PHY~\cite{granjal2015security}, and the data link / MAC address is used to uniquely identify single units~\cite{bachir2010mac,rajandekar2015survey,sotenga2020media}.
These terms originate from the OSI model and refer to the physical and data link layers between devices~\cite{iso1983basic}.
Originally, a link would be a copper or fibre cable.
However, wireless transmission has adopted the same terminology, so PHY usually refers to some kind of antenna that uses a pattern of signals to encode information~\cite{kothari2022multi}. 
To enable a more efficient use of bandwidth and potentially allow for more devices, Cognitive Radios allow secondary users to use communication channels while primary users are idle~\cite{chen2014machine,aijaz2015cognitive,andreev2015understanding}. 
The main trade-off for the PHY connection is between data rates, connection distances, and energy consumption.
Currently, technology can be approximately grouped into four application areas: Body Area Networks (BAN), Personal Area Networks (PAN), Local Area Networks (LAN), and Wide Area Networks (WAN) (see Table~\ref{tab:phy:technologies} for an overview).

\begin{specialtable}[!htbp]
	\caption{Communication technologies for PHY layer} 
	\centering
	\resizebox{\columnwidth}{!}{
    	\begin{tabular}{r*{3}{c}}
    		\toprule
    		Scope & Abbreviation & Technologies & Reviews\\
    		\midrule
    		Body Area& BAN & BLE, Zigbee & \cite{akyildiz2010internet,akyildiz2015internet,gravina2017multi,miraz2018internet,singh2021wireless}\\
    		Personal Area & PAN & UWB, Bluetooth, WiFi & \cite{arslan2006ultra,tozlu2012wi}\\
    		Local Area& LAN & WiFi, LTE & \cite{tozlu2012wi,ratasuk2016overview}  \\
    		Wide Area& WAN & LoRaWAN, Sigfox,  NB-IoT & \cite{raza2017low,nolan2016evaluation,oliveira2011routing,bor2016lora, petajajarvi2017performance, georgiou2017low, gruebel2021feasibility}\\
    		\bottomrule
    		\label{tab:phy:technologies}
    	\end{tabular}
	}
\end{specialtable}

The large standardising bodies, IEEE and IETF, approach protocols in constrained environments based on expansion of the IEEE Standard 802.15 ~\cite{palattella2013standardized, kurunathan2018ieee} and the IETF LLN protocol~\cite{sheng2013survey, ishaq2013ietf} to handle IP.
In addition, there are technological niches that have not been fully covered by those standardising organisations such as the Low Power Wide Area Networks (LPWAN) where protocols vary and are highly incompatible, representing a low level of industry cooperation~\cite{raza2017low,nolan2016evaluation,oliveira2011routing,bor2016lora}.
However, several consortia have developed around the different core technologies associated with each protocol, and only subsequent years will show how prevalent each WAN technology has become.

\begin{figure*}[htbp!]
	\centering
	\includegraphics[width =0.95\textwidth]{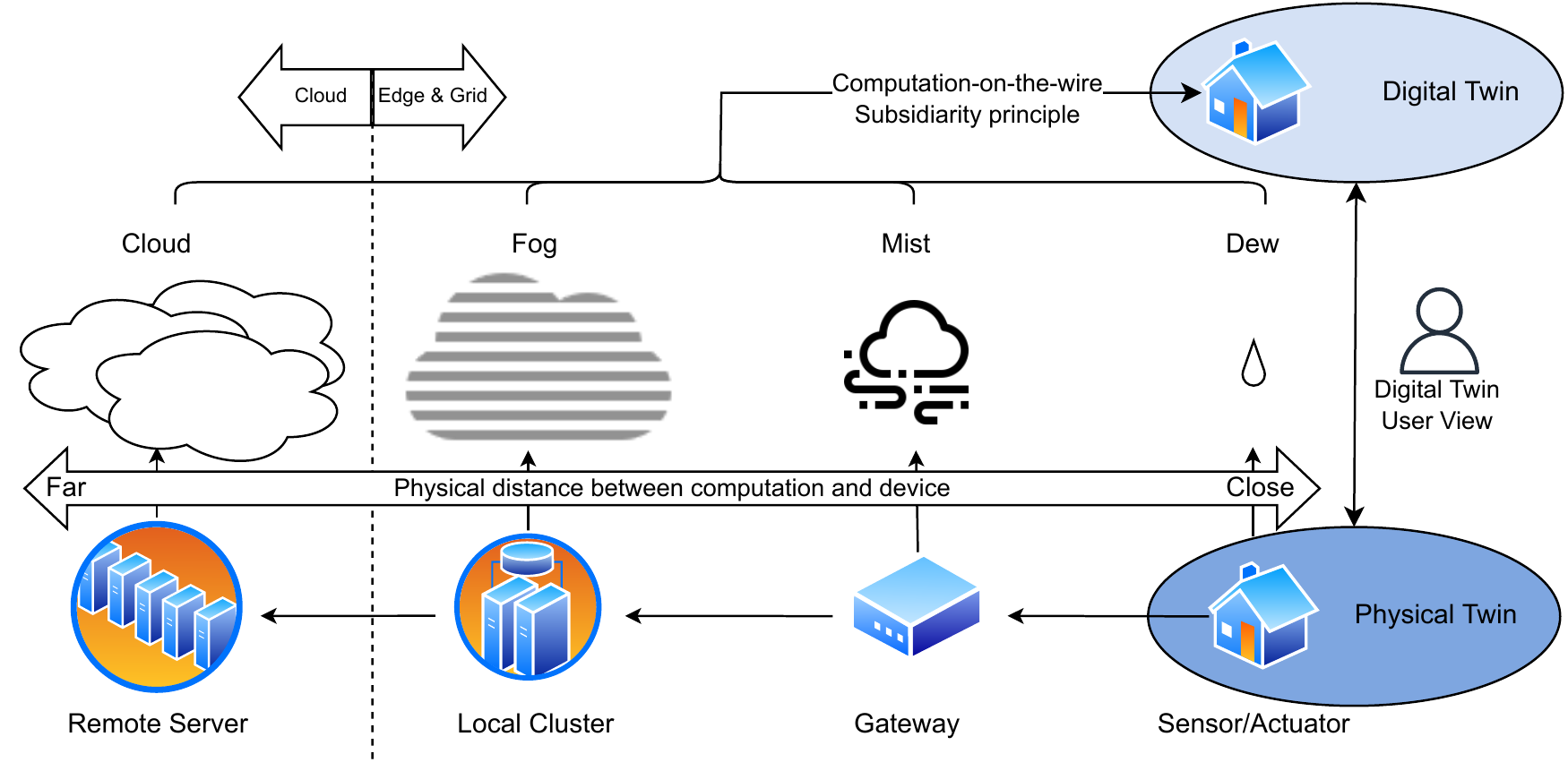}
	\caption{\textbf{Computing Hydrology of Computation-on-the-wire}.
		Computing power can be located locally at the sensor/actuator or more remotely at the gateway, a local cluster, or remote server.
		The computations form the basis for a DT.
		Computing in IoT and smart cities is often conceptualised with a hydrological terminology (i.e., cloud, fog, mist, and dew) that maps to the computing location and encapsulates different trade-offs in computational intensity, traffic, latency, accuracy, energy consumption, and security~\cite{flugel2009scientific,roman2011securing}.
		These trade-offs are best evaluated under the subsidiarity principle to decide where to place the computation physically.
		Edge or grid computing~\cite{shi2016edge,hossain2018edge} is another terminology to covers all non-cloud computing.
		DTs rely heavily on these different data and computing sources while at the same time abstracting away their existence in the end users' view. Generated with \url{http://draw.io}.}
	\label{fig:computing:hydrology}
\end{figure*}

The commonality among all protocols is that they are simplified to move data from the constrained devices to some kind of translation unit that transfers the data into the IP protocol, making it widely available on the Internet~\cite{hui2008ip, hong2010snail, rodrigues2010survey}.
This translation is necessary to cope with the reduced computing capacity of constrained devices as well as questions of how to transmit data and scale local IoT networks for use in a DT.
A difficulty arises for these protocols because the translation unit of constrained devices do not only send data (such as with sensors) but may also receive data (such as with actuators).
This aspect requires a bidirectional translation protocol that can uniquely address every node from the outside~\cite{imani2018comprehensive}.

\subsubsection{Computation-on-the-wire / Dew / Mist / Fog / Cloud and Edge Computing}
\begin{quote}
	Computation-on-the-wire computes all necessary transformations on data at the best possible point in the network according to the \emph{subsidiarity principle} to deliver Ubiquitous Computing for CPSs such as DTs.
\end{quote}
The trend to locally compute, distribute, and store data in end-user devices and local clusters rather than centralised data centres~\cite{bonomi2012fog,bonomi2014fog} is a reaction to the third Generation of IoT's \emph{cloudification}~\cite{atzori2017understanding}.
At the same time, the trend away from cloudification is the foundational moment for the fourth Generation of IoT.
Furthermore, sensors generally pose a trade-off between computational power, latency, cost, accuracy, energy consumption, and security~\cite{flugel2009scientific,roman2011securing,stojmenovic2014fog,jalali2016fog}.
To balance these trade-offs, additional computational power was introduced physically \emph{closer} to the sensors, which also limited data transmission and reduced the load on the network.
The placement of computational units in the physical world to provide the necessary computing capacity for Ubiquitous Computing produced a set of \emph{hydrologically} inspired terms to describe different trade-offs.
The division of remote computing into cloud~\cite{soliman2013smart,botta2014integration,fortino2014integration,cavalcante2016interplay,singh2016twenty,alam2017c2ps,bittencourt2018internet}, fog (also edge or grid)~\cite{bonomi2012fog,aazam2014fog,satyanarayanan2015edge,shi2016edge,chiang2016fog,mao2017survey,naha2018fog,stojmenovic2014fog,vaquero2014finding,giang2015developing,cavalcante2016interplay,dastjerdi2016fog,aazam2018offloading,bittencourt2018internet,sarkar2018assessment,hossain2018edge,negash2019towards,alam2019autonomic}, mist~\cite{liyanage2016mepaas,dogo2019taking}, and dew~\cite{hong2013mobile,liyanage2016mepaas,rindos2016dew,ray2017introduction,gushev2020dew} is an attempt to follow this principle (see Figure~\ref{fig:computing:hydrology}).
While the analogy may have been taken a bit too far, this simile is quite adept at equating the distance from the observer to the water droplet and the distance from the device to the computation location.
At the same time, the high level of differentiation and the unclear boundaries between these concepts---sometimes used interchangeably---makes the discussion more complicated.
For example, fog, mist, and dew computing are often used interchangeably across papers, and edge and grid computing often fit partial definitions of the hydrological terminology.

Another, less metaphorically laden, view focuses on \emph{Computation-on-the-wire} to describe the same phenomena~\cite{bhattacharyya2018wait}.
This perspective assumes that data needs to be manipulated between the data generation point (i.e., a Thing at a physical location) and the data consumption point (i.e., a user inquiring about something related to the PT via its DT).
To be the most cost-effective, the computation is required to take place as close to the generation and consumption points as sufficient and as far away as necessary.
Therefore, the computation-on-the-wire concept is essentially the application of the subsidiarity principle.
Taken from political science, subsidiarity implies that a central authority should only perform tasks that cannot be performed more locally~\cite{oxford2018subsidiarity}.
The principle is most famously applied in the European Union and should be taken as the guiding principle for IoT as well.
On a more abstract level, it implies that actions regarding an entity should be taken \emph{a)} as close as possible to an entity and at the same time \emph{b)} as far as necessary to successfully execute the action.
Applied to Computation-on-the-wire and DTs, this principle ipmlies that computations for DTs should occur \emph{a)} as close to a device, a requester, or the PT as possible and \emph{b)} as far into the cloud as necessary to successfully compute the output. 
The concept elegantly joins the hydrological simile and the underlying trade-offs by applying the subsidiarity principle~\cite{oxford2018subsidiarity}.

\subsubsection{Remote Sensing}
\begin{quote}
	Remote sensing has two meanings. First, it provides observations recorded by a distant measurement device with some form of camera.
	Second, remote sensing refers to distant access to a local measurement device.
	Both can be processed to represent the physical environment around a CPS (such as a DT) and to possibly interact with the physical environment.
\end{quote}
Remote sensing can be easily misunderstood as either exclusively meaning distant measurements based on some kind of camera~\cite{campbell2011introduction} or using remote access to sensing units~\cite{viani2012pervasive,sutaria2013making} to measure local properties from afar such as in agriculture~\cite{triantafyllou2019precision}, healthcare~\cite{ruiz2017empowerment}, smart cities~\cite{demetri2019automated}, or everywhere~\cite{abdelwahab2014enabling}.
Traditionally, remote sensing exclusively refers to camera-based, often airborne, approaches~\cite{campbell2011introduction}.
However, the secondary meaning of remote sensing as a shorthand for ``remote access to local sensing units'' has become prevalent across nearly all applications that do not apply traditional remote sensing. 
Whereas there is a clear difference in the meanings of remote and sensing in both interpretations, recent applications combining the two such as sensor networks on Unmanned Aerial Vehicles (UAV) often linked to ground stations and the Internet~\cite{ismail2018internet,fraga2019review} may make the difference irrelevant in the long term.
More recently, for the context of the DT, the precise sourcing of information from the Physical Environment is not important.
Instead, the accuracy of the measurements must be reported to allow for a decision-making process that accounts for uncertainty.
Throughout this review, we understand remote sensing in both its meanings as the main source for information from the Physical Environment.
Furthermore, we assume that any form of traditional remote sensing can be made available in real-time as a virtual sensor in an IoT network~\cite{srivastava2005virtual,wang2014research,xiao2018towards}.



\section{The Fused Twins Concept}
\label{sec:fused:twins}

\begin{figure}[hbpt!]
	\centering
	\includegraphics[width =\columnwidth]{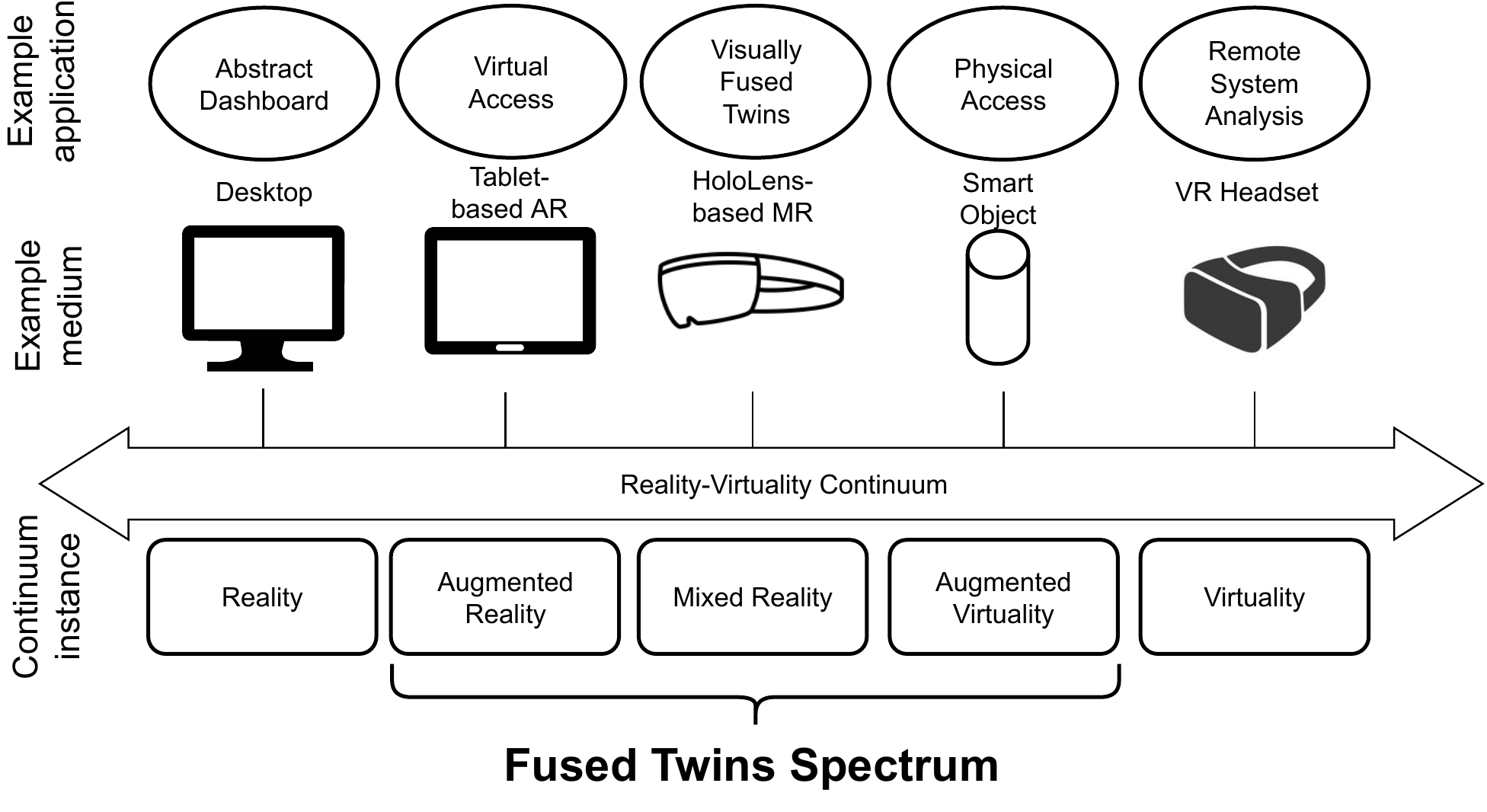}
	\caption{\textbf{Devices to access Digital Twins on the Reality-Virtuality-Continuum}~\cite{milgram1994taxonomy}. On the Continuum, FTs span instances of Extended Reality (XR) input modalities and media from Augmented Reality (AR) via Mixed Reality (MR) to Augmented Virtuality (AV). Typical media include tablets, phones but also holographic displays and smart objects. Reality-embedded access to DT focuses on abstract dashboards whereas Virtuality-embedded access focuses on remote analysis of the DT. The FTs Spectrum covers direct interaction with the PT that is fused with augmentations from the DT either displayed via AR on tablet, MR on HoloLens or haptically through AV smart objects. Generated with \url{http://draw.io}.}
	\label{fig:digital:twin:continuum}
\end{figure}

DTs require visualisation to unfold their potential.
Naturally, it is the Virtual Environment~\cite{gruebel2021feasibility, gruebel2016eve, prouzeau2020corsican} of a DT through which information be accessed.
To make sense of the different forms of visualisations, we use the Reality-Virtuality-Continuum~\cite{milgram1994taxonomy}, see Figure~\ref{fig:digital:twin:continuum} to classify a broad range of human interface devices. 

Commonly used implementations of DTs make use of visualisation types on both ends of the continuum which includes the Dashboard and Virtual Reality (VR)-based digital models.
Dashboards~\cite{batty2015perspective} reside on the Reality end of the spectrum because they display information independent of the physical world on screens which display numeric visualisations\footnote{Note that reality here is the continuum instance in which the user is located and not the medium through with the application is experienced.}.
The visualisation is not embedded directly in reality and is accessed in a non-immersive context.
On the other end of the spectrum, remote monitoring~\cite{bednarz2020digital,spielhofer2021physiological} is located in the Virtuality end and can be implemented both as VR and non-local AR.
The information usually has a digital model that is displayed in the virtual space with no direct relations to the physical space in which the data is consumed.

The FTs concept captures the situation where the DT is queried from within the PT allowing for a fused \emph{in situ} representation.
A media is used to connect the PT with the DT through the display of Situated Analytics~\cite{fleck2022ragrug}.
Possible media range from tablet-based (AR) and HoloLens (MR) to SOs (AV).
AR and MR devices can be used to overlay immersive analytics of the PT with the DT.
In contrast, SOs~\cite{sanchez2012adding} are interactables in the environment that allow users to manipulate the state of both the PT and the DT. 

Despite the similiarites, FTs are different from situated visualisations~\cite{thomas2018situated}.
FTs require a fully-fleshed DT behind the situated visualisation that includes all five types of environments and the ability to compose the underlying DT with other DTs that may be spatially or logically connected. 
FTs use the situated visualisation as an entry point to interact with all systems that are connected to that particular point in the PT to provide access to the whole PT through the DT. 
In contrast, many situated visualisations focus on local applications such as a smart switches for IoT devices~\cite{hua2021arciot}.
However, at first sight, the situated visualisation is often indistinguishable from FTs for end users.
Nonetheless, continuous interaction with the tool eventually reveals the limited functionality.
Thus, any situated visualisation can be expanded to implement the FTs paradigm by integrating it into a DT model.

This section will outline the DT Requirements to implement the FTs paradigm.
The particular features of the main environments are discussed, but the Data Environment and the Connection Environment are not discussed in detail.
The Connection Environment has not been implemented as interoperability between DTs or their components is as of yet uncharted territory.
Consequently, the Data Environment is usually a proprietary implementation that requires no further discussion here.
An open standard for DTs would open up the research on the Connection Environment and the Data Environment.
Lastly, while discussing the Virtual Environment, we reflect upon current situated visualisations that are close to implement the full FTs concept.

\subsection{The Physical Environment of the Digital Twin}

\begin{figure*}[htbp!]
	\centering
	\includegraphics[width =\textwidth,trim={0 0 0 0},clip]{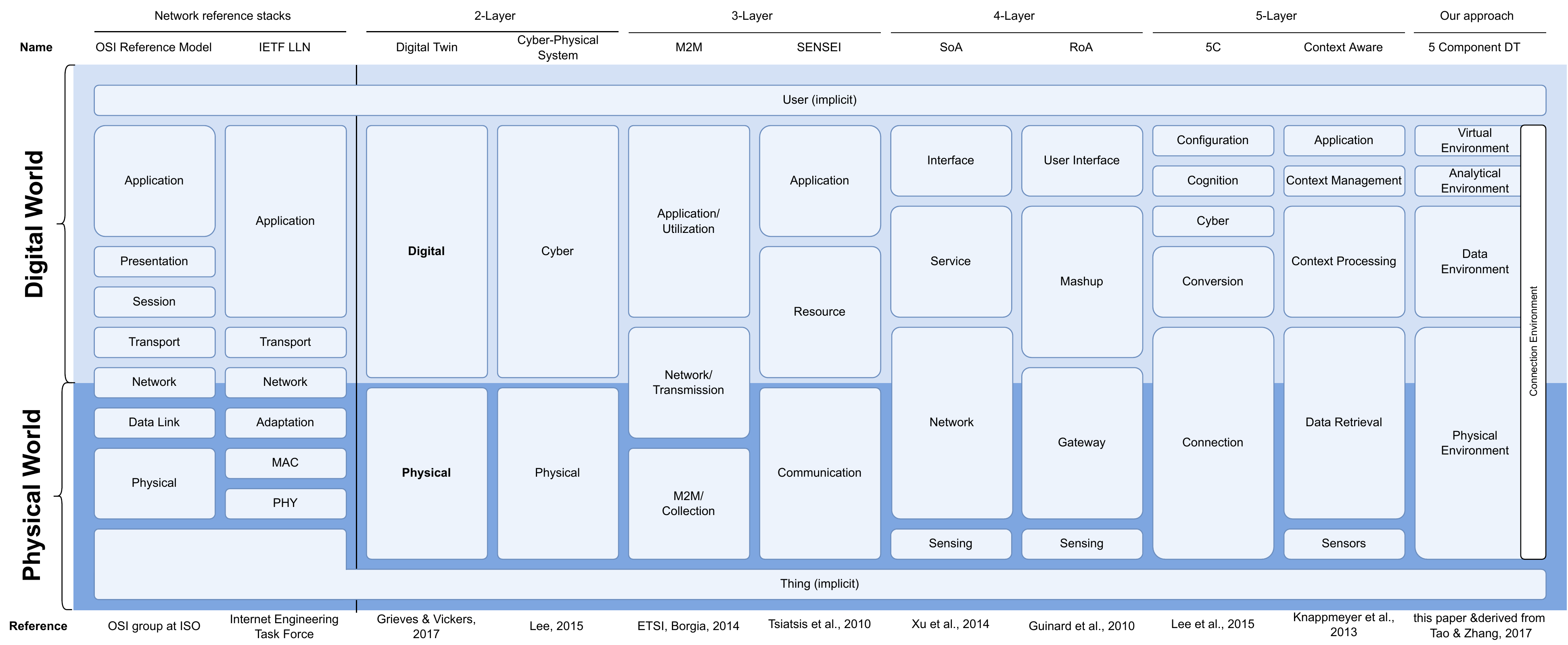}
	\caption{\textbf{Comparison of Digital Twin Architectures}.
		Data flows from the bottom (Thing) to the top (User).
		The functionality of each layer is displayed in comparison to other architectures where possible by spawning each layer vertically such that it contains all functionality that similar layers in other architectures provide.
		For example, Presentation and Session in the OSI Reference Model are subsumed in Application in the IETF LLN Stack.
		The labels at the top provide the name of the models, and the labels at the bottom references the origin of the respective model.
		\emph{From left to right:} OSI Reference Model, Variation of IETF Internet Model~\cite{kurunathan2018ieee}, Digital Twin~\cite{grieves2017digital}, CPS~\cite{lee2015past}, ETSI M2M~\cite{borgia2014internet} \& SENSEI~\cite{tsiatsis2010sensei}, SoA~\cite{xu2014internet}, 5C~\cite{lee2015cyber}, and Context Aware~\cite{knappmeyer2013survey}.
		Generally, management and security columns spawning all layers are not shown.
		Our own definition is mapped to these architectures in the last column.
		Generated with \url{http://draw.io}.}
	\label{fig:iot:architectures}
\end{figure*}

The Physical Environment encapsulates the infrastructure to cast the digital shadow~\cite{schluse2018experimentable} of the PT (see Figure~\ref{fig:shadow:model}) into the Data Environment.
When assembling an IoT or remote sensing system for a DT, it is important to be able to determine the dimensions which matter most for the functioning of the DT and choose an architecture accordingly.
Most research on architecture has been done in the context of IoT.
However, as the time progressed, architectures focused more on how to deal with the data and consequently today, these architectures are also used in DT cognates and DTs themselves.

International institutions have developed IoT reference architectures~\cite{weyrich2016reference,di2018internet,lynn2020internet}.
Architectures are usually presented as layers.
However, multiple architectures exist with a similar number of layers but different functional goals.
Going into detail of how each layer is supposed to work sometimes contradicts other architectures with the same number of layers (see  Figure~\ref{fig:iot:architectures}).
On the one hand, there is no unified architecture and due to the sheer endless number of application scenarios across different branches of research.
On the other hand, architectures are sometimes compatible, and we provide a mapping of the different components in Figure~\ref{fig:iot:architectures}.
These architectures may appear as fully-fleshed DTs at first glance but on closer inspection will reveal themselves to be mostly concerned with the Physical Environment and the Data Environment. However, depending on the overall goal of the architecture, the upper layers may actually contain features of the Analytical Environment and the Virtual Environment.
IoT architectures are the basis for most current DT applications and will probably morph into fully-fleshed DTs in the coming decade.

Architectures aim to be somewhat self-explanatory and ideally each layer covers exactly one functionality on transforming the data.
Realistically, layers help conceptually to assign tasks to different parts of the real system.
However, most implementations rely on a messier, less clean interaction that takes place between the components.
To choose the layer, it is typical to consider aspects of security, energy consumption, data collection density (spatial and temporal), maintenance, availability (real time vs long-term evaluation), interactivity (explicit reaction versus implicit reaction), and heterogeneity~\cite{flugel2009scientific,roman2011securing,rawat2016cognitive, raza2017critical}.
Lastly, management of the infrastructure and security concerns relate to every layer and are often represented as vertical columns to the side of the layers~\cite{borgia2014internet,atzori2010internet,javed2018internet}, but they are not shown in Figure~\ref{fig:iot:architectures} for brevity.
When comparing this to our DT design, it is those management layers that map to the Connection Environment that would ultimately allow to compose DTs given appropriate permissions. 

While the number of architectures is infinite, we will focus on a few of the most common approaches in more detail here.
The simplest architecture divides the world into two layers, the Physical layer and some abstract layer such as Digital~\cite{grieves2017digital} or Cyber~\cite{lee2015past,alam2017c2ps}.
Mapping these realms to our DT model, the Physical covers the Physical Environments and the Digital/Cyber covers the Data, Analytical, and Virtual Environments.
The Connection Environment maps most closely to the Management and Security columns in previous architectures but the focus is more on interoperability while also satisfying management and security requirements.
This architecture has no representation of how data is collected, transported, transformed, and attributed in the DT.
Nonetheless, it is a powerful vision that is easily understood despite the technological intricacies necessary to make such a vision possible.
This perspective strengthens ontological approaches to describe processes and qualities of objects in such a way that allows for interaction.
This concept is often applied in Industrial IoT (IIoT) where it is usually used as an end-user perspective on a Service-Oriented Architecture (SOA)~\cite{xu2014internet}.

For practical implementations, often a three-layer architecture is used that describes the detailed work in the middle~\cite{tsiatsis2010sensei,borgia2014internet}. These architectures are already able to address how the DT arises and as such a third layer is brought to attention: the middleware.
The physical world is sensed in a perception layer mapping to the Physical Environment.
Then, the data is structured into semantically meaningful units in the middleware layer mapping to the Data Environment and the Analytical Environment and finally presented to end-users in an application layer mapping to the Virtual Environment.
This architecture belongs to the earlier models and is focused more on the process of how to create basic IoT infrastructure.
The perception layer describes the process of data acquisition usually provided from sensors, actuators, and tags.
The middleware layer aggregates the data and provides APIs to access necessary information.
When comparing the three-layer model to the two-layer model, middleware is the hidden layer that converts the real world's PT into its DT.
However, most real systems stop short of complex analysis based on simulation and often limit itself to present raw data or aggregates. 

The application layer offers useful services on top of the data.
In a sense, those services are offered in the Virtual Environment of the DT (see Figure~\ref{fig:digital:twin:components}), but they heavily rely on the Analytical Environment to produce interesting services.
Naming conventions for the three layers vary, but their basic functionality remains the same such as in ETSI M2M~\cite{borgia2014internet}, SENSEI~\cite{tsiatsis2010sensei}.
Additional layers always focus on very specific features and mostly subdivided the Application layer for Service-oriented Architecture~\cite{xu2014internet}, Resource-oriented Architecture~\cite{guinard2010resource} business applications~\cite{lee2015cyber}, or Context Awareness~\cite{knappmeyer2013survey}.



\subsection{The Analytical Environment of the Digital Twin - Models and Services}\label{sec:service}

The Analytical Environment is often considered to be the DT itself~\cite{schluse2018experimentable} because it simulates the PTs behaviour. However, throughout this review paper, we have outlined more specific tasks for the Analytical Environment such as running simulations, producing aggregations, and providing services. We also have elaborated why a DT is more than just a simple simulation with real-world data. Nonetheless, it is the Analytical Environment that usually gives DTs and their Cognates the Smart label such as in Smart City.

The word \emph{smart} generally allures to the superiority of a (connected, digitised) Thing compared to its dumb (unconnected, analogue) counterpart.
Since connectivity and digitalisation on their own do not make anything \emph{smart} but only provide accessibility, the smartness implies easy interaction with the Thing, new insights into the Thing, and new applications of the Thing.
To attain any of these goals, models are required to represent reality, predict the future, and monitor ongoing processes.
For DTs, this means that services~\cite{de2011service,asghari2018service} are provided on top of the data through the Analytical Environment to enrich a PT's functionality and simplify interactions. 
In general, most Things will require a DT to become smart.

However, the term smart has been miss-used.
The previous description of \emph{smart} is applicable to any meaningful service. The use of ``\emph{Smart}'' as a prefix for services has proliferated to the point where it has become superfluous.
Following the above argument, the prefix has been attached to anything that could have a DT or has implemented features that are indistinguishable from a DT.
In the simplest case, a Smart Object provides Awareness, Representation, Interactivity, and Identity~\cite{kortuem2010smart,sanchez2012adding}. 
Smart Objects are concrete implementations of a DT (and even the FTs paradigm), but they limit the Thing to be a small-scale object with a fixed use. 
Smart Objects are also a form of data physicalisation~\cite{dragicevic2020data} that provides the opportunity to interact with data beyond a visual medium.

As objects increase in size and become places, the terminology adapts to Smart Environment~\cite{bonomi2014fog}, Smart Home~\cite{hui2017major,laput2019exploring}, Smart Office~\cite{laput2019exploring}, Smart Museum~\cite{chianese2014designing}, Smart City~\cite{su2011smart,suciu2013smart,soliman2013smart,vlacheas2013enabling,zanella2014internet,krylovskiy2015designing,rathore2016urban,puiu2016citypulse,hui2017major,alavi2018internet,kaur2020convergence} and even Smart Regions~\cite{matern2020smart} and Smart Communities~\cite{li2011smart,sun2016internet}. 
Conceptually, these Smart Places~\cite{glaeser2006smart} do not differ from Smart Objects~\cite{sanchez2012adding} but only generalise it to sets of Things.
This scaling aligns with the composability of DTs and allows for these concepts to be nested.

However, there is a difference to be made between physical Things and non-physical Things such as processes.
The Thing in this context is a non-physical entity but interacts with physical entities in a defined way allowing to define smart processes such as Smart Governance~\cite{meijer2016governing}, Smart Manufacturing~\cite{shrouf2014smart,lu2020smart}, Smart Construction~\cite{kochovski2018supporting}, Smart Health Care Services~\cite{rohokale2011cooperative}, Smart Mobility~\cite{faria2017smart}, and Smart Experiences (for touristic activities)\cite{kontogianni2020smart}. 

With everything labelled smart, it becomes more meaningful to talk about the kind of actions a Smart Thing can provide. Interestingly, it is these interactions that are also sought after in the immersive analytics context~\cite{marriott2018immersive}. We locate actions on a scale of how independently they can react to external input:
\begin{enumerate}
	\item Sensing: Only produce data
	\item Reacting: Use threshold \emph{on collected data} to trigger action
	\item Interacting: Use \emph{additional input} beyond collected data to change behaviour
	\item Analysing: Use a \emph{model to interpret collected data} when choosing actions
	\item Proposing: Use a \emph{decision model} to propose actions based on data models
	\item Deciding: \emph{Autonomously perform} all necessary action to interact with data.
\end{enumerate}

The action space of Smart Things is constantly expanded and we cannot enumerate everything here.
However, all of these developments can be understood as the Analytical Environment of a DT. 
Research has covered many approaches to provide the Analytical Environment needed to transform data into insights (see Table~\ref{tab:iot:reserach:ongoing}). 

\begin{specialtable}[!htbp]
	\caption{Research Areas for Digital Twin Analytics} 
	\centering
	\resizebox{\columnwidth}{!}{
		\begin{tabular}{rl*{2}{c}}
			\toprule
			Domain & Description & Review\\
			\midrule
			Ontologies& Make semantic sense of the data and produce knowledge and context&~\cite{song2010semantic,hachem2011ontologies,wang2012comprehensive,szilagyi2016ontologies,de2020context,knappmeyer2013survey,perera2014context}\\[5pt]
			Middleware & Infrastructure to identify, process, store, secure and present information&\cite{ngu2017iot,razzaque2016middleware,mineraud2016gap,fremantle2017survey,almusaylim2019review}\\[5pt]
			Data preprocessing & Data aggregation and data fusion &~\cite{rajagopalan2006data, wang2011networked, li2013compressed, zhang2010outlier,alsheikh2014machine,bagaa2014data, gravina2017multi,al2017information}\\[5pt]
			Data mining & Cloud-based Big Data analysis on unstructured data&\cite{bin2010research, tsai2014data, ganz2015practical, chen2015data,ahmed2017role,shadroo2018systematic,ge2018big, hadi2018big}\\[5pt]
			AI\&ML& Machine learning and deep learning &~\cite{kulkarni2011computational,alsheikh2014machine,al2020survey,atitallah2020leveraging}\\[5pt]
			Blockchain & Security, trust, and privacy through smart contracts &\cite{christidis2016blockchains,zheng2017overview,dorri2017blockchain,khan2018iot,mistry2020blockchain}\\[5pt]
			HCI&Human-Computer Interaction for improved access&~\cite{kranz2010embedded,koreshoff2013approaching,atzori2011siot,atzori2014smart,chatain2020digiglo,genay2021being,pei2022hand}\\[5pt]
			ABM&Agent-based modelling  Objects&\cite{vinyals2011survey,wu2014cognitive,kataria2021towards,gath2020cogarch}\\
			\bottomrule
			\label{tab:iot:reserach:ongoing}
		\end{tabular}
	}
\end{specialtable}


\subsection{The Virtual Environment of the Digital Twin - Access, Interaction and Representation through the Fused Twins Paradigm}


The last environment we discuss provides the Human-Computer Interaction~\cite{jacko2012human,helander2014handbook} with the DT. 
When implementing the FTs paradigm for a DT, the main focus lies on AR interactions with the DT.
AR is only beginning to arrive in the world of SOs, the IoT, remote sensing, and DTs but has already been accepted as a crucial feature in the near future~\cite{chettri2019comprehensive}.
AR is of interest for FTs as it explicitly supports embodied interaction: interaction that is not agnostic to its physical and social context~\cite{dourish2004action,gervais2016introspectibles}.
AR can also reduce negative split-attention effects~\cite{mayer1998split,sweller2011split}, and appropriate visualisations of data can lower cognitive load~\cite{scaife1996external} and enhance cognition~\cite{willett2021superpowers}.
The expectation is that AR will offer a new avenue to simplify interaction with SOs, the IoT, and DTs~\cite{gruebel2021fused}.
The FTs paradigm is a formalisation of the interaction between AR and DTs in a situated context.

Because the combination of technologies is only emergent now, most published material towards this end is either a prototype~\cite{jo2016ariot,park2019iot,hua2021arciot,prouzeau2020corsican,gruebel2021fused} or a proposal for future research directions~\cite{michalakis2018visualizing,carneiro2018bim,jo2019ar,chen2021development}.
The main reason is that applying AR on top of the already complex combination of SO, IoT, remote sensing, and DT stacks requires many software systems to run smoothly that are currently only being prototyped.

The Fusion of the the digital representation and the physical entity lies at the core of the FTs paradigm~\cite{gruebel2021fused} but also of the more general Situated Analytics~\cite{thomas2018situated}.
Interestingly, other research on the topic of Situated Analytics has often implemented FTs to some degree but without the explicit mentioning of the DT aspect~\cite{jo2016ariot,park2019iot,hua2021arciot}. 
A notable exception is the `Corsican Twins''~\cite{prouzeau2020corsican} implementation of a FTs platform with an explicit DT powering the Situated Analytics.
Short papers on future research directions in Situated Analytics clearly point towards FTs-like implementations~\cite{michalakis2018visualizing,carneiro2018bim,jo2019ar,chen2021development}. 
We summarise ongoing and future research that is already covering a wide variety of topics (see Table~\ref{tab:fused:twins:instances}).
Some of the research explicitly mentioned the DT Approach or an Situated Analytics approach but often simply uses the available technology to demonstrate that their problem can be addressed~\cite{madsen2013interactive,chen2021development, carneiro2018bim}, implicitly developing a DT and applying Situated Analytics.

\begin{specialtable}[!htbp]
	\caption{Recent and Ongoing Research Applying the Fused Twins Paradigm} 
	\centering
	\setlength\extrarowheight{2pt}
	\resizebox{\columnwidth}{!}{
		\begin{tabular}{*{2}{r}*{5}{c}}
			\toprule
			&&\multicolumn{5}{c}{Environments}\\
			\cmidrule(lr){3-7}
			Application Type&Year&Physical&Data&Analytical&Virtual&Connection\\
			\midrule
            Infrastructure Management~\cite{schall2009handheld} & 2009 & * & \checkmark & * & \checkmark & -  \\[5pt]
			City Information~\cite{white2009sitelens}           & 2009 & * & \checkmark & \checkmark & \checkmark & -  \\[5pt]
			History~\cite{madsen2013interactive}                & 2013 & * & \checkmark & - & \checkmark & -  \\[5pt]
			City Information (concept)~\cite{xu2014building}    & 2014 & * & * & * & * & *  \\[5pt]
			Shopping~\cite{elsayed2015situated}                 & 2015 & * & \checkmark & \checkmark & * & -  \\[5pt]
			IoT device interaction~\cite{jo2016ariot}           & 2016 & \checkmark & \checkmark & - & \checkmark & *  \\[5pt]
			City Information~\cite{zollmann2016visgis}          & 2016 & * & \checkmark & * & \checkmark & -  \\[5pt]
			Shopping~\cite{abao2018design}                      & 2018 & * & \checkmark & \checkmark & * & -  \\[5pt]
			Smart Campus~\cite{dave2018framework}               & 2018 & \checkmark & \checkmark & \checkmark & \checkmark & \checkmark \\[5pt]
			Evacuation~\cite{lochhead2018communicating}         & 2018 & * & - & - & \checkmark & -  \\[5pt]
			Medicine~\cite{pratt2018through}                    & 2018 & \checkmark & \checkmark & - & \checkmark & -  \\[5pt]
			Facility Management~\cite{stojanovic2018towards}    & 2018 & \checkmark & \checkmark & \checkmark & \checkmark & - \\[5pt]
			Rehabilitation~\cite{lee2019novel}                  & 2019 & \checkmark & \checkmark & * & \checkmark & -  \\[5pt]
			IoT device visualisation~\cite{park2019iot}         & 2019 & \checkmark & - & - & \checkmark & -  \\[5pt]
			Seat Selection~\cite{guarese2020augmented}          & 2020 & * & \checkmark & \checkmark & * & -  \\[5pt]
			Equipment Maintenance~\cite{mourtzis2020real}       & 2020 & * & \checkmark & - & \checkmark & -  \\[5pt]
			In situ visualisation~\cite{prouzeau2020corsican}   & 2020 & * & \checkmark & * & \checkmark & *  \\[5pt]
			Education~\cite{vidal2020analysis}                  & 2020 & \checkmark & \checkmark & \checkmark & \checkmark & *  \\[5pt]
			Training \& Assistance~\cite{vidal2020creating}     & 2020 & * & * & - & \checkmark & *  \\[5pt]
			Firefighting~\cite{chen2021development}             & 2021 & - & - & \checkmark & \checkmark & -  \\[5pt]
			General~\cite{gruebel2021fused}                     & 2021 & \checkmark & \checkmark & \checkmark & \checkmark & *  \\[5pt]
			IoT device interaction~\cite{hua2021arciot}         & 2021 & \checkmark & \checkmark & \checkmark & \checkmark & - \\[5pt]
			General~\cite{fleck2022ragrug}                      & 2022 & \checkmark & \checkmark & \checkmark & \checkmark & *  \\[5pt]
			\bottomrule
			\multicolumn{1}{l}{Note: }&\multicolumn{6}{r}{* Partially implemented or insufficient information}
			\label{tab:fused:twins:instances}
		\end{tabular}
	}
\end{specialtable}

Since the FTs concept in particular and DT concept in general is not well defined yet, we analysed the literature for traces of it.
To construct Table~\ref{tab:fused:twins:instances}, we analysed the papers for the presence of implementations of any of the five components. For the Physical Environment, we considered it as fully implemented (i.e., we gave it a check-mark [\checkmark] in Table~\ref{tab:fused:twins:instances}) if active sensor data collection was reported, or we considered it as partially implemented (i.e., we gave it a star [*] in Table~\ref{tab:fused:twins:instances}) if passive sensors or markers were reported (e.g., QR code stickers).
For the Data Environment, we considered it fully implemented when explicit data collection and storage was reported (e.g. SQL databases or BIM models), and we accepted partial implementation if data was streamed but not permanently stored.
For the Analytical Environment, we considered it fully implemented if there was an element of only analysis, and we considered it partially implemented if simple data transformation were performed.
For the Virtual Environment, we considered it fully implemented if some digital model was reported (e.g., 3D models of rooms or machines) or partially implemented if a digital representation could be inferred but was not explicitly reported (e.g., spatial representation of labels).
For the Connection Environment, we considered it fully implemented if it was possible to add or compose elements in different environments such as adding new sensors, changing the virtual environment, or using different analysis and partially implemented if it was theoretically extendable or if the reports hinted to extendability of a single component (e.g. adding more sensors).


Across all partial implementations of the FTs paradigm, we find the recurrent theme of how to tightly couple the PT and the DT.
All DTs require an effective representation of the PT to convey meaningful information to the user.
For the built environment, this commonly involves Building Information Models (BIM)~\cite{dave2018framework,carneiro2018bim,tang2019review,gruebel2021fused,chen2021development}, or City Information Models (CIM), and Geographic Information System (GIS)~\cite{carneiro2018bim, xu2014building}.
In a city context, usually, BIM/CIM is used for the local model, whereas GIS provides a global reference frame.
However, if BIM/CIM is not available, remote sensing offers methods to produce placeholder models~\cite{stojanovic2018towards}.
Both GIS and BIM/CIM have a long research tradition and have been used in a variety of ways in industry, academia, and governance.
It could be argued that they provided DTs before the word was created.
However, in the DT literature, a consensus has built up that they rather provide a digital model~\cite{schluse2018experimentable,sepasgozar2021differentiating} and are at best the basis for a DT. 

The reason that these opinions have developed is that BIM/CIM and GIS are often seen as a rather static representation of DTs and that, for dynamism, it is required to continuously measure from sources such as IoT~\cite{jo2016ariot,carneiro2018bim,gruebel2021fused,chen2021development}.
These measurements are often called the Digital Shadow~\cite{schluse2018experimentable,sepasgozar2021differentiating}.
Additionally, the raw data provided by sensors is often classified as Big Data~\cite{ang2016big,boubiche2018big}. Visualising the data becomes nearly impossible without
analytical tools.
This highlights the strong interaction between the Analytical Environment and the Virtual Environment.
At the same time, a DT is characterised by the bi-directional interaction between the PT and DT. Therefore, there is an expectation for actuators~\cite{jones2020characterising} to be able to be triggered from within the Virtual Environment to implement the full functionality of a DT.

Beyond an accurate model, data acquisition, and actions, FTs require a mapping between the virtual world and the physical world.
Many approaches are available, and they commonly identify features in the physical world that can be used as anchors for the virtual world. 
These location services are based on different technologies such as Bluetooth-based~\cite{hua2021arciot}, image-marker-based~\cite{prouzeau2020corsican}, vision-based~\cite{jo2016ariot, gruebel2021fused}, or signal-based~\cite{park2019iot}.
Consequently, it is possible to display content in the Virtual Environment as an overlay in physical world.
Depending on the medium, it is possible to interact with the augmented content through poses, gestures, mid-air interfaces, tangibles, touch, voice, face, or gaze~\cite{zhou2008trends,nizam2018review,kim2018revisiting}.
Immersive and Situated Analytics~\cite{marriott2018immersive,thomas2018situated} provide the research field in which most of the visualisations that will be used in FTs will be developed. 

\begin{figure}[htbp!]
	\centering
	\includegraphics[width =\columnwidth]{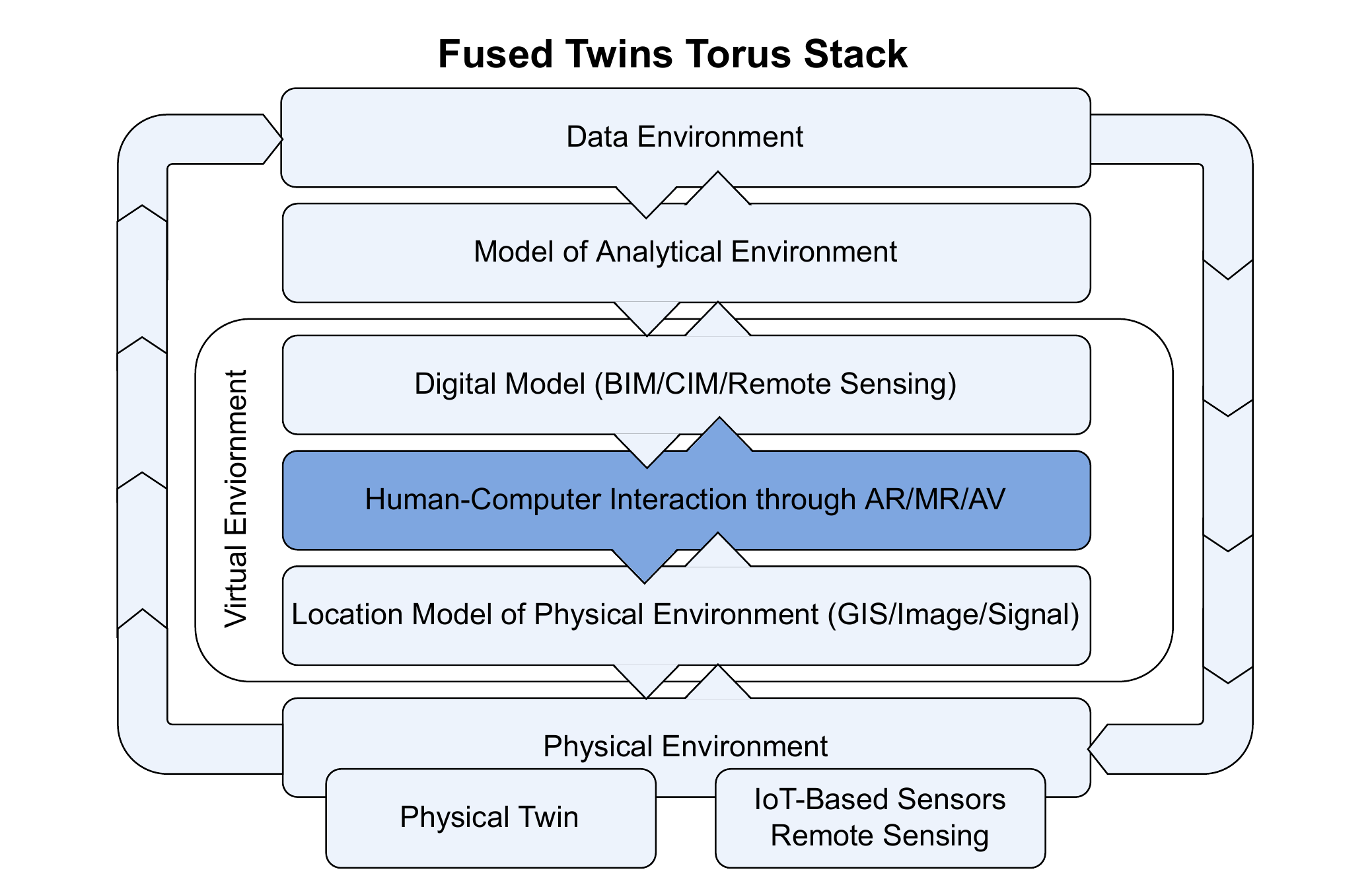}
	\caption{\textbf{Fused Twins Torus Stack}.
		Generalised technology stack for the FTs Paradigm based on~\cite{gruebel2021fused}. The stack implements a full DT and expands the Virtual Environment to accommodate the FTs. Data flow is shown through the arrows. The stack becomes a torus because the chain of technologies and environments requires circular links of the data. Changes to a DT in the Virtual Environment flow back to the PT through the Data Environment. Generated with \url{http://draw.io}.}
	\label{fig:fused:stack}
\end{figure}

Implementing FTs requires some expansion of an existing DT. In the Virtual Environment, services for the digital model, the Human-Computer Interaction, and the location model must be integrated to be able to effectively display the measurements from the IoT and remote sensing in situ. The services first locate data in relation to the BIM/CIM~\cite{dave2018framework,tang2019review} and then locate the BIM/CIM in the physical world through Computer Vision and GIS~\cite{liu2017state}. The result is displayed in AR (see Figure~\ref{fig:fused:stack}) and offers Situated Analytics to the observer. 

\section{Discussion}
\label{sec:discussion}

To understand FTs, we have reviewed three distinct but connected literatures.
First, this review has focused heavily on how to understand DTs today.
The fuzziness around their definition has contributed to the continued vagueness of nearly any ICT being labelled or associated with IoT and DTs.
However, with the rising practicability of DTs, no aspect of life will be beyond digitalisation for good and bad.
Second, we looked at AR technology as a mean to interact with information and Situated Analytics in particular to cover the interaction with a DT in situ of its PT. This combination is forming the basis to fuse the twins.
Third, we explored the tightly connected topic of Smart Cities which often has been a canvas for testing ICT, IoT, DT, and AR applications. We expect the FTs concept to be applied often in this context, even under a different name or no name at all.
We have enumerated a large number of FTs applications that partially qualify already today to give an idea of what is to come.

\subsection{Smart Cities Discussion}
\label{sec:discussion:smart:city}

Smart Cities are still a far way from being smart. In particular, because most ``smart services'' are not accessible.
DTs offer an organisational paradigm that could make the complex Analytical Environment of cities accessible by flattening the hierarchy from the user perspective.
The DT abstracts the heterogeneous technological structure of a city into semantically meaningful units that can be queried for information extracted from the underlying networks.
At the same time, the DT also offers an opportunity to overcome the heterogeneity of sources from IoT to remote sensing without loss of accuracy.
Lastly, DTs building on top of heterogeneous sources can represent abstract concepts that have no physical counterpart such as the livability and sustainability of a city \cite{helbing2021ethics} or the spatial movement of traffic jams \cite{gazis1992moving}.

A fully implemented DT would impact the management, planning, and use of a city.
It would allow its economy to become more efficient, markets to be more transparent, processes to be explorable and problems possibly to be more addressable.
If a DT of a city is made accessible to the general public, it could even serve as a participatory tool and integrate the city's population.
At the same time, concerns of privacy and security won't be easily resolved and a delicate balance between gains and losses has to be struck to realise DTs of cities.

\subsection{Situated Analytics Discussion}
\label{sec:discussion:situated:analyitcis}

AR combines real and virtual elements registered in three dimensions and provides interaction in real-time.
Despite decades of development, immersive analytics are still in their infancy, and no standard user interfaces have been established~\cite{fonnet2019survey, bressa2021s}.
Recent reviews have further found that, for Situated Analytics, there is still a lack of an overarching definition, data representation is not unified, and the target audience is not defined~\cite{bressa2021s}.
Broad adoption is only now within reach, and the future will bring many new developments.

Interestingly, the description of how to assemble a Situated Analytics system maps closely to our layout of a DT (see Figure~\ref{fig:situated:analytics}).
There is a synergy between AR and the FTs concept: the FTs concept offers one of the most promising applications of AR, while AR supports an implementation of FTs that reduces split-attention effects and cognitive load.
In this context, the FTs paradigm combining a DT of Smart Cities with Situated Analytics in AR integrates the physical and social aspects of the DT and thus aligns with the theories of embodied interaction, embodied cognition, situated cognition, and socially situated cognition~\cite{dourish2004action,wilson2013embodied,brown1989situated,smith2004socially}.
Therefore, we argue that the FTs concept may become the largest application of Situated Analytics to be seen, used, and even embodied, by most people on the planet.

\subsection{Digital Twins Discussion}
\label{sec:discussion:digital:twin}

We have reviewed the history of the DT to disambiguate the concept from its cognates.
There is a fine line between DTs and CPSs, M2M, WSANs, SOs, Ubiquitous Computing, IoT, Computation-on-the-wire, and Remote Sensing.
In the context of this review, we have shown that most cognates are actually underlying technological approaches or perspectives that constitute components of the DT.
The sole exception is a CPS which is a more general approach with the DT being a specialised instance thereof.
We believe that DTs have the potential to subsume current cognates similar to how the IoT has been previously considered the most encompassing term in this area.
At the same time, each cognate carries an important meaning that is worth maintaining, and we strongly recommend to understand these differences in order to gain a deeper understanding of DTs.

Most DTs ``in the wild'' are prototypes that were created as novel approaches to particular problems but whose theoretical background is underdeveloped.
Indeed, DTs lack theoretisation beyond the assembly of the system. 
This is beginning to change with the theoretical work of \citeauthor{tao2018digital} that first developed a more sound differentiation of DT components. 
We follow the initial ideas from \citeauthor{tao2017digital} but diverge in some details.
We generalise the theoretisation into five general components (see Figure~\ref{fig:digital:twin:components}) that we rename the Physical, Data, Analytical, Virtual, and Connection environments of the DT. 
Ideally, these environments work in isolation and are interoperable to enable the composition of DTs on the level of components as well as nesting.
Furthermore, our definition covers most DTs and DT-like applications by assuming partial or implicit implementations of its components.
Throughout this review, we have gathered evidence that cognates, other DTs models, Smart Cities, and even Situated Analytics have theoretical correspondences to the components in our DT model.

At the same time, it is difficult to assemble the different environments.
The servicisation of IoT has provided an interesting foundation upon which DTs could be quickly assembled with very little overhead.
However, Iaas, PaaS, and SaaS are not sufficiently separate in the real world and, under current implementations, would require a Connection Environment that leans towards a hypervisor to compensate for the unreliable interfaces between services.
It would be preferable to develop DTs and their components in a more standardised way via isolated and interoperable environments.
If  Iaas, PaaS, and SaaS were optimised to strictly map to DT components, then DT components ``as a service'' could democratise the use of DTs by enabling people to quickly bring DT services together.
Two open questions that remain are whether the services should be provided in a centralised or decentralised manner and whether the services are provided as a government service or a market product.
These services could be more accessible with an open standard for DTs that clearly define how the different components interact.

\subsection{Fused Twins Discussion}
\label{sec:discussion:fused:twins}

Without appropriate Human-Computer Interaction, the data in the DT will remain insurmountably complex and thus incomprehensible.
The composability of DTs underpinned by interoperable IoT and remote sensing will enable a smoother interaction between people and the environment.
In this context, the FTs concept will allow for an intuitive interface to access data, information, knowledge, and understanding where it is most needed and useful.
The FTs Paradigm attains this with AR/MR/AV interaction between the DT and the PT to identify opportunities for simple interactions on the complex data \emph{in situ} where the data is generated and collected.
It is paramount to align the user interfaces, data formats, general interoperability, and creative services that eventually converge towards real Smart Cities for its users.

The idea underlying FTs has been around for some time, albeit under a different or no specific name.
DTs and their cognates Ambient Intelligence and Smart Objects, as well as sometimes the more general Things, can all be understood as different embodied FTs.
At the same time, FTs and Situated Analytics are very similar, which opens the question of whether a new term adds clarification.
We do believe that a new term is necessary because we are currently observing the confluence of two topics, and the new terms expand beyond each individual conceptualisation.
These two topics are the aforementioned research in AR that does not necessarily rely on DTs (but may possibly implement FTs without mentioning it) and the expanding applications of DTs in nearly every research area and field.
The necessity to visualise the DT, and the capability to do it in augmented space rather than on a display, ushers in a new era in which data can be made visually and spatially accessible at the time of creation and across its consumption.
FTs play a special role in this context where the data is fused to its spatial context which differs from classical AR that merely uses the real world space as an empty canvas devoid of any context (in most cases). 

FTs are not unique to Smart Cities, but they are especially applicable in Smart Cities.
Some application areas for FTs include healthcare and industry.
For example, in healthcare, human DTs have been proposed to improve individual diagnosis and care.
Monitoring patients with FTs allow practitioners to locate medically relevant information on the human body.
This may be especially helpful in the context of surgery during which representations of bones and other tissues may be superimposed to facilitate the surgeon's decision-making.
In an industrial context, the inner workings of machines can be overlaid onto the real machines and augmented with analytical information.
While healthcare and industrial application are most useful for specialists, FTs in a Smart Cities context may be an accessible form of FTs for the average citizen.
For example, the augmented reality maps from Google already embed navigation information in the physical city and arguably employ a DT in the background, making it a simple and accessible example of FTs.
Similarly, in the context of tourism, museums have created DTs of specific exhibits but could be expanded to included the entire exhibition to create an FTs.
In the future, FTs could also be expended to communicate city planning initiatives to the public.
 


\section{Conclusion}

In this review paper, the relevant technologies and terminology behind the FTs concept for interacting with the DT of Smart Cities have been presented.
Suitable examples of FTs applications across different literatures have been highlighted (see Table~\ref{tab:fused:twins:instances}.
Nonetheless, a stable and applicable out-of-the-box solution to fulfil the promised FTs for Smart Cities remains to be developed.
DTs are also remain to be well-specified, which may cause delays in developing FTs.
Here, we provided a thorough review of DTs and their cognates to address open points and a concrete definition that can be applied in retrospect to nearly all DTs.
Compared to previous research that employed dashboards and other interfaces for Non-situated Analytics, FTs represent a special application of the Virtual Environment of a DT to provide Situated Analytics.
The key technologies behind the FTs Paradigm for DTs are IoT, remote sensing, BIM/CIM, GIS, and Location services.
The IoT, remote sensing, BIM/CIM, and GIS are used to assemble a DT and provide analytics, and Location services are used to place the Situated Analytics in the PT.
Each of these technologies on its own is well-developed, but only now are they being combined into prototypes that can be classified as FTs.
While many researchers have theorised about similar systems that could be built, few FTs full-fledged systems have been implemented.
Given the plurality of fields that have converged on this general idea, we expect that the FTs paradigm will be a common theme in future technologies and may eventually be as common as a smartphone today.


\vspace{6pt} 



\authorcontributions{Conceptualisation, J.G.; methodology, J.G.; validation, J.G., T.T.; investigation, J.G., L.A., M.M., J.C.; resources, J.G.,C.H.,V.S.,R.S.; writing---original draft preparation, J.G.; writing---review and editing, J.G., T.T., L.A., J.C., M.M., V.S., R.S., C.H.; visualisation, J.G., M. M., J.C.; supervision, V.S., C.H., R.S; project administration, J.G., V.S., C.H.; funding acquisition, T.T., V.S., C.H., J.G. All authors have read and agreed to the published version of the manuscript.
}

\funding{This research was funded by ETH Zurich grant number ETH-15 16-2 and J.G. was supported by an ETH Zurich Doc.Mobility Fellowship.}

\institutionalreview{Not applicable}

\informedconsent{Not applicable}

\acknowledgments{The authors thank Johanna Beyer and Zhu-Tian Chen for the feedback on the AR section. All diagrams were generated with templates from \url{http://draw.io}. And don't forget your towel.}

\conflictsofinterest{The authors declare no conflict of interest.} 

\abbreviations{The following abbreviations are used in this manuscript:\\

\noindent 
\begin{tabular}{@{}ll}
IoT & Internet of Things\\
DT & Digital Twin \\
PT & Physical Twin \\
FTs & Fused Twins \\
XR & Extended Reality\\
CPS & Cyber-Physical System\\
UC & Ubiquitous Computing\\
SO & Smart Object\\
WSN & Wireless Sensor Network\\
SN & Sensor Network\\
SAN & Sensor and Actuator Network\\
WSAN & Wireless Sensor and Actuator Networks\\
MR & Mixed Reality\\
AR & Augmented Reality\\
VR & Virtual Reality\\
AV & Augmented Virtuality\\
IaaS & Infrastructure as a Service\\
PaaS & Platform as a Service\\
SaaS & Software as a Service\\
SenaaS & Sensing as a Services\\
SemaaS & Semantics as a Service\\
RaaS & Robots as a Service \\
RSaaS & Remote Sensing as a Service \\
RFID & Radio Frequency Identification\\
OS & Operating System\\
ICT & Information and Communication Technologies\\
M2M & Machine-to-Machine\\
BAN & Body Area Networks\\
PAN & Personal Area Networks\\
LAN & Local Area Networks\\
WAN & Wide Area Networks\\
LPWAN & Low Power Wide Area Networks \\
BIM & Building Information Model\\
CIM & City Information Model\\
GIS & Geographic Information System\\
UAV & Unmanned Aerial Vehicles\\
\end{tabular}}

\appendixtitles{no} 
\appendixstart
\appendix
\end{paracol}

\reftitle{References}


\externalbibliography{yes}
\bibliography{refs}

\section*{Short Biography of Authors}
\bio
{\raisebox{-0.35cm}{\includegraphics[width=3.5cm,height=5.3cm,clip,keepaspectratio]{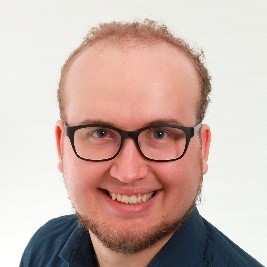}}}
{\textbf{Jascha Gr\"ubel} is a doctoral student in Computer Science at ETH Zurich and a visiting fellow at Harvard University's Visual Computing Group. He is working at the intersection with Cognitive Science and City Science at the Chair of Cognitive Science at ETH Zurich. In his thesis, he is investigating human behaviour in public spaces with DT. In addition, his thesis is focusing on the visualisation of data in VR and AR such as through FTs and the analysis of Big Data as presented in this article. Jascha is also maintaining a framework to run VR experiments for scientific purposes called EVE that is intended to be intertwined in the future with this work to enable augmented reality experiments. This summer, Jascha will start work as a senior researcher at the Center for Sustainable Future Mobility at ETH Zurich on a Digital Twin of the Swiss Mobility System.}

\bio
{\raisebox{-0.35cm}{\hspace{3.5cm}}}
{\textbf{Tyler Thrash} is currently a PhD student in Biology at Saint Louis University. Previously, he was a postdoctoral research in cognitive science and geography at ETH Zurich and the University of Zurich, respectively.  His work has focused on spatial cognition and navigation from a largely ecological/Gibsonian perspective. With this approach, he attempts to explain higher-level cognition (e.g., biases in spatial memory) in terms of lower-level, perceptual processes (e.g., visual exposure to environmental structure). Dr. Thrash also develops simulations for interactive virtual environments and mathematical models for understanding human behaviour. }

\bio
{\raisebox{-0.35cm}{\includegraphics[width=3.5cm,height=5.3cm,clip,keepaspectratio]{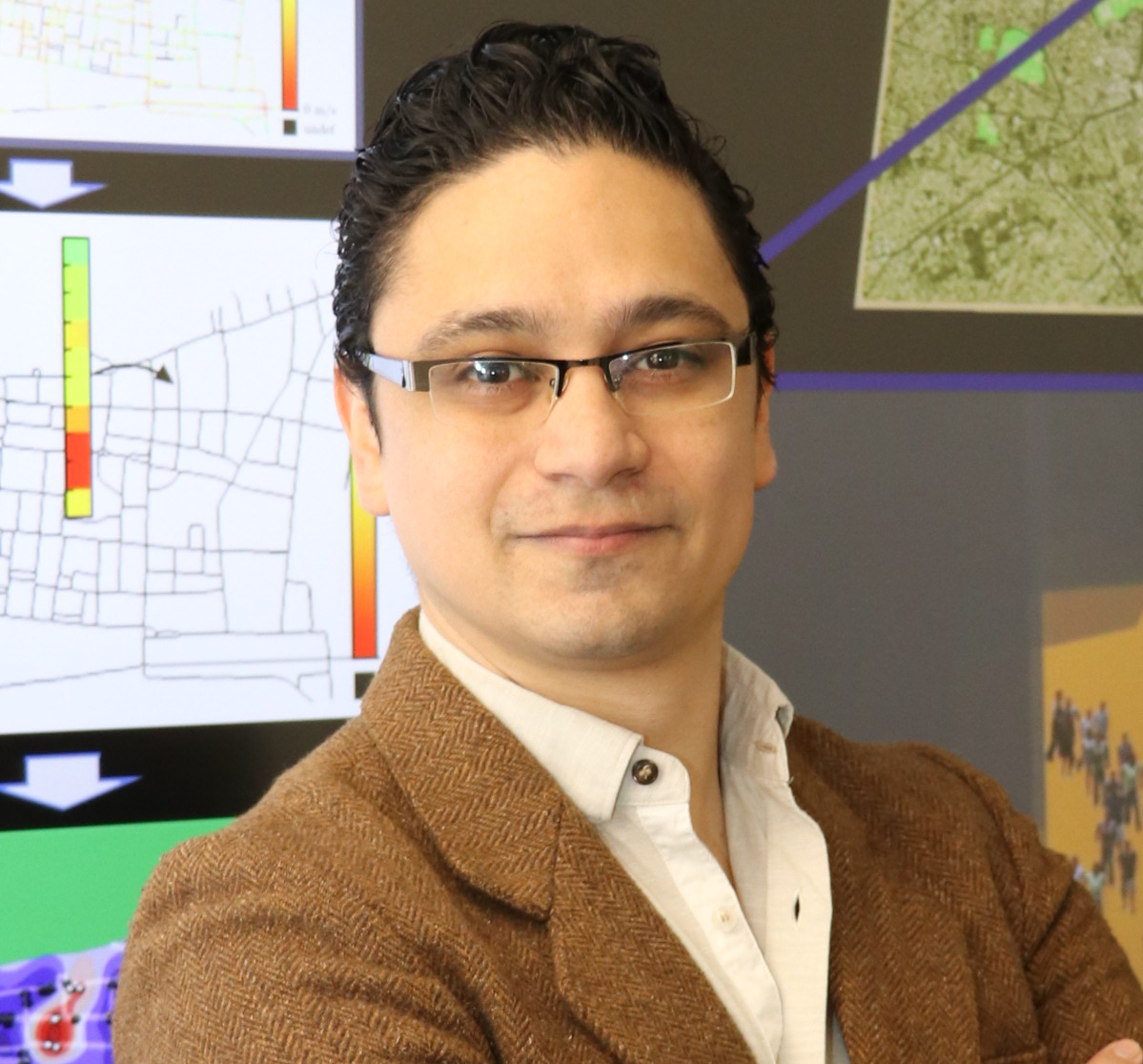}}}
{\textbf{Leonel Aguilar} is a Lecturer and Senior Researcher in Applied Artificial Intelligence at the Data Science, Systems and Services laboratory (DS3), ETH Zurich. He held postdoctoral appointments at the Cognitive Science (COG) and Computational Social Science (COSS) groups, ETH Zurich, and the Earthquake Research Institute of the University of Tokyo, Japan. Leonel obtained his PhD at the Computational Science and High-Performance Computing Laboratory at the University of Tokyo. His research focuses on modelling, simulating, and analysing social phenomena and the development and deployment of AI/ML-based systems to support this. He has worked on quantifying human behaviour through experiments and has leveraged these insights in AI-driven software for applications such as evacuation planning and education.}

\bio
{\raisebox{-0.35cm}{\includegraphics[width=3.5cm,height=5.3cm,clip,keepaspectratio]{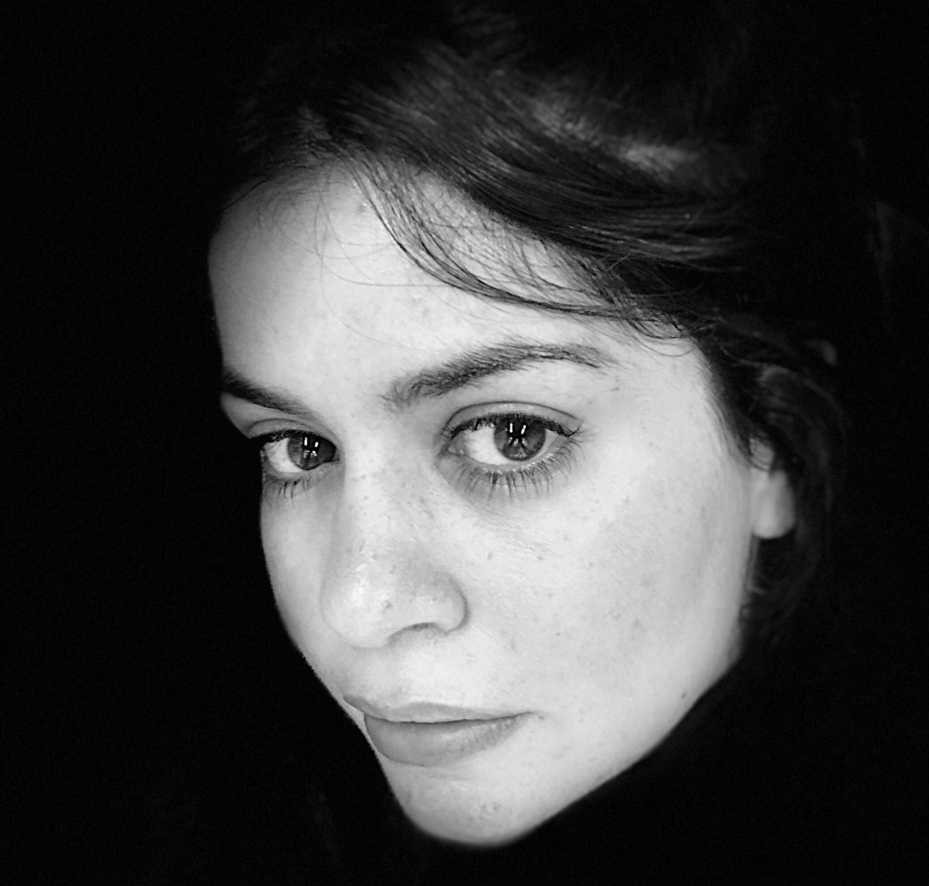}}}
{\textbf{Michal Gath-Morad} is an SNSF postdoctoral fellow at the University of Cambridge and at University College London, as well as a lecturer at the Chair of Cognitive Science at ETH Zurich. Michal is an architect by training and a cognitive scientist with a doctoral degree from ETH Zurich.  Michal's interdisciplinary research and teaching seeks to develop, critically evaluate, and creatively experiment with methods of evidence-based co-design in architectural design pedagogy and praxis to promote health and well-being by architecture.}

\bio
{\raisebox{-0.35cm}{\includegraphics[width=3.5cm,height=5.3cm,clip,keepaspectratio]{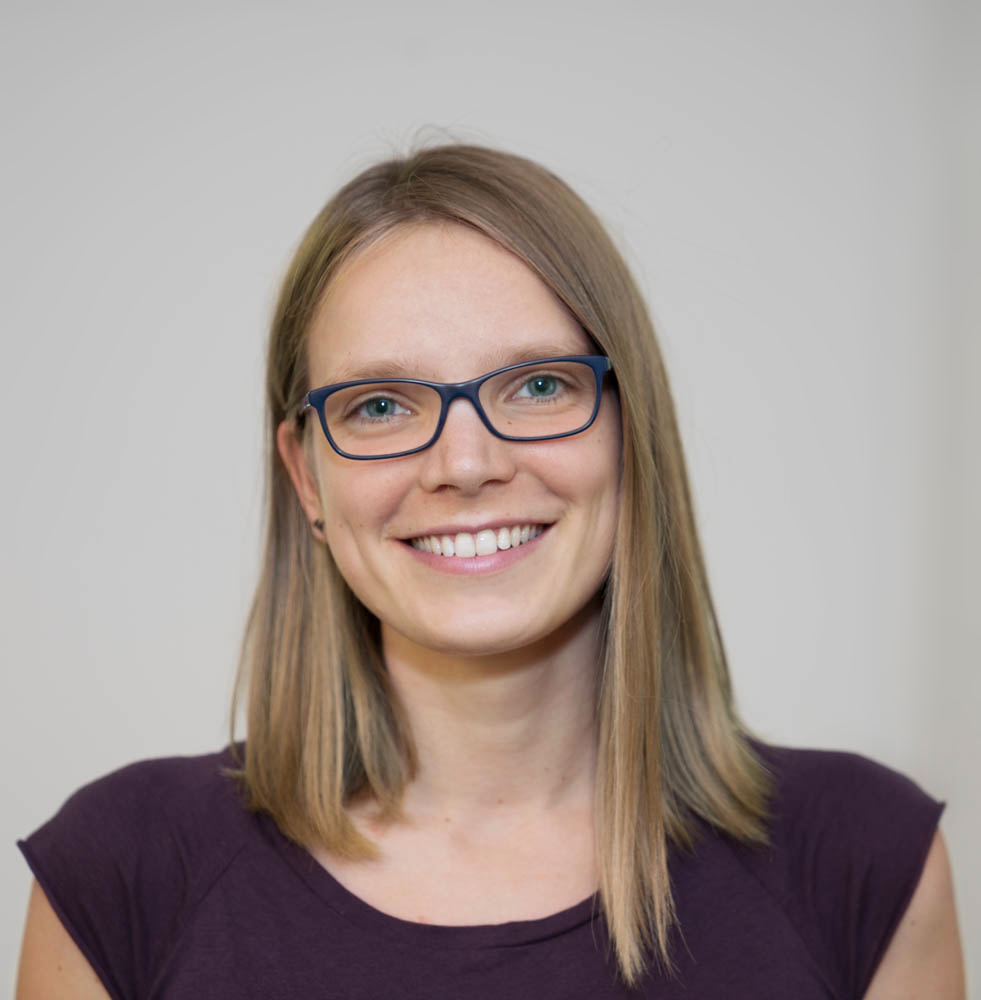}}}
{\textbf{Julia Chatain} is a doctoral student at ETH Zurich with the ETH Game Technology Center and the Professorship for Learning Sciences and Higher Education. Her work focuses on embodiment, and, in particular, the role of embodied interaction and embodied concreteness in learning abstract concepts and making mathematics graspable. In this context, she explores AR and VR as means to implement embodied interaction supporting embodied cognition. Her work is supported by the Future Learning Initiative of ETH Zurich.}

\bio
{\raisebox{-0.35cm}{\includegraphics[width=3.5cm,height=5.3cm,clip,keepaspectratio]{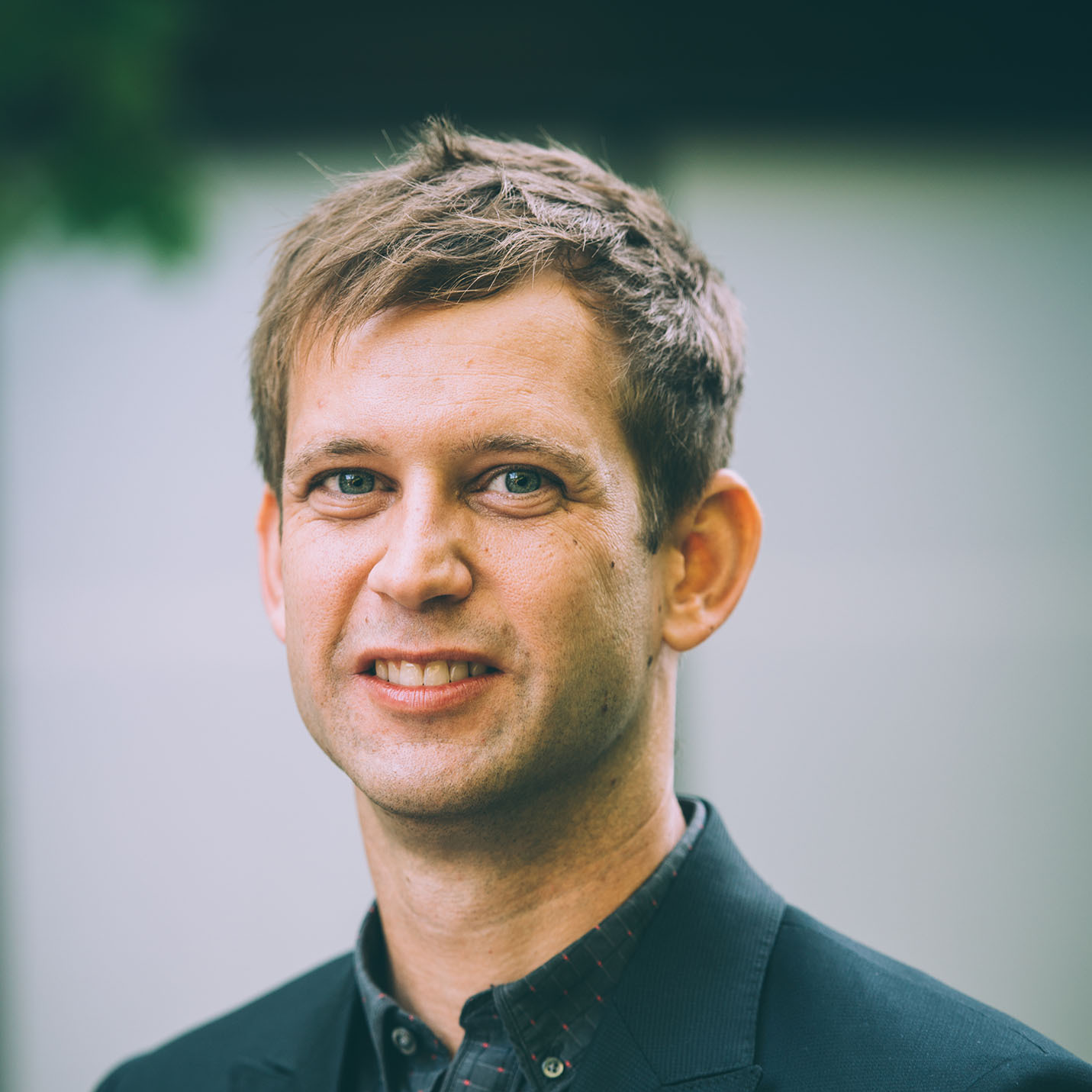}}}
{\textbf{Robert W. Sumner} is the Director of Research and Development at the Walt Disney Studios and an Adjunct Professor at ETH Zurich. Robert received a B.S. degree in computer science from the Georgia Institute of Technology and his M.S. and Ph.D. degrees from the Massachusetts Institute of Technology. At ETH, Prof. Sumner teaches a course called the Game Programming Laboratory in which students from ETH and the Zurich University of the Arts work in small teams to design and implement novel video games. In 2015, Prof. Sumner founded the ETH Game Technology Center, which explores the unique way game technology can advance ETH’s mission in research, education, and outreach. Prof. Sumner was featured on BBC Click and Ars Technica for his work on Unfolding the 8-Bit Era as well as Reuters for his augmented reality research.}

\bio
{\raisebox{-0.35cm}{\includegraphics[width=3.5cm,height=5.3cm,clip,keepaspectratio]{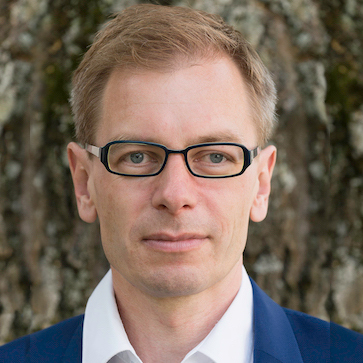}}}
{\textbf{Christoph H\"olscher} is Full Professor of Cognitive Science in the D-GESS at ETH Zurich since 2013, with an emphasis on Applied Cognitive Science. Since 2016 Christoph is a Principal Investigator at the Singapore ETH Center (SEC) Future Cities Laboratory, heading a research group on ‘Cognition, Perception and Behaviour in Urban Environments’. Christoph is the Program Director of Future Resilient Systems FRS at the SEC since 2019, leading the current FRS 2 phase (2020-2025). He holds a PhD in Psychology from University of Freiburg, served as honorary senior research fellow at UCL, Bartlett School of Architecture, and as a visiting Professor at Northumbria University Newcastle. Christoph has several years of industry experience in Human-Computer Interaction and usability consulting.}

\bio
{\raisebox{-0.35cm}{\includegraphics[width=3.5cm,height=5.3cm,clip,keepaspectratio]{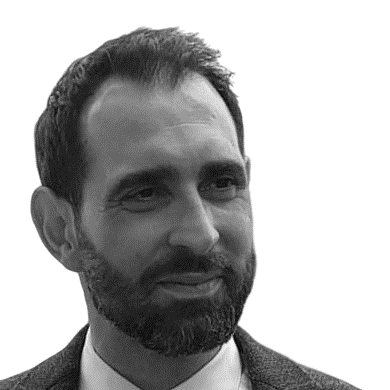}}}
{\textbf{Victor R. Schinazi} is Assistant Professor at the Psychology Department of Bond University and PI of the Early Detection of Health Risks and Prevention module in the Future Health Technologies programme in Singapore. Previously, Victor was a Senior Lecturer at the Chair of Cognitive Science and the GeoGazeLab at the ETH Zurich. He also served as co-PI for the Cognition, Perception and Behaviour in Urban Environments project at the Future Cities Laboratory in Singapore. Victor holds a PhD from the Centre for Advanced Spatial Analysis (CASA, UCL). He also was a Postdoctoral Fellow at the Center for Cognitive Neuroscience (UPenn) and served as Chief Science Officer for Strategic Spatial Solutions, Inc. Victor has published high impact articles and book chapters in psychology, cognitive neuroscience, geography and computer science.}


\end{document}